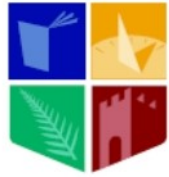

NUI MAYNOOTH

Ollscoil na hÉireann Má Nuad

# A Framework for Enabling Distributed Applications on the Internet

by

Mark Anthony McLaughlin

A thesis presented for the degree of

## Master of Engineering Science

National University of Ireland, Maynooth

Department of Electronic Engineering

October 2006

Head of Department:     Dr. Frank Devitt
Supervisors:            Dr. Tomás Ward
                        Thomas J. Naughton

# Abstract

The last five years have seen the rapid rise in popularity of what we term internet distributed applications (IDAs). These are internet applications with which many users interact simultaneously. IDAs range from P2P file-sharing applications, to collaborative distributed computing projects, to massively multiplayer online games (MMOGs). Currently, there is no framework that combines IDAs collectively within a single context. We provide a basis for such a framework here.

In considering IDAs collectively, we found that there was no generic description that had been applied to them as a group. We have therefore put forward such a description here. In our description, IDAs are functionality separated into three logic layers, which are designed and built individually. Each layer is represented by functionality on the software client running on each participating computer, which together comprise the overall IDA.

The core contribution of this work is a framework, called the Internet Distributed Application Framework (IDAF), which outlines how IDAs can be designed, built and run. The IDAF outlines a set of constraints that each implementing software system must abide by. To verify the IDAF, we have built a system prototype implementation called the Internet Distributed Application System (IDAS). The IDAS includes an implementation of the IDAF layer model, which specifies IDAs are built. The IDAS also includes a generic software client that is capable of simultaneously running and managing arbitrary IDAs. We provide sample IDAs and demonstrations to verify both that the IDAF is implementable and that the IDAS is a workable usable system.

## Declaration

I hereby certify that this material, which I now submit for assessment on the programme of study leading to the award of Masters in Electronic Engineering Science is entirely my own work and has not been taken from the work of others save to the extent that such work has been cited and acknowledged within the test of my work. This thesis contains no material that has been accepted for the award of any other degree or diploma in any university.

Mark McLaughlin

# Dedication

I would like to thank Tomás and Tom for giving up their time to supervise me for this masters project. I would doubly like to thank them for correcting my thesis as it evolved over many months. In doing so, they helped me to bring my technical writing skills up to the required standard.

I would also like to express my appreciation to the institution that is NUI Maynooth for giving me the opportunity to pursue further study, and the Electronic Engineering dept. (and the fine people that staff it) for supplying me with the resources that I needed.

Aside from that, there was only one person who lost sleep over this, and I will be rewarding him personally with a well deserved holiday very soon!

# Table of Contents

# Chapter 1   Introduction

## 1.1  Motivation

The initial impetus for this project arose out of the question, "how can I facilitate the development and proliferation of applications that run across many computers on the internet, in which all computer users can participate?" I termed the applications in question, 'internet distributed applications' (IDAs)[1].

As with other research projects, I was also keen to investigate new ways of harnessing the untapped resources of the multitude of computers on the internet in order to provide useful applications and services. This investigation was kept deliberately general: I did not enumerate the resources that we could make use of, and I did not have any particular applications in mind. I wanted to enable as broad a range of applications as possible.

The approach taken was to examine existing applications with a view to establishing commonalities between them, and, if there were any, to arrive at a generic interpretation of an internet distributed application that most existing applications, and other applications I could conceive of, would conform to.

My first task was to draw parallels between technologies that had often been treated as separate since their inception. In some cases, the primary impetus into their development had been given by the academic community (e.g. distributed computing applications, see Section 2.2); in other cases by industry (e.g. massively multiplayer online games (MMOGs), see Section 2.4). Often, few attempts had been made to formalise common nomenclature, or to establish potential generic development platforms that a wider range of applications could be built upon. Even in the area of distributed computing itself, it wasn't until the introduction of BOINC [Anderson '04] that very similar distributed computing applications began to use a common network layer[2] rather than a custom implementation. Ultimately, I came to the conclusion that there is a sufficiently high degree of commonality between existing applications that a generic interpretation of an internet distributed application can be formalised, and subsequently built.

---

1   No suitably general and unambiguous, established term could be found. This term is defined formally in Section 1.5.

The motivation for this work is to investigate the viability of a generic development platform on which a wide range of internet distributed applications could be developed. Such a platform would be extremely useful as an aid to conceiving and developing such applications. Duplication of effort could thus be avoided, as has been shown by BOINC for a class of distributed computing applications.

## 1.2   Objectives and Approaches

The objective of this thesis is to develop a development platform which would facilitate the conception and creation of a wide range of internet distributed applications. Further, it is one of the aims of this work to produce a prototype platform, and sample applications that can be run on this platform, in order to prove the concept. This investigation must necessarily involve a comparison of existing applications in order to identify common building blocks that could be used and re-used to develop as wide a range of applications as possible. It is intended, firstly, that such building blocks are formally identified, so that applications can be described in terms of them; and secondly, that sample applications be built from these building blocks, and run on the developed platform. These applications must be chosen carefully, to show the diversity of applications that can be accommodated.

## 1.3   Contributions

The main contributions of this thesis are a framework, called the Internet Distributed Application Framework (IDAF), which develops a description of internet distributed applications, and a platform for running them; and system called the Internet Distributed

---

2   'network layer' is a term that we use provisionally to encapsulate that functionality common to all SETI-like/BOINC applications, which handles dispatch and retrieval of work units to and from participating clients. We will develop this concept fully in Chapter 4.

Application System (IDAS), which allows applications described by the framework to be built and run. Other contributions are as follows:

- An analysis of internet distributed applications and the commonalities between them.
- A model for the partitioning of IDA development into logic layers of functionality which allows specialist programmers or computer scientists to focus only on the problems that are relevant to them[3].
- An outline of the potential of increased 'edge-peer' participation in the internet's service offering, in terms of a far greater utilisation of computing resources available on the internet, and a greater provision of 'local' services available to ordinary 'peers'. (See Chapter 7.)

## 1.4  Outline

This thesis is organised into seven chapters. Chapter 2 gives background into current developments in the general area of internet distributed computing[4]. Chapter 3 develops the background for the IDAF. Chapter 4 describes the IDAF. Chapter 5 describes the IDAS prototype that was developed. Chapter 6 details examples and demonstrations of the system (and hence the framework). Chapter 7 gives conclusions, describes a vision of an internet with significantly larger edge-peer participation, and discusses possible future work.

## 1.5  Important Definitions

For the sake of clarity, it was deemed necessary to establish definitions for certain core terms as early as possible, as they will be discussed throughout the thesis and so ought to be clear and concise from the start.

### 1.5.1  Internet Distributed Applications

This term was introduced in Section 1.1. We define it formally here as an umbrella term to describe applications which have the following four characteristics:

---

3   This is expressed within the IDAF design in Chapter 4, but the the idea goes beyond any particular application enabling platform

4   We use the term 'internet distributed computing' to describe the area of study concerned with the development of 'internet distributed applications', which is defined in Section 1.5.

1. Applications that are considered to span, or 'run across', multiple computers. That is, applications that have an executable software component on each participating computer which contributes a portion of the functionality of the overall application.

2. Applications that are capable of running on the internet[5], are available over the internet to the public[6] and are suitable for use by the public[7].

3. Applications that offer some benefit to the users or owners of the computers that participate in the hosting and running of the application.[8]

4. Applications that tend to increase in their utility as the number of participating computers increase, and can scale to accept the participation of an arbitrarily large number of computers. (Subject to the vagaries of the topology[9] structure.)

Throughout this thesis I will often use the acronym 'IDA' to refer to internet distributed applications.

It is true that 'service' or 'web service' [Berners-Lee '06], 'system', or 'network', may often be more appropriate terms than 'application' to describe particular IDAs, however these labels have difficulties of their own, and are no less ambiguous. I use 'application' at all times in order to be consistent, except in a few selected instances where clarity demands that technology specific jargon be used.

### 1.5.2   Topology Structure

In order to discuss IDAs in general terms, so that they can be compared and contrasted, I need a term to describe (usually in terms of a graph[10]) how computers are inter-connected in the context of these applications. The term I will use for this description is 'topology structure'.

---

5   This is in view of the various technical network issues associated with the internet as opposed to maintained, administered networks. (e.g. Computers with low uptimes, bad connections, and so on.)
6   Perhaps subject to restrictions (social, organisational or cost of use), but inherently public.
7   Some applications may be suited only for private use on 'trusted' networks due a lack of security features.
8   If the user derives no benefit from the software component running on their computer, then this component can hardly be seen as an 'application' from the user's standpoint. (e.g. In Section 2.2 Condor is ruled out as an IDA on this basis, whereas this does not apply to SETI@home.)
9   The concept of the 'topology', as intended here, is developed in Chapter 4.
10  As in graph theory [Wikipedia '06a]. Topology structures can also be described by a relation (see Section 4.1.1).

The related, but separate term, 'topology', refers to the pluggable design component (developed in Chapter 4) in the IDAF topology layer. 'Topology structures' refer (only) to structural descriptions of collections of connected nodes. As part of its design, a 'topology' specifies a 'topology structure'.

## Chapter 2   Developments in Internet Distributed Applications

We will consider, in this chapter, the most prominent of the IDAs that currently exist, with a view to establishing commonalities between them later on in the thesis (particularly in Chapter 3). These are existing examples of the applications whose creation we would like to facilitate. Although we will use the umbrella term, 'internet distributed applications', to describe them, in the literature they tend, individually, to be classified by their architecture, such as 'client-server' or 'peer 2 peer' (P2P), or by the actual application itself, such as Bittorrent [Bittorrent '06]. In the IDAF, developed in Chapter 4, in which IDAs can be described, such classifications refer only to different 'topologies'[11], or to applications built on those topologies, but not to entirely separate technologies, as is currently the case in the literature.

IDAs come in numerous different forms. Those that are currently popular range from file-sharing applications to games (MMOGs). However, it is our claim in this thesis that the network level functionality of IDAs is similar enough so that a common infrastructure can be built to support them. The development of the IDAF, in Chapter 4, is predicated on this case being valid. ([Singh '06] gives a service oriented 'reference architecture' for the P2P domain that describes P2P applications.) For this reason, we give close scrutiny to the underlying network infrastructure of IDAs as we examine them (see Chapter 3).

The  consideration of the following factors during our analysis should help us establish those aspects of the network infrastructure that are common to most or all applications:

> Factor 1.  The method by which computers 'discover'[12] the application and by which they participate in the running of it.

---

11  The structure called 'topology' is developed in Chapter 4.

Factor 2. The way in which computers interact with each other during the normal operation of the IDA. We note particularly the topology structure (see Section 1.5.2).

## 2.1 Client-server Applications

Client-server[13] applications are those that have a centralised software component on the 'server' computer, which is accessed by an arbitrary number of software components on connecting 'client'[14] computers (see Figure 2.1).

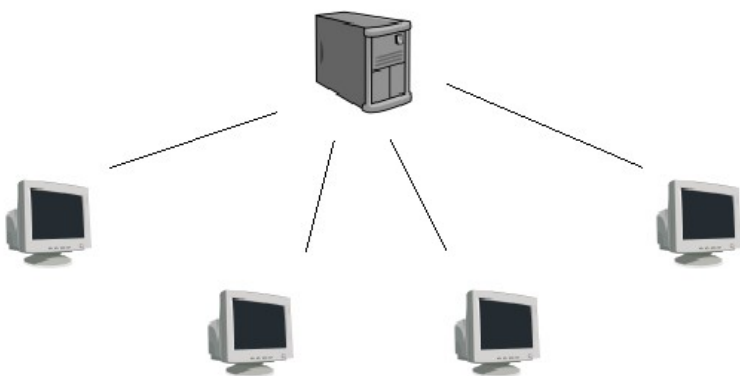

*Figure 2.1.: Client-server: Many clients connect to one server.*

Clients interact with the application by establishing a connection with the server. This connection is used to send data, usually in the form of a 'request', to the server. The server then establishes a connection with the client in order to send back data, usually in the form of a 'response'. Prominent examples of such applications are World Wide Web (WWW), or web, browsing, via the HTTP protocol [W3C '06], and 'e-mail', via the POP [RFC1939 '96] and SMTP [RFC0821 '82] protocols. In the case of web browsing, the server side component of the application is the web server software that responds to requests for WWW content. The client side component is the web browser, which requests the content and allows the user using the client computer to view and interact with this content.

---

12 Discovery, in this context, refers to the way in which an IDA comes to the attention of the user. For example, the user may hear about the IDA through word of mouth, or there may be some automated mechanism which allows software on the user's computer to automatically discover them from the network.

13 What is meant by a 'client-server' (architecture) can vary depending on the application. A working definition differing slightly to the one we use can be found in [Schollmeier '02] (p.2) where it is contrasted with P2P.

14 The terms 'client' and 'server' refer to executable software, but often they are also used to refer to the computers that host the client and server software, respectively.

Since we are interested only in applications, which run across many computers on the internet (Section 1.5.1 (1)), we are not concerned with client-server applications that merely persist between one client and one server. (Such as simple 'web surfing' where a static page is served to a browser client.) In cases where one client may cause an alteration of some state within the server application component that can be further interpreted and modified by an arbitrary number of other clients, these do constitute IDAs. Examples include internet discussion boards such as phpBB [phpBB '06] and Wikis [Wikipedia '06b].

Below we discuss the network infrastructural factors that we mentioned in the chapter introduction (p. 6) with regard to client-server architectures.

Factor 1: Clients participate in client-server IDAs via the server. Servers are 'discovered' through the knowledge of its well-known internet address[15], and contacted via the client (see above). Clients need not have a well-known or permanent address of their own, and do not connect to each other directly.

Factor 2: These IDAs are composed of many client computers and one server computer. We have termed the topology structure that best describes such network interactions a 'star': many points (clients) radiate outward from a central point. (The server) See Figure 2.2.

---

15  The user learns of the internet address or Universal Resource Locator (URL) of the server and directs the client to it. Alternatively, the user procures client software which has the server URL built in to it.

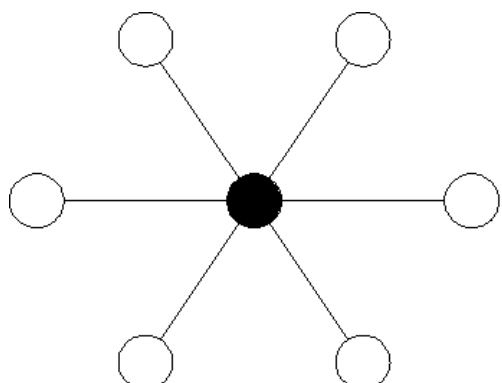

*Figure 2.2.: Graph representation of a (sample) 'star' (1 server, 6 clients).*

Other prominent examples of client-server IDAs are SETI@home [Sullivan '97], see Section 2.2, and MMOGs, see Section 2.4, which are explained further in these sections.

## 2.2  Distributed Computing Applications

Distributed computing is concerned with "computer systems in which several interconnected computers share the computing tasks assigned to the system" [IEEE '90].

One of the first, and certainly the most successful, distributed computing applications was Condor [Litzkow '88]. Systems such as Condor utilise the 'spare' processing cycles of idle 'workstations'[16] to participate in solving computational problems. (Recent systems improve on Condor in several important respects. For example, [Keane '03] is open source, cross-platform without requiring re-compilation and offers inbuilt security.) The simplest distributed computing applications contain a software component on a centralised, 'master' computer, called the scheduler, and a software component on each of a number of other 'slave' computers, which perform tasks at the behest of the scheduler. Other scheduling strategies involve a distributed scheduling approach, in which each computer in the system plays a role in scheduling[17], but these need not concern us here[18].

---

16  Condor uses 'workstation', but hereafter we will be consistent with our own terminology and use 'computer' (which are equivalent for our purposes).

17  Condor offers both centralised and decentralised scheduling mechanisms [Litzkow '88].

18  A discussion of the various scheduling mechanisms is beyond the scope of this work. The centralised scheme, as outlined, describes all current distributed computing IDAs (such as BOINC IDAs).

The scheduler is responsible for dividing the overall computational task required to solve a given problem into manageable work units, and decides how best to distribute these work units among a number of (slave) computers. When all of the work units have been completed, the individual results (output of each work unit) are collected by the scheduler and combined to form an overall result. Systems such as Condor are not IDAs because they do not confer any benefit to the users of the participating computers (Section 1.5.1 (3)), which act merely as slaves for the scheduler. The benefits of the computation are only of interest to the scientist whose problems are solved. In addition, they are often not suitable for use on an open, public network, such as the internet, (Section 1.5.1 (2)) due to security concerns[19].

SETI@home [SETI@home '06], a project which analyses signals coming to earth from outer space in an attempt to discover evidence for the existence of extra-terrestrial life, was the first distributed computing application of note which a) used the internet as its network, allowing anyone to participate in the running of the application, and b) offered a service to participants: users could aid in the search for extra-terrestrial life, be updated on the status of the project, and peruse a wide variety of statistics. These two points fulfil the remaining criteria for a distributed computing application to be considered an IDA (Section 1.5.1 (2) (3)), since all distributed computing applications fulfil the first and fourth criteria (Section 1.5.1 (1) (4)). Another distributed computing application which is structurally similar to SETI@home is climateprediction.net [climateprediction.net '06], which allows users to participate in forecasting the climate of the 21$^{st}$ century: climate change and the weather being topics that are of interest to many. Both of these applications utilise the vast public computing resources available on the internet to solve computationally intense problems, while offering something to the participating users.

SETI@home and climateprediction.net were built separately, on entirely separate code bases. However, it is clear that both applications are quite similar, in that a) both require an infrastructure for accepting the participation of many internet users, and b) both require a scheduler to divide the workload between them and to gather and analyse results. Responding to these needs, the BOINC [Anderson '04] [BOINC '06] software was developed, which provides this functionality, and allows developers of such applications as SETI@home and climateprediction.net to focus their software development efforts on the analysis of extra-terrestrial signals, or climate/weather data, respectively. Both of these applications, and many more, have therefore migrated to BOINC. At the time of writing, the BOINC statistics website [BOINC! STATS '06], maintained statistics on 20 projects, which combined boasted a number of participants in excess of 764,737.

---

19  Applications designed for private LANs do not have to be secured against the activities of malicious users. E.g. Such individuals may attempt to poison results using 'trojan' clients, or expose vulnerabilities in the software in order to hack into participating computers, and so on.

BOINC projects are client-server: the user of the 'client' computer bootstraps the client software, usually in the form of a screen saver which only processes tasks when the computer is idle, and allows it to connect to the server, which hosts the scheduler. Client and server will thereafter interact autonomously.

Below we discuss the network infrastructural factors outlined in the chapter introduction (p. 6) with regard to distributed computing IDAs (such as BOINC IDAs):

Factor 1: Servers hosting schedulers of distributed computing IDAs are discovered by clients in the same manner as other client-server applications (see Section 2.1)[20].

Factor 2: As with other client-server applications, the topology structure that best describes the network interactions of distributed computing IDAs is the 'star'.

## 2.3  Grid Computing

Grid computing [Foster '01] is increasingly being seen as the next stage in distributed computing. Foster defines the 'grid problem' as one of facilitating "flexible, secure, coordinated resource sharing among dynamic collections of individuals, institutions, and resources—what we refer to as *virtual organizations*" [Foster '01]. He considers that, "in such settings, we encounter unique authentication, authorization, resource access, resource discovery, and other challenges". These are the sort of issues that grid computing aims to address that were not fully addressed in traditional distributed computing. The term 'grid computing' is also used (incorrectly) to describe distributed computing on a large scale.

---

20  Slaves/clients are supplied with a URL which they use to contact the master/server.

Grid computing systems, in their current incarnation, are not IDAs. We mention grid computing here because it is an important, emerging, related distributed technology which, in some sense, distributed computing IDAs can be compared with.

Grids are intended [Foster '01] to be computing resources that are available to private (probably paying) groups, called virtual organisations (VOs). Although a VO could theoretically be extended to encompass the entire world, this is not what is currently intended by its proponents. Also, although grid computing emphasises the goal of aggregating geographically distributed computing resources to comprise one system, the scenarios outlined by Foster [Foster '01] involving the composition of VOs, indicate that what is usually intended is that geographically distributed computing 'farms'[21] be linked together, rather than a huge collection of heterogeneous computers across the public internet. For these reasons, grids cannot be considered IDAs (Section 1.5.1 (2)). Furthermore, it is not clear, in any case, what benefits would accrue to ordinary internet users who join their computer to a grid, except perhaps the option of submitting a problem of their own to the system (Section 1.5.1 (3)).

## 2.4  Massively Multiplayer Online Games

Wikipedia defines a MMOG as "a computer game which is capable of supporting hundreds or thousands of players simultaneously, and is played on the internet. Typically, this type of game is played in a giant persistent world." [Wikipedia '06c][22]. The first true MMOG appears to have been Meridian 59 [Meridian 59] [Kent '03], released in 1996. Today there are an estimated 12.5 million continuously subscribed gamers participating in games worldwide [MMOGCHART '06] with World of Warcraft [World Of Warcraft '06] having a 52.9% market share [MMOGCHART '06] at the time of the writing.

MMOGs are (almost exclusively) designed by commercial game designers and are hosted on powerful servers, which accommodate thousands of concurrent connections. Game players use the game client to connect to these servers. The game 'world' is maintained by software on the servers, while a graphical representation of this world is rendered for the user by the game client software. Each player controls a character or 'avatar' in the game world. This character interacts with the environment, including other characters, in pursuit of certain goals which vary depending on the game.

---

21  Processor farms (or server or compute farms [Webopedia '06]) are a network of computers at one location that can combine to work on a distributed computing problem.

22  The term MMOG, and its exact meaning, has evolved informally between gamers and the games industry, and therefore the current wikipedia definition is used here in place of a formal definition.

MMOGs are IDAs because they have all of the characteristics of IDAs, as outlined in Section 1.5.1.

Below we discuss the network infrastructural factors outlined in the chapter introduction (p. 6) with regard to MMOGs:

Factor 1: Users purchase the game client software. Upon bootstrapping, this client automatically discovers the MMOG game world servers using internet addresses hard coded into the client.

Factor 2: As with other client-server applications (Section 2.1), the topology structure that best describes the network interactions of an MMOG is the 'star'.

Although MMOGs use, predominantly, a client-server architecture, this may not be the case in the future. Players must pay a monthly subscription to the games company to participate in an MMOG. Due to the high costs of game development and hosting, there are no popular, free MMOGs [MMOGCHART '06] since costs must be recouped. It may be that hosting costs could be reduced by changing from the current centralised client-server architecture, to a decentralised architecture, such as P2P (see Section 2.5). Because characters and objects are distributed geographically throughout the game world, and it takes time to travel between distant locations[23], a case can be made for decentralising the 'ownership' of portions of the game world from the current bank of centralised servers to a larger number of smaller servers, hosted by peers on a P2P network [Knutsson '04] (We suggest an alternative in Section 3.3.) Therefore, it is not clear what architectures future MMOGs might have.

---

23 In the MMOG subcategories MMORPG, MMOSG, MMORTS [Wikipedia '06c], at least, game characters do not rapidly traverse large parts of the game world.

## 2.5 Peer 2 Peer Applications

Peer 2 Peer (P2P) applications are those which have a software component on each participating computer that contributes to the overall application, subject to the following:

· All computers, or 'peers'[24], are 'equal'[25] and not subject to, or reliant on, a centralised server for the functioning of the application (see Figure 2.3).

· Since peers are intended to be, primarily, ordinary computers connected to the internet, which have no well-known internet address; peers use a distributed mechanism for discovering resources, and routing messages between them.

This is the definition of a P2P application that we will use. Many others have been advanced [Milojicic '02] [Schollmeier '02] [Graham '01], however there is no standard definition.

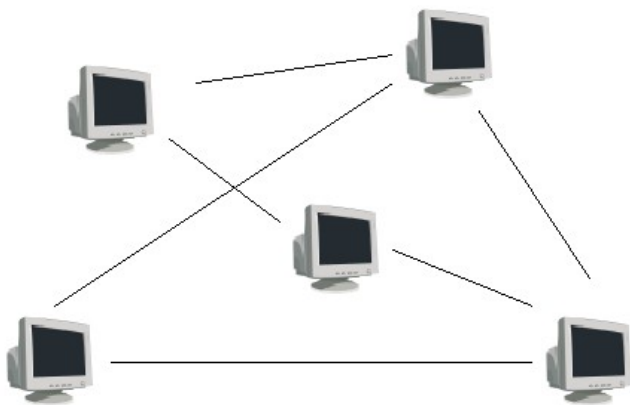

*Figure 2.3.: Peer 2 Peer: Peers connect to each other directly. (No central server.)*

P2P applications are inherently IDAs, since P2P clients are run by users, of their own volition, in order to participate in the running of particular applications. In practice all P2P applications adhere to our definition (see Section 1.5.1).

---

24 Computers (really the client software) participating in a P2P application are known as peers.
25 The notion of equality arises from the fact that client-server applications are often seen to be 'unequal', because the many clients are thought to be dependent on one server, often maintained by a commercial interest, for the hosting of (a vital component of) the application. P2P is often thought of as 'equal' because no single computer enjoys a special role in the hosting of the application.

P2P is one of the few internet technologies that is solely aimed at providing utility to ordinary internet users: users host the entirety of the application equally across their computers, and therefore there is no requirement for another entity to provide infrastructure that users might have to pay for[26].

Popular examples of P2P applications[27] are file-sharing clients that use protocols such as Gnutella [Kirk '06] or Kademlia [Maymounkov '02] (such as the original Gnutella client or eMule [eMule '06] respectively). The P2P file-sharing client is a software component that runs on each participating computer; collectively they form a file-sharing network. Users can expose files on their computer hard drives to the network for the purpose of sharing them with others. The network facilitates the routing of search queries for files throughout the system, and when files are located, the ability to download them from the computer on which the file resides.

Aside from P2P file-sharing, other P2P applications are starting to emerge, such as Skype [Skype '06] which provides direct peer to peer telephony, instant messaging and file transfer[28]. Skype not only uses a distributed, P2P search mechanism, allowing users to discover other users, but also relays communications through intermediate peers when any of the participants in the communication are behind a Network Address Translation (NAT) interface and/or a firewall [Baset '04].

Below we discuss the network infrastructural factors outlined in the chapter introduction (p. 6) with regard to P2P architectures.

Factor 1: The P2P IDA is 'discovered' by retrieving the P2P client software from a website hosted at a well-known address. The client software contains hard coded mechanisms that allow it to discover other computers participating in the running of that application.

---

26  In practice, a well-known internet address must be referenced at least once in order for each peer to discover at least one other peer already participating in the running of an application. Thereafter, other peers are discovered automatically. This first interaction requires a reliably reachable contact (perhaps hosted at a fixed IP address).

27  Napster [Barlow '00] is considered by many to be the first popular P2P file-sharing application, however it was not actually P2P because it used centralised servers.

28  Skype services are evolving all the time. It is also possible to call to and from traditional telephone numbers using Skype by subscribing to a paid service.

Factor 2: Computers interact with each other in two important ways when participating in the running of a P2P IDA. These are outlined as follows:

1. Peer discovery, resource discovery and message passing require a pro-active and re-active framework for maintaining a peer's 'view'[29] of the network and for making communication between peers possible. This is the P2P overlay network [Doval '03][30]: that is the network (or P2P layer) that sits on top of the Internet Protocol (IP) [RFC791 '81] internet, that provides the additional P2P mechanisms that P2P applications require in order to function. It is the overlay network that compensates for the lack of a centralised server to co-ordinate client activities. (See Section 2.7 for discussion on JXTA, which implements an overlay network. See Section 4.2.1 for the case for a JXTA-like overlay network in the IDAF.)

2. According to the specific P2P application, peers interact with each other in different configurations. File-sharing applications allow users to upload and download files to and from arbitrary peers. Skype establishes communications between two or more known peers. Thus, P2P architecture itself prescribes no topology structure: this is dependent on the P2P application. (The overlay network prescribes only an underlying topology structure, not an application level topology structure.)

Figure 2.4 gives a combined illustration of the application level and the overlay network level topology structures. In this thesis we will usually be concerned with application level topology structures.

---

29  Each peer must be aware of the existence and location of a number of other peers, and often must maintain a record of which peers certain resources or application elements belong to. This is called the 'peer view'.

30  The authors define 'overlay network' here somewhat more narrowly than we do. In [Doval '03], overlay networks are contrasted with P2P 'flooding' networks, whereas overlay networks can be seen as encompassing any networks, P2P or otherwise, that provide a functional layer above the IP internet.

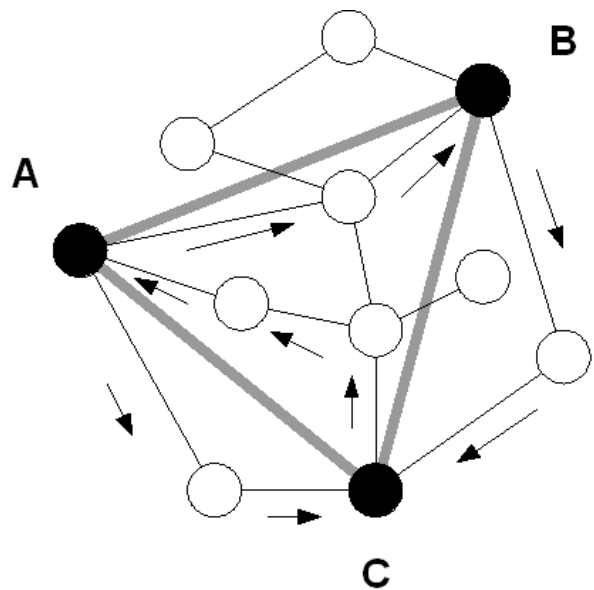

*Figure 2.4.: P2P IDA with peers A, B & C
participating. Thin lines represent overlay network
connections (and topology). Arrows represent possible
traffic flow. Thick lines represent application level
topology structure.(Depending on the overlay network,
traffic may pass between any two peers via
intermediary peers. Intermediary peers may form a
separate part of the same structure, or, if a single
overlay network is used to facilitate many IDAs (see
Chapters 3 and 4), they may not be participating in the
same IDA. These are the unlabelled peers shown.)*

## 2.6  Bittorrent

Bittorrent [Bittorrent '06] is a file distribution application that allows files to be downloaded in a decentralised fashion from numerous peers rather than from a single server. This is facilitated by a software client component which runs on each participating computer.

A single server cannot serve files to an arbitrarily large number of downloaders at a given fixed transfer rate to each, because ultimately the server's bandwidth will become saturated as the number of downloaders becomes very large. Bittorrent addresses this problem by dividing a file into pieces, and by allowing downloaders to download those pieces from the server(s), or 'seed(s)'[31], as well as from other downloading peers, which have already downloaded those particular parts. While initially all content must be downloaded from the original seed(s), soon peers begin to download more and more from each other, as they receive and host more pieces of the file. In this way, the required download bandwidth for all peers to receive a copy of the file is decentralised from being wholly provided by a central server's bandwidth, to being the aggregate upload bandwidth of all participating peers, collectively known as the 'swarm' (plus that of the original seed(s) if still connected). Hence no one internet connection becomes saturated, regardless of the number of downloaders, i.e. the size of the swarm. Once a peer has downloaded all pieces of the file, it becomes a seed itself until it is removed from the swarm by the user.

---

31  Peers which host the file in its entirety. Bittorrent jargon terms are described in [Bittorrent '06].

In order to download a file, one obtains the .torrent file[32] and opens it in a Bittorrent client. The .torrent file contains a description of the file, a summary of the pieces that it has been divided into, and the location of a 'tracker' server. The tracker server continually informs the client about other peers and seeds. The client then begins to download the file pieces from other peers and seeds. Clients must collectively upload as much (of the file) as they download, by mathematical necessity. A tit-for-tat choking algorithm [Cohen '03] is used on each client to help insure that, individually, users upload about as much as they download, thus ensuring the quickest download for all clients collectively[33].

Bittorrent is an IDA because it is has all of the characteristics of an IDA as outlined in Section 1.5.1.

Below we discuss the network infrastructural factors outlined in the chapter introduction (p. 6) with regard to Bittorrent.

Factor 1: .torrent files are (usually) distributed via a (centralised) website. The client uses the .torrent file to discover the internet address of a tracker server. The tracker server then directs the client to a list of peers and seeds hosting the file parts. The client then downloads file parts directly from these peers and seeds.

Factor 2: Although Bittorrent is often regarded as a P2P application, and uses 'peer' to refer to connecting clients, it is in fact a compound client-server application. It is essentially composed of two main client-server applications: a torrent file hosting (website) application and a tracker server application. In addition, clients download directly from each other, adding another client-server element to the IDA. The apparent P2P-like interconnectivity of the swarm, is actually no more than a series of client-server interactions. True P2P applications cannot function without a P2P overlay layer, which Bittorrent does not use; and do not rely on centralised servers, as Bittorrent does. Bittorrent's network interactions, as with other client-server applications, are described by a 'star'. (Although the aggregate of these client-server interactions require a more complex description.)

---

32 Usually from a centralised source, i.e. a website.
33 A discussion of co-operation and cheating in Bittorrrent, which has a direct affect on collective download rates, can be found in [Hales '05].

Recent developments have muddied the waters somewhat between Bittorrent clients and P2P file-sharing clients. Many Bittorrent clients (e.g. [uTorrent '06] [Azureus '06]) now have the ability to substitute a tracker server query for a distributed P2P Distributed Hashtables (DHT) [Pairot '06] overlay network search. Thus tracker servers are no longer required by these clients to discover swarm peers (a decentralised P2P lookup can be used instead). It is also a simple matter to distribute the .torrent file via P2P file-sharing. In this way, provided clients handle .torrent searches and downloads in a consistent way, these modified Bittorrent clients could essentially become peers proper in a P2P network, since P2P overlay network(s) replace all centralised elements.

Simultaneous to these developments, newer P2P file-sharing clients (e.g. [eDonkey '06] [eMule '06]) have improved on P2P file downloads by not only conducting simple (single) peer to (single) peer transfers, but also, if many peers are downloading a particular file simultaneously, by making use of the Multisource File Transmission Protocol (MFTP) [MFTP '06], which allows a tit-for-tat, Bittorrent style, swarm download mechanism.

In this way, Bittorrent clients and P2P file-sharing clients are becoming technologically very similar.

## 2.7  JXTA

The goal of the Project JXTA[34] [JXTA '06] is "to explore a vision of distributed network computing using peer-to-peer topology, and to develop basic building blocks and services that would enable innovative applications for peer groups." [JXTA FAQ '06]. To this end, JXTA provides a fully functioning overlay network, and a high level, software development kit (SDK) on which P2P applications can be built [Gong '02]. (This can be downloaded free from the website [JXTA '06].) See Figure 2.5.

---

34  Short for Juxtapose, however "Project JXTA" is trademarked by Sun Microsystems.

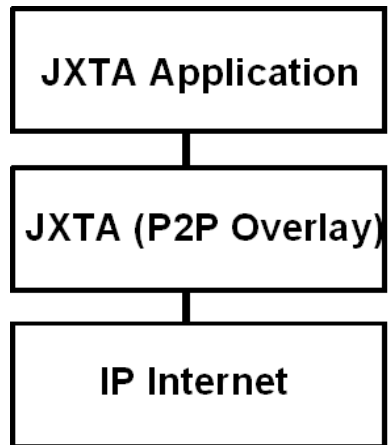

*Figure 2.5.: JXTA Stack*

At the time of JXTA's inception, many technologists and journalists considered P2P to be synonymous with file-sharing, (e.g. [Bricklin '00]), and much of the work done on P2P over the years has been focussed on the development of P2P overlay networks intended to facilitate file-sharing, such as those mentioned in Section 2.5. JXTA, released in 2001, was one of the first P2P technologies that focussed on enabling the development of P2P applications, rather than providing a specific, end user application. JXTA is not an application in itself but a platform for building applications.

For this reason, we do not introduce JXTA here as an example of an IDA, or a type of IDA, as we do with the other technologies that we have discussed in this chapter; but as a platform that enables developers to build P2P applications. In Chapters 3 and 4, we will show that such a platform can play a major role in the development of the IDAF. Therefore, in order to facilitate further discussion on JXTA (and other possible platforms) later on in the thesis, we give a brief description of it here.

JXTA provides a set of constructs within its SDK that can be re-used by application developers in their applications to refer to, and to interface with, the underlying 'P2P layer'. Most prominent among these constructs are 'Peer', 'PeerGroup', 'Service', 'Pipe' and 'Message' [JXTAv2.0 '06]: these represent aspects of JXTA's P2P platform, reflecting peers, groups of peers (with access to common resources), P2P applications[35], inter-peer communication 'pipes'[36] and inter-peer messages, respectively. In order to make productive use of these constructs, JXTA implements various protocols; the standard protocols are listed here: 'Peer Discovery Protocol', 'Peer Information Protocol', 'Rendezvous Protocol' and 'Pipe Binding Protocol'. (JXTA Specification, [JXTAv2.0 '06]) These protocols are used by the application developer in their applications (usually implicitly) to allow the application to interact with the P2P network analogously to the way the IP and other protocols allow client-server, internet applications to interact with the client-server internet. For example, the routing of peer discovery packets (Peer Discovery Protocol) and data packets (Pipe Binding Protocol) between peers is handled by JXTA, just as computers are 'discovered' (referenced) using the IP protocol and data packets are routed between peers using protocols such as TCP and UDP [Kessler '04] in a client-server network.

A platform with a sufficiently flexible and powerful overlay network and development SDK, could allow all current and future P2P applications to be build and run on it, thus saving application developers a great deal of programming time. JXTA has yet to fully prove itself such a platform, since none of the dominant P2P applications built since 2001 are built on it[37]. However, this may be because many of the most popular P2P applications are proprietary applications that are closed source in order to maintain a competitive advantage (e.g. Skype). It may also be that the general, application agnostic overlay layer used by JXTA is considered by developers to be too general for their specific applications, in its current incarnation. Certainly, JXTA must incorporate the latest advances in overlay layers gleamed from the best file-sharing and messaging applications, and to allow such specific applications, of at least equal quality, to be developed on it, in order for it to merit future consideration.

In any case, assuming that a generic platform containing a P2P overlay network with its own SDK is a viable entity (since it is a desirable one), as Sun Microsystems and many others (e.g. [Li '03]) do, then the overlay network at the heart of the platform can be continually improved (as indeed JXTA has been between the 1.0 and 2.0 specifications).

---

35 What we term 'applications', the JXTA project (usually) terms 'services'.
36 Encapsulated streams of data transmitted between two or more peers, often via intermediary peers.
37 However, there are indications that adoption of JXTA for scientific work is becoming prevalent, e.g. [Jan '06].

It is only of critical importance, therefore, that the SDK, specifically the application programming interface (API) that the developer must interface with, is well designed, such that it is rarely changed, if ever. This is important, since it allows successive generations of overlay layers to be introduced into the platform without causing any changes to applications built on it.

As an enabling P2P infrastructure, one of primary goals of the JXTA project is to "support multiple platforms and languages, micro-devices to servers" [JXTA vision '06]. Computer platform independence, in terms of language, hardware, and operating system, and JXTA's open specifications and open source SDK, means that there is no impediment, either practical or legal, to adopting it as a universal platform for developing applications for any purpose or to run in any environment.

JXTA is currently the most advanced P2P development platform, and although not yet in heavy use within IDAs, it has been thoroughly investigated by the academic community over the past 5 years. It has been analysed with respect to performance and benchmarking [Halepovic '03], a comparison with other P2P overlay networks [Airamo '05], mobility [Bisignano '03], use within a grid [Matossian '03], and large scale deployment [Antoniu '04]. It is thus a relatively mature platform.

In this chapter we discussed a range of current IDAs. In the next chapter, we will extend this analysis further by investigating a framework capable of accommodating them.

# Chapter 3    Background to an IDA Framework

In this chapter, we will look at the primary considerations of an IDA framework. We give here the background to the decisions and constraints that inform the IDAF, which we present in Chapter 4.

We will begin, in Section 3.1, by examining the discovery and participation of computers in the running of IDAs, and the network issues arising from this examination. We consider here how software components on participating computers (i) discover applications, (ii) participate in the running of applications, and (iii) route and pass data to and from other application software components running on other computers. When investigating the network infrastructure of existing IDAs, in Chapter 2, we considered the first two points with regard to specific IDAs (the 'Factor 1' as introduced in the Chapter 2 introduction, p. 6). The third point was discussed, briefly, in relation to the two dominant network infrastructural models used by most[38] of the IDAs described, namely client-server and P2P. We will extend this analysis of IDAs with regard to these three points here.

In Section 3.2 we make the case for using P2P as a basis for the network infrastructure of the IDAF. In Section 3.3 we examine potential topology structures that the IDAF ought to accommodate.

## 3.1   IDA Participation and Network Infrastructure

### 3.1.1   IDA Discovery and Participation

It is our goal to develop a framework that facilitates arbitrary IDAs (see Section 1.2). For this reason, a generic software component that runs on each computer that is capable of discovering and participating in the running of arbitrary IDAs is a requirement of the IDAF. That is, a component that runs prior to the discovery of IDAs that can, after an IDA has been discovered, participate in the running of that IDA.

---

38  Bittorrent, as a counter example, uses a more complex model (see Section 2.6).

Examples of such generic software components that can participate in the running of arbitrary IDAs are web browsers, for WWW IDAs, and JXTA, for P2P IDAs. All of the other IDAs that we discussed in Chapter 2 use their own dedicated clients. For example, eMule requires an eMule client, and World of Warcraft requires a World of Warcraft client. Dedicated clients, of course, make the question of application discovery a superfluous one[39]. However, this question must be tackled when a generic software component is used which has no knowledge of any application. In this section, we will discuss the problem of IDA discovery and participation by a generic software component running on a user's computer.

IDA discovery is the process by which a software component on a single computer, executed by a user, learns of the existence of a running IDA. IDA joining is the process by which the software component connects to other software components that are already participating, in order to negotiate its own participation.

Both of these processes are necessary for generic software components to participate in the running of IDAs: the first because knowledge of an IDA is essential for participation; the second because an IDA is but the sum of its parts, and it is only through negotiation with one or more of these IDA software component parts (on other computers) that a new part can be added.

We will discuss the discovery and join processes below, and their implications for the IDAF.

In considering the discovery process, we note that in the case of all of the IDAs examined in Chapter 2 the computer user learned of the application through his or her own efforts: that is, the user learned of the existence of the application, and either fed this knowledge into a client or obtained a client with this knowledge hard coded into it (e.g. a WWW discussion board).

---

39  The user has 'discovered' the game through word of mouth or advertising. The client, once obtained, need not 'discover' the game. (See below for more on IDA 'discovery'.)

It will always be useful for users to discover IDAs themselves, whether it is through advertising, or through their own inquiries; however, with emerging P2P application platforms, such as JXTA, there is now also the possibility of discovering IDAs via an automated P2P mechanism. If the IDAF were to be based on P2P, it would be natural to use the underlying P2P network to publish and to discover the existence of IDAs. This would additionally allow ordinary internet users, for the first time, to host IDAs of their own, without having to incur a hosting cost. Other users could then discover these IDAs using a distributed search, or by having their generic software component listen over the network for the publication of IDAs that may be of interest. The possible benefits of decentralised advertising and discovery of IDAs are discussed further in Chapter 7. For now it will suffice for us to say that it would be desirable to have the option of discovering IDAs via some automated, P2P mechanism, in addition to traditional methods, provided the use of a P2P architecture were otherwise warranted.

In considering the join process, we note that the IDAs discussed in Chapter 2 use varying mechanisms for admitting the participation of applicant software components. Client-server IDAs (Section 2.1) require interaction, and often authentication, with the central server. Bittorrent (Section 2.6) uses client-server mechanisms for torrent downloads and tracker server access. In P2P file-sharing applications (Section 2.5) software components on peers apply to other, arbitrary peers, rather than to a centralised server.

In order for the IDAF to allow the full range of IDAs to be described by it, both client-server and P2P IDAs, applicant software components (having already discovered an IDA) must support both the 'apply to well known server' scenario and the 'apply to automatically discovered peers' scenario. A P2P solution naturally suggests itself, in the latter case (although Bittorrent style centralised indexes (see Section 2.6), rather than decentralised P2P mechanisms, could be used instead to track 'peers').

In analysing IDA discovery and participation we see that there are definite advantages to adopting P2P as an underlying infrastructure for the IDAF. We examine the case for a P2P infrastructure in Section 3.2.

### 3.1.2  Overcoming Network Issues

The current incarnation of the internet is based on the Internet Protocol (IP) v4 (IPv4) [RFC791 '81], which provides the primary network layer[40] [Zimmerman '80] protocol. The IP specifies that each computer connected to the internet is reachable by a 32 bit identifier, known as an IP address. The transport layer [Zimmerman '80] protocols facilitate data transfer over the IP internet, with the Transmission Control Protocol (TCP) [RFC793 '81] and User Datagram Protocol (UDP) [RFC768 '80], between them, accounting for the movement of most of the internet's traffic: more than 95% [Fomenkov '03][41].

The IP, TCP and UDP protocols were designed over 25 years ago, and over this time the internet has declined as a medium for hosting (a variety of) IDAs, due to the fact that host to host communication has become problematic for various practical reasons. We discuss the technical difficulties encountered by the IDAs today, and their implications, in this section.

The first difficulty is due to the insufficiently large IP address space. It was assumed that all of the world's computers could be addressed within the 32 bit address space specified for IP addresses, which allows for a maximum of 4.3 billion addresses. This belief, while reasonable at the time, did not foresee the future popularity of the internet, and the ubiquity of internet capable devices, such as PCs, internet capable phones, and Personal Digital Assistants (PDAs). At the time of writing, [Geohive '06] indicates that the number of PC's connected to the internet is over 772 million worldwide. [Gantz '04] indicates that there may be a similar number of other, non-PC, internet capable devices in existence. The IDC predicts that the total number of internet capable devices will rise to 6 billion by 2012 [Gantz '04]. It is clear, therefore, that the 32 bit IP address space is insufficient to cater for future needs. In addition, the fact that IP addresses are currently not allocated equally (the US department of defence has a larger IP address allocation than all of Asia [Patrizio '06]), means that certain IP address allocations are already insufficient to cater for local demands.

---

40  As defined in the OSI Model. (See reference.)
41  Although the study period was conducted between 1998 to 2003, the authors showed that the proportion of traffic carried over the TCP and UDP protocols tended not to vary over time, despite the increase in prevalence of P2P file-sharing.

The scarcity of IP addresses has led to the proliferation of technologies such as Network Address Translation (NAT) [RFC3022 '01]. These technologies allow a private network of computers to be represented on the internet by a single IP address. Internally, all computers have private IP addresses which are not routed by the internet [RFC1918 '96]. One of the undesirable consequences of the increase in NAT usage is that fewer and fewer hosts on the internet can act as a server as well as a client. This severely curtails host to host communication on the internet.

NAT, and similar solutions, cause major problems for applications such as host to host telephony (see the host to host telephony example below) and P2P filesharing. Two solutions to this problem have been proffered. The first solution is to establish servers on the internet that computers behind a NAT device can connect to, to act as a relay or proxy, in order to simulate a host to host connection[42]. (The problem is that these servers are expensive and can expect to receive large amounts of traffic for popular applications.) The second solution is to establish a P2P network, where, instead of a few central servers relaying the traffic, many ordinary peers, which happen not to be behind a NAT device, act as relays and proxies for those peers which are. The second solution is actually a decentralised version of the first.

Host to host telephony, or voice over IP (VoIP) [Wikipedia '06f], has shown that a P2P network can provide a good solution to the NAT problem. In January, 2004, John Walker announced the end of life of his Speak Freely application which conducted host to host telephony [Walker '04]. One of main reasons for his doing so was that he felt that the internet was no longer an hospitable environment for host to host applications, given the proliferation of NAT. He also pointed out that, for a free application, it would be impossible for him to fund a high traffic server to relay voice communications. Six months after the death of Speak Freely, Skype [Skype '06] v1.0 was released. It used a P2P network to overcome these problems. Although it appears[43] that Skype also uses servers to stabilise communications, the application simply could not function without the participation of a large number of ordinary peers that are not behind a NAT device relaying traffic for the peers that are.

---

42  Two computers behind a NAT device can connect as clients to a proxy/relay server. The server can then pass traffic sent to it from one client back to the other. As a result, neither computer need be addressed directly.

43  The Skype protocol is proprietary and remains undisclosed, but has been studied in [Baset '04].

JXTA also uses a P2P network that transparently bypasses NAT devices. P2P simulates direct host to host communication whether it would otherwise be possible or not. P2P is not strictly necessary to overcome difficulties with NAT, but some overlying 'substrate'[44] is currently required to enable transparent, universal host to host communication. Instead of relay peers being discovered in a P2P manner, relay hosts could be obtained from a centralised index, similar to a Bittorrent tracker server. However, P2P overlay networks that provide such a substrate lend themselves toward the solution of the NAT problem.

The second difficulty that the current incarnation of the internet presents IDAs is concerned with ports. Both the TCP and UDP protocols use port (numbers) through which data can be sent and received between application software components and the internet. Most applications always use the same, well advertised ports for communications. As the internet has evolved, difficulties have arisen with regard to the transparent usage of ports.

Network firewalls [Cisco '02] often intercept and block traffic destined for certain ports, on entering or leaving a network. This is usually done to enhance security or to restrict internal access to certain web applications. While it may often be appropriate to do so, there is a cost. The internet, or at least the part of it that has its traffic restricted by a firewall, has been damaged or partially disabled. This is done because it is easier to disable parts of the internet's infrastructure than it is to restrict the applications that are run 'inside' (the internal network protected by) the firewall, or to ensure that all network services running on all computers on the internal network are protected from threats from without. Unfortunately, this often unpredictable disabling of lines of internet communication (combined with the NAT issues outlined above) make for a complicated environment in which to run IDAs. Often it is unclear to the user whether or not certain IDAs can be run, or if so, what manual port settings[45] or network configuration might be required to enable software components on their own computer to participate in the running of a given IDA.

---

44  We use the term 'substrate' here to mean a layer that masks the problematic IP internet, providing more simplified and more transparent access to it.

45  Many applications allow users to set their application port(s) manually (e.g. Bittorrent clients).

Many current (mainly file-sharing) IDAs resist attempts to restrict them using port blocking alone (numerous ports are tried). (The difficulties encountered by an enterprise in restricting IDAs is given in [Sears '06].) Nonetheless, port blocking constitutes an obstacle to IDAs (where P2P substrates are not used).

The network problems we have discussed have had the unintended side effect of giving additional motivation to the adoption of P2P overlay networks (or in particular P2P substrates) in IDAs. We have seen that Skype uses such a P2P substrate to defeat the problems of NAT. A more complete substrate and overlay network is available in JXTA ([Neto '05] uses JXTA for its ability to bypass firewalls), which makes all IP structures (and therefore issues) irrelevant and allows IDAs to run in a purely P2P environment. (Peer IDs replace IP addresses, the notion of the JXTA Service replaces TCP and UDP ports.) All that is necessary is for a JXTA client to attach to another JXTA client, via any port, proxy or relay, and an IDA software component on any peer can interact with any other, freely and transparently, permitting the running of arbitrary IDAs.

In conclusion, although the internet was intended to facilitate arbitrary host to host communication, in practice, this is often not possible. The difficulties facing certain IDAs running on the internet, as it functions today, lends impetus toward the use of substrates, and in particular, P2P substrates, which are already prevalent.

## 3.2 A P2P Network Infrastructure for the IDAF

In Chapter 2, we outlined the two main architectures that can be used to describe most or all, current IDAs: that is, client-server and P2P. It is clear that the IDAF ought to allow both architectures to be described by it. Here we will make a case for adopting a P2P infrastructure, such as JXTA, as a basis for the IDAF. (This infrastructure is integrated into the IDAF design as described in Section 4.2.1.) We will begin by isolating the special features and drawbacks to both architectures, then we will determine if both can be accommodated within a single framework. (Other ad-hoc non-P2P architectures, such as Bittorrent, can be described in terms of client-server interactions, see Section 2.6.)

P2P offers a facility that was previously unavailable to application developers using a client-server architecture for their applications: namely, the ability to discover other computers (peers) and resources, such as files, automatically from the network. Notwithstanding the technical issues, dealt with in Section 3.1.2 (by a P2P substrate), P2P does not add anything more, technologically, to the IP based internet. Any host can communicate with any other host on the internet, just as any peer can communicate with any other peer on a P2P based network, provided all of the host IP addresses are pre-known in the former case.

Early P2P overlay networks could not guarantee that a search for a particular peer or resource would be successful. The Gnutella Protocol [Kirk '06] v0.4, for example, simply flooded the network with search queries and hoped for the best. It had seemed, initially, that in exchange for the facility of being able to discover hosts automatically, P2P had rendered itself unable to deterministically contact any particular peer by name or ID: the search query might simply return no result even if the sought peer were on the network. In other words, P2P, although capable of discovering arbitrary peers, could not (deterministically) contact a particular peer, which would rule out client-server like, or in particular host to particular host, communication. However, with the development of Distributed Hashtables (DHTs) [Pairot '06], this changed. DHT overlay networks can deterministically route to any given peer (if a unique peer identifier is known) on the network[46], within a maximum number of peer hops $O(\log N)$ (Pastry [Rowstron '01], Chord [Stoica '01] and Kademlia [Maymounkov '02] are examples), where N is the number of peers on the network, and 'hops' is the number of intermediary peers that the search query must propagate through.

For these reasons, the IDAF could be based on P2P, without prejudicing itself against client-server IDAs (with a centralised named server). The 'server' in a P2P based IDAF, is simply another peer with a known, unique peer reference, such as a peer id (PeerID in JXTA). Therefore, P2P can technologically cater for both P2P and client-server applications and could form a basis for the IDAF.

Although the use of P2P does not affect the range of IDAs that can be described by the IDAF, it is not a technology that is generally used without good cause. This is because P2P overlay networks can be demanding in terms of bandwidth. The Gnutella protocol [Gnutella '06], for example, must propagate a number of 'messages' throughout the network (considerably less post v0.4) to discover other peers (namely PING and PONG messages) and to route queries for files (namely QUERY & QUERY_HIT), along with other, less frequently used 'signalling' messages. These messages incur an ambient bandwidth cost above and beyond that which is required for the application itself, i.e. file download in the case of Gnutella. Signalling traffic will be generated for any P2P application. The signalling bandwidth generated by Gnutella, an open and well analysed protocol, has been examined in [Ilie '04]. While there is very little hard data on bandwidth requirements on the individual user using file-sharing clients, where no files are being shared, i.e. the ambient bandwidth cost to the user (to maintain a peer view and to help propagate queries); it is heavily dependent on the network, the client used and the structure of the P2P overlay network used. That said, users frequently report high bandwidth utilisation and a reduction in network responsivity [Kedroskey '06] (Skype) [Gnutella Forums '06] (Gnucleus) where clients do not provide a facility for controlling bandwidth consumption. Some applications are capable of consuming as much bandwidth as is available to them [Bearshare '06] [Skype '06].

---

46  To a high degree of reliability determined by the specific fault tolerant mechanisms used by the overlay network for coping with the disappearance of peers (that can provide valuable routing data).

It is often the case that a P2P client will use most of a user's available bandwidth unless there is an option to restrict bandwidth utilisation in the client. It is therefore not an attractive proposition to run two (or more) P2P clients side by side, both of whom will compete for the users bandwidth and will likely adversely affect the bandwidth capacity and network responsivity (latency) of the user's other applications. It would seem, therefore, that basing all IDAs described by the IDAF on a P2P architecture would be unworkable in practice, even if it would be a good option in theory.

However, an IDAF based on P2P would need only one common P2P substrate or overlay network to service all IDAs. Only one 'peer view', and one medium to propagate search queries, is required. In the case of client-server IDAs, the overlay network would not incur any additional 'query' traffic, since no queries are propagated.

The ambient bandwidth cost, therefore, of building the IDAF on a P2P architecture is the bandwidth incurred by one continuously running P2P overlay network. If none of the IDAs running make use of the automatically generated 'peer view', then, certainly, bandwidth is wasted. However, with a large proportion of current IDAs already based on P2P, it is likely that a user running a few IDAs (and we foresee users running many more IDAs in the future, see Chapter 7) will run at least one P2P IDA, in which case the inclusion of a P2P overlay network is justified. If the user wishes to run two or more P2P IDAs there will be economy in maintaining a single peer view and in the caching of search queries for resources that may be common to more than one IDA. In any case, as with many, modern, P2P clients, e.g. [Bearshare '06], the bandwidth (of the substrate, available to the overlay network) ought to be adjustable by the user, to ensure good network performance and responsivity for other applications.

We have shown that all of the IDAs introduced in Chapter 2 can be accommodated within a P2P architecture, even if they currently use a client-server architecture. We have also shown that the bandwidth cost of a P2P overlay network, and therefore the use of a P2P architecture, although real, can be limited, and may scale to simultaneously support a higher number of concurrently running P2P applications. In Section 3.1.1 we saw that P2P would give the user the additional option of automated IDA discovery and hosting. In addition, in Section 3.1.2, we showed that a P2P substrate, such as that which is used in JXTA, allows us to overcome many of the problems endemic in the modern internet, making universal, host to host communication possible (which is necessary in order for the full range of IDAs to function). For these reasons, we use a P2P overlay network and substrate as the basis for the IDAF.

See Section 4.2.1 for a discussion on the inclusion of a P2P overlay network in the IDAF.

## 3.3  Topology Structures

The unstructured P2P network provided by a P2P overlay network, as it stands, is not directly amenable to running IDAs other than those composed of a set of random one to one connections, such as P2P file-sharing IDAs. Every other IDA examined in Chapter 2 requires additional structure across the constituent computers (peers) in order to run the IDA. We use the term 'topology structure' (see Section 1.5.2) to describe this additional structure.

In this section we will give examples of topology structures that we have observed in existing IDAs (see Chapter 2) as well as other possible structures that may be of potential use in future IDAs. We use graph theory [Wikipedia '06a] terminology to describe these structures. (A formal analysis of the applicability of graphs to modelling topology structures is made in Section 4.1.1.)

The Tree

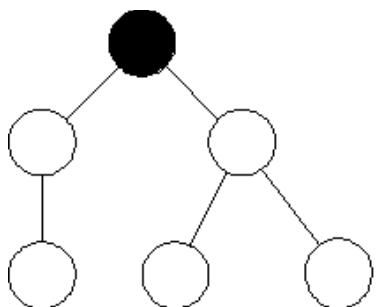

*Figure 3.1.: Tree topology*

In a tree, nodes are interconnected such that there is only one path between any two nodes. Rooted trees are hierarchical, that is, trees that are centred around a root node have an implied structure: other nodes are said to be closer or further away from the root, on the tree.

Trees are used in distributed computing to provide 'hierarchical task stealing', which provides load balancing, and decentralised traffic patterns for distributed computation, e.g. [Baldeschwieler '96] [Nieuwpoort '01]. Trees are also ideally suited to data dissemination, since data transmitted from the root node is guaranteed to reach all nodes, e.g. [Kim '03] (although high throughput media streaming is best done at the overlay network level, rather than our application level[47]).

---

47 It is even more efficiently conducted using IP multicast, however, there is currently poor support for IP multicast on internet routers.

## The Star

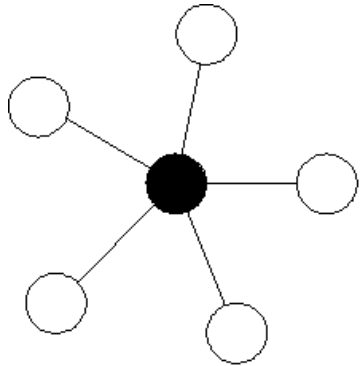

*Figure 3.2.: Star topology*

A star is a special type of tree. In a star, many nodes connect to one central node.

Prominent examples are BOINC based IDAs such as SETI@home and climateprediction.net. See Chapter 2 for examples of client-server IDAs, which conform to this structure.

## The Mesh

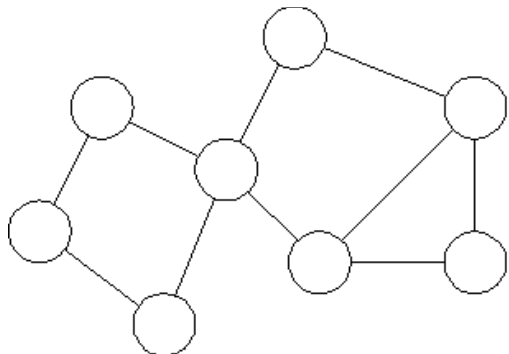

*Figure 3.3.: (Planar) Mesh topology*

In a mesh nodes are interconnected such that at least two nodes have two or more paths between them. (In Figure 3.3 nodes are connected to form a planar graph of degree 4.[48])

Mesh topology structures have been proposed for overlay networks to distribute media content [Magharei '05] [Guo '03], although, at the 'application level', examples of meshes in use are not forthcoming. However, we propose one here: a topology structure

---

48  The degree is <4 for nodes at the exterior of the graph.

as a basis for a range of MMOGs. Although suggestions for non-client-server based, P2P MMOGs have been put forward [Knutsson '04], popular commercial MMOGs currently use a star topology structure (see Section 2.4). The use of a star topology structure does not seem particularly well suited to MMOG IDAs, and was probably a de-facto choice due to the prevalence of client-server, and therefore star, based IDAs. We list here the potential advantages and disadvantages of adopting a regular planar mesh topology, rather than a star topology, as a basis for a range of MMOGs:

Advantages:

1.  The geography of the game world can be mirrored by the geography of the mesh. (Each node can take responsibility for a contiguous portion of the game world.)

2.  No single point of failure. Nodes share responsibility for maintaining the game world.

3.  The game is inexpensive to run since computing power, bandwidth and electricity is provided by the players.

4.  Bandwidth utilisation may be decentralised if intelligence in the overlay network can be used to allow computers that are close together – using the same Internet Service Provider (ISP), for example – to host portions of the game world that are geographically close.

Disadvantages:

1.  As with all P2P IDAs where the IDA is run jointly by all nodes, there are numerous issues of security and network resilience.

2.  Centralised hosting of the game world makes it easy for a commercial entity to maintain the game infrastructure. (Assuming it is possible to scale to cater for the number of players.)

3.  The mesh topology hosts contiguous portions of the game world at contiguous nodes on the mesh. It would less well suited to MMOGs where characters moved through the game world quickly. (Requiring a chain of nodes to be traversed rapidly.)

None of these disadvantages ought to be fatal to the case for running MMOGs on a mesh topology. Indeed a hybrid client-server P2P, or super-peer topology may also prove to be a good choice, such as that analysed in [Knutsson '04]. A sample MMOG IDA that was used to test the IDAF/IDAS in given Chapter 6.

## Complex or ad-hoc structures

We have looked at topology structures that can be described by well-analysed graphs. However, topology structures can be arbitrarily ad-hoc or complex. We give examples below.

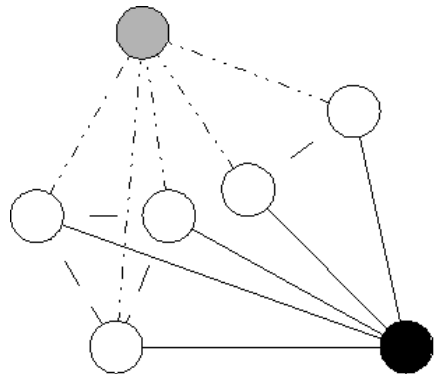

*Figure 3.4.: Bittorrent topology representation*

In Figure 3.4, we provide an interpretation of Bittorrent's topology structure. The grey node represents the web server that hosts torrents, the black node represents the tracker server, and the white nodes represent downloading clients[49] (see Section 2.6 for discussion on Bittorrent). (Considering that downloading nodes only contact the grey node once, to obtain the torrent, the grey node and all connecting edges could be removed from the above structure, since it is our primary goal to model continuous IDA operation.)

---

49 Dashed lines are used for edges/connections between downloading client nodes for the sake of clarity. (Although dashed lines take on additional significance in Section 4.1.) In the figure, two groups of nodes download two different torrents (tracked by the same tracker server).

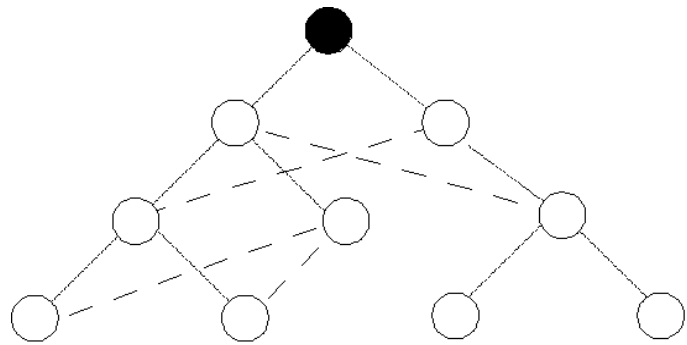

*Figure 3.5.: Rooted, binary tree topology with redundant connections*

In Figure 3.5 we give a rooted binary tree which has 'extra' edges/connections, indicated by dashed lines. One of the vulnerabilities of a tree structure is that if any connection is broken, an entire sub-graph of the overall tree (also a tree) becomes disconnected from the rest. In applications using hierarchical task stealing (see discussion on 'Tree' above, p.33), it is very important that nodes further away from the root be able to pass results and tasks to the root and to nodes along the way. If a node should disappear from the topology structure, in a real world scenario such as the disconnection of a computer from the network, tasks and results from the sub-graph connected to the main graph through this node would not be able to propagate toward the root, unless there is additional redundant connectivity. In the sample scheme outlined in Figure 3.5, connections are made not only between child and parent nodes, but also between child and sibling of parent nodes on the binary tree[50], in order to provide this redundant connectivity. (Two sibling nodes would have to disappear simultaneously in order to cause a complete disconnection.)

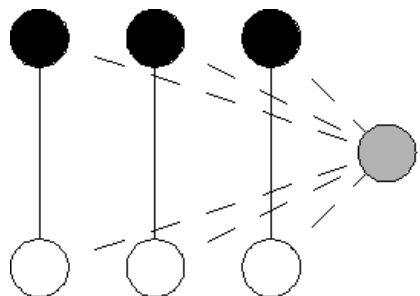

*Figure 3.6.: White & black nodes connected as bipartite graph with each node connecting to grey node.*

---

50 For an n-tree (where n>2) the redundant connection could be made to an arbitrary sibling, or perhaps connections could be made to a number of siblings, making the tree even more robust.

Thus far we have discussed topologies which aid computation or file transfer, however, many other types of IDA, and therefore topology, are possible. In Figure 3.6 we give a balanced, bipartite graph connecting black and white nodes, except that each node is also connected to a grey node (these edges are dashed for clarity). We imagine an IDA that implements a mechanism similar to the Distributed Proofreaders [Proofreaders '06] project used by Project Gutenberg [Project Gutenberg '06] that distributes a page for proof-reading from a central point (the grey node) to a node (say a white node). The user at that node then proof-reads the page. When they are finished with the page, it is then sent to a node on the other side of the bipartite graph (in this case a black node) so that the user at that node can proof-read it, and examine previous edits. Then the page can be sent back to the first-proof reader again, and ultimately back to the central point (the grey node) after the desired number of iterations.

In this chapter we examined modern day IDAs and the modern day internet and arrived at general conclusions on an IDA framework. In the next chapter, we introduce the IDAF itself.

# Chapter 4   An Internet Distributed Application Framework

In this Chapter we will develop a framework for building IDAs, called the Internet Distributed Application Framework, or IDAF. The Oxford dictionary defines 'framework' as 'a supporting or underlying structure' [Oxford '01]. The IDAF will describe the underlying structure of IDAs. It will allow us to describe IDAs with logical, structural components, and it will allow us to build IDAs from these components. Ultimately, it will enable us to build SDKs for the systematic development of IDAs (see the IDAS in Chapter 5).

In Section 4.1, we will develop the concept of the 'topology'. Topologies incorporate a topology structure (defined in Section 1.5.2) (in terms of a graph such as those examined in Section 3.3) as well as additional logic to maintain and optimise this structure in a real world environment. In this section we give a formal description of topologies in the IDAF.

In Section 4.2 we will give an overview of the IDAF Layers, which comprise the core to the IDAF design. In Section 4.3 we will outline the additional constraints of the IDAF.

## 4.1   IDAF Topologies

### 4.1.1   Topology Definition

In Section 3.3, we discussed a number of topology structures. In this section we formally introduce the concept of 'topology', in the context of the IDAF.

We define the 'topology' (of an IDA within the IDAF) as a relation (or a graph) describing how nodes interconnect in the IDA, and also the logic required to maintain, refine or evolve this interconnection of nodes[51].

We define the 'configuration' as the dynamically changing actualisation of the topology structure. (In other words, a 'live' interconnection of nodes that is dictated by the topology structure.)

IDAs require logic to maintain the configuration in the event of failure or structural degradation. On an unstable network such as the internet, nodes can be viewed as appearing and disappearing at random, as computers are booted up, shut down, connected and disconnected. For this reason, nodes cannot simply be formed into a configuration dictated by a topology with the expectation that the same computers will occupy the same nodes on the configuration at some arbitrary time in the future. For this reason, there must be logic in place (translating to algorithms in an IDAF implementation) that monitor the configuration and take steps to ensure that it remains viable.

It may be desirable (depending on the topology) to refine the configuration. A refinement, or optimisation, of the configuration is any change to node positions or node inter-connections on the configuration that make it more 'efficient' (an example of how a configuration might become more efficient is given below). However, no such refinement may violate the topology structure describing how nodes interconnect in the IDA. Taking a spanning tree topology as an example, such a topology would continually attempt to reduce the number of edges on a spanning tree configuration. That is, to refine the structure from a less than minimum spanning tree [Nesetril '00] (one with more edges than a minimum spanning tree) to a structure that is, or is closer to, a minimum spanning tree. (If such a refinement is deemed to improve the performance of the IDA then the configuration is deemed to have become more efficient.)

It may also be desirable (depending on the topology) to evolve the configuration. Certain application conditions could trigger a change in the topology structure itself. Such evolutions can be as varied and as complicated as any two 'before' and 'after'

---

structures that can be imagined. However, it is not merely a case of specifying a set of graph state descriptions that a configuration/topology structure might evolve through, the logic governing the actual transitions between these states must also be specified. In practice, this process could be very difficult to implement. (We discuss the evolve process further in Sections 4.3.1 and 5.4.3.)

The task of maintaining, refining or evolving a configuration must ultimately fall to the individual computers participating in the running of an IDA. Code must execute on each computer, such that the overall affect on the configuration is achieved. Furthermore, any logic that is to maintain, refine or evolve a configuration must operate on the basis of local knowledge only. In our definition of the IDA (Section 1.5.1 (4)), we said that IDAs can benefit from, and can scale to admit, the participation of an arbitrary large number of computers. This being the case, a complete knowledge of the entire network of participating nodes cannot be assumed, since a survey may be impractical. The discrete and localised nature of topology structures call for a decentralised, node oriented, iterative mechanism for specifying such logic. (We will investigate further implications of this in Section 4.1.3 in relation to building configurations from topologies.)

Nodes may perform different roles in a topology. In the topology structures that we examined in Section 3.3, we highlighted these different roles using colour: for example the star had a coloured root node (black), to indicate that it performed a separate role to the leaf nodes (white). It is convenient to use colour to illustrate this graphically, however, we will also use the term 'NodeType' to allow us to describe these roles by function, e.g. 'root NodeType'.

Similarly, connections (edges) between nodes may be of different types. We will use the term 'ConnectionType' to describe these types of connection. ConnectionTypes may specify various properties about the nature of a connection between two nodes. (e.g. whether the data stream is continuous, based on a HTTP-like request/response sequence, or whether the connection is UDP or TCP.)

We will use the term 'direction' to describe the direction that a TopologyConnection 'points in'. The concept of direction is only applicable to rigidly structured topologies, discussed in Section 4.1.2, p.47, and in the context of a topographical framework such as that used by the IDAS Locator class in Section 5.3.2. Intuitively, it is the direction in which a TopologyConnection 'points', e.g. 'north', (coordinates) (0,1,4), 'N20°W', 'towardsTheOrigin', and so on.

We give here the relation that describes a topology structure configuration, and show its equivalence to a graph.

We begin by defining the following sets. N is the set of computers participating in an IDA. T is the set of possible NodeTypes. O is the set of possible ConnectionTypes. D is the set of possible directions. Also, let n1, n2 ∈ N, representing (two) nodes; t1,t2 ∈ T, representing (two) NodeTypes; o ∈ O representing a ConnectionType; and d ∈ D representing a direction. Let C be a 6-tuple that describes a connection between two nodes of two given NodeTypes of a given ConnectionType, on a configuration.

$$C = (n1, n2, t1, t2, o, d) \tag{1}$$

Let P be the set of all connections. P is defined as P=N×N×T×T×O×D where x denotes Cartesian product. Let the set of connections in a particular real-world configuration be denoted as P', such that

$$P' = \{C : C \in P \text{ and } C \text{ is a connection on this configuration}\} \tag{2}$$

P' is a subset of P. We say that P' is a relation over the sets N1,N2,T1,T2,O, D, where the set of all nodes appearing in the first position in the tuples in P' is denoted N1. Similarly, the sets of all nodes appearing in the second, third, and fourth positions in the tuples in P' are denoted N2, T1, and T2, respectively. Obviously, N1,N2 ⊆ N and T1, T2 ⊆ T.

P' can be incorporated into a 7-tuple G as follows:

$$G = (N1, N2, T1, T2, O, D, P') \tag{3}$$

The 7-tuple G is a graph, and can be expressed graphically as illustrated in Section 3.3. (d can only be conveniently illustrated graphically on paper for planar topology structures.)

### 4.1.2   Describing the Topology Structure

We have shown that a relation (and therefore a graph) can be used to describe a configuration. However it remains to be shown, a) how a topology structure is specified, and b) how a specified topology structure gives rise to a configuration.

We discuss a) in this section. b) is discussed in the Section 4.1.3.

## A Topology Structure Description

From (1) (Section 4.1.1, p.42), each connection in a configuration is an element of P'. Given a connection (n1,n2,t1,t2,o,d), let (n1,n2) describe the two nodes, and (t1,t2,o,d) describe the nature of the inter-connection between that node pair. Let two sets I and S be defined as

$$I = \{ \ (n1,n2) : (n1,n2,t1,t2,o,d) \in P' \ \} \tag{4}$$

$$S = \{ \ (t1,t2,o,d) : (n1,n2,t1,t2,o,d) \in P' \ \} \tag{5}$$

I is a relation over the sets N1, N2. S is a relation over the sets T1, T2, O, D. I describes pairings of two nodes on a configuration. S specifies how two nodes on a configuration can be paired. I is meaningful only within the context of a given configuration, since it refers to pairs of real nodes, however S is meaningful for all possible configurations, and thus is indicative of the topology structure. Since S specifies the set of all possible types of connection between nodes that exists in a particular configuration, we regard it as a template for legal connections C on a topology. We call S the 'topology structure relation' and each element of S a 'TopologyConnection'. Let TS be the set describing all possible topology structures, and let $k = 2^{|T1xT2xOxD|}$ be the cardinality of TS:

$$TS = 2^{T1xT2xOxD} \tag{6}$$

From (6), legal connection combinations on a given topology are given by an element of TS. The example of a Bittorrent (Section 2.6) topology (see Section 3.3, p. 36, Figure3.4) template (or element of TS) is given below.

| | t1 | t2 | o[1] | d |
|---|---|---|---|---|
| | | | ***Bittorrent*** | |
| Grey_to_Node | grey | node | Once | null[2] |
| Node_to_Grey | node | grey | Once | null |
| Node_to_Node | node | node | Lasting | null |
| Black_to_Node | black | node | Intermit. | null |
| Node_to_Black | node | black | Intermit. | null |
| Black_to_Grey[3] | black | grey | Lasting | null |
| Grey_to_Black[4] | grey | black | Lasting | null |

[1] Here we use notional values for o (ConnectionType), that describe Bittorrent connections. 'Once' indicates a TopologyConnection that connects nodes for one operation (obtaining the torrent); 'Lasting' indicates a TopologyConnection that connects nodes for a continuously running operation (such as continuous downloading); 'Intermit.' indicates a TopologyConnection that connects nodes for intermittent operations (referral to a tracker server).

[2] There are no directed edges (and hence TopologyConnections) on this topology.

[3] [4] These TopologyConnections are not illustrated in Figure 3.4. They provide a link between the grey and black nodes on the configuration. This link is provided because the IDAF does

not currently allow disconnected topology (graph) structures; if the grey and black nodes appear on the configuration first, this would be the case, were they not connected as shown. (See Section 4.1.3 for the process of inductively generating configurations.)

If the topology structure relation S is obeyed, no two nodes n1 and n2 can be connected in an illegal fashion. This description is sufficient to describe the structural component of any configuration that has been generated according to a topology structure specified by S.

However, we cannot build configurations from S, since S does not specify how nodes assume a NodeType, when nodes make a connection corresponding to a TopologyConnection, and a number of other factors. We extend the description of a topology structure relation below to account for these factors. In Section 4.1.3, we use this expanded description to show how configurations are constructed from topology templates.

## Extensions to the Topology Structure Description

- Choosing a NodeType

Every node, once it has been allowed to participate in a configuration, must adopt a given NodeType. This NodeType will be negotiated between the joining node and one or more pre-existing nodes on the configuration, and perhaps changed via some process at a later stage. How this occurs is dependant on the topology.

- Re-usable TopologyConnections

Re-usable TopologyConnections are those which can describe more than one connection on the configuration. On a star or tree topology, for example, it is desirable to specify a 'root/parent to leaf' TopologyConnection that specifies a number of actual connections, since a number of leaves may attach to a root/parent node. Re-usable TopologyConnections are particularly useful when there is no limit on the number of actual configuration connects that can be made (since in this case, specifying x separate TopologyConnections is impossible). We extend S to include the concept of re-usability by adding a 'maximum number of uses' element.

We introduce the set Q that contains all of the possible options for this parameter. Q is defined as $Q=\{ \ \mathbb{N}-\{0\} \ \}\cup\{\infty\}$ where $-$ denotes set subtraction. The new topology structure S is thus defined as (a more rigorous definition is not given; the reader is referred to the previous rigorous definition for further detail)

$$S \text{ is a relation over the sets } T1, T2, O, D, Q \tag{7}$$

- NodeType Masks

It may be useful in certain topologies to specify 'NodeType masks' in place of NodeTypes for t1 and t2. A NodeType mask specifies one or more NodeTypes: it is a filter (such as a regular expression for strings) that specifies a set of NodeTypes. For example, in a tree topology, it may be desirable to specify that a leaf node can connect to either the root node *or* to another leaf node (its potential parent). In this case, t1 and t2 are sets of nodes, and T1 and T2 are sets of sets of nodes, though otherwise S remains the same.

It should be pointed out that this (re-)definition of S is not meaningful within the context of P', from which it was originally derived, since each t1 and t2 in P' ((5) p.43) refer to one source and one destination node. However, a single NodeType is ultimately matched to the NodeType mask, and so the mask is best thought of as 'any one matching NodeType' (and therefore legal in an expanded definition of P').

- Required and Optional TopologyConnections.

Some TopologyConnections are 'required' to be satisfied by a topology: for example, on a star or a tree topology, the leaf must satisfy its TopologyConnection that specifies a connection to the root or parent node. However, for the root or parent, it is not essential for them to connect to a leaf in order to join the configuration: therefore TopologyConnections specifying these connections are 'optional'. Each NodeType has two sets of TopologyConnections, REQ and OPT, which correspond to TopologyConnections that are either required or optional (respectively) for nodes of a given NodeType to satisfy.

$$REQ = \{(s, f) : s \in S \text{ and } f \text{ is true}\} \tag{8}$$
$$OPT = \{(s, f) : s \in S \text{ and } f \text{ is false}\} \tag{9}$$

We can expand S to include a flag 'f' that indicates whether a TopologyConnection is required or optional. Let F = { true, false } and f ∈ F.

*Then S is defined to be a relation over the sets T1 , T2 , O , D , Q , F* (10)

(Since we extend S to include f, REQ and OPT become, REQ=(S)=(t1,t2,o,d,q,f) where f is true, and OPT=(S)=(t1,t2,o,d,q,f) where f is false.)

This is sufficient for a topology specification within the IDAF in the case where TopologyConnections can be dealt with in isolation: in other words each TopologyConnection in REQ or OPT for each node on the configuration can be considered and satisfied in isolation to the others. We call topologies where this description is adequate, 'unstructured'. (See (a) in Figure 4.1 and examples below (p.48).)

- Contingent TopologyConnections and Structured Topologies

For some topologies we want to specify that in order for a given TopologyConnection to be satisfied, one or more other TopologyConnections must also be satisfied simultaneously. Such a TopologyConnection is said to be contingent on other TopologyConnections.

We introduce the notation $CGT_i$ to denote a particular subset of topology connections that are contingent on one another. In sets of TopologyConnections, $CGT_i$ , where $CGT_i \subseteq REQ \ or \ CGT_i \subseteq OPT$ [52], each TopologyConnection in $CGT_i$ must be satisfied simultaneously. If all cannot be satisfied, then none are satisfied. Where $CGT_i \subseteq REQ$ all TopologyConnections must be satisfied, regardless of whether there are nodes on the configuration to satisfy them or not; where $CGT_i \subseteq OPT$ all TopologyConnections must be satisfied, provided nodes exist in order to satisfy them. In both cases, we say that TopologyConnections in these sets are 'contingent' upon each other, or are 'mutually contingent'[53]. We call topologies which require sets of contingent connections to be defined, 'structured'. (See (b) in Figure 4.1 and examples below (p.48).)

---

[52] CGTi is a subset of REQ or OPT, not a subset of the union of both, because it makes no sense for a REQ TopologyConnection to be contingent on an OPT TopologyConnection, or vica versa, since both cause a logical conflict in the definitions of REQ and OPT TopologyConnections.

[53] Intuitively, we could have asymmetrically, rather than mutually, contingent TopologyConnections, however, it is not clear whether this more complicated description would yield any additional meaningful topologies.

In structured topologies, where there is a defined relationship between contingent TopologyConnections in a set $CGT_i$, such that a path can be traced leading away from a given node in the direction (d ∈ D) of one of its TopologyConnections, via other nodes, and back to the given node in the opposite direction of one of its other TopologyConnections, we say that these topologies are 'rigidly structured'. (See (c) in Figure 4.1 and examples below) This path, or circuit, is indicative of the fact that a node's mutually contingent neighbours, which are only one edge away from the node, must be addressable and contactable from every other neighbour, so that a node, in connecting to one neighbour, can also connect to the correct 'contingent neighbours'. (Neighbours are chosen based on 'location' as well as NodeType.) This concept is illustrated in an example below (p.48), and also in the worked example of a planar mesh on page 52.

The concept of TopologyConnection contingency can be incorporated into S by giving each TopologyConnection in a contingent set a reference to the relevant $CGT_i$ it is a member of.

Let R be defined as $R \subset \mathbb{N}$, where each r ∈ R is a unique reference for each $CGT_i$. Now S is redefined as

$$S \text{ is a relation over the sets } T1, T2, O, D, Q, F, R \qquad (11)$$

This is the definition of S that we will use in Section 4.1.3 and from now on.

- Summary and Examples of unstructured, non-rigidly structured and rigidly structured topologies.

In this thesis we will refer to unstructured non-rigidly structured and rigidly structured topologies. The reader will note that in order for a topology to be a non-rigidly structured, just one set $CGT_i \subseteq REQ$ or $CGT_i \subseteq OPT$ is required: there may be any number of other TopologyConnections which are not contingent on any other, and therefore it may have unstructured and structured elements. Similarly, in order for a topology to be a rigidly structured, just one set is $CGT_i \subseteq REQ$ _or_ $CGT_i \subseteq OPT$ required, where members of $CGT_i$ have a defined (rigid) relationship between them. There may be other sets $CGT_i$ where this is not the case, and other TopologyConnections that are not mutually contingent at all. For simplicity, the examples below do not contain the 'structural mix' that is possible for structured topologies (rigid or otherwise). We will often use such 'simple' examples for clarity. See Figure 4.2 for an example of a 'mixed' structure.

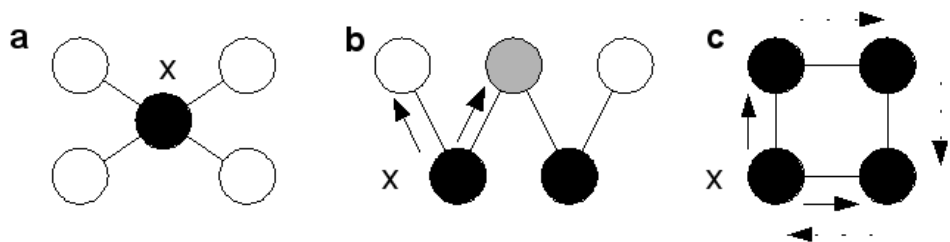

*Figure 4.1.: Shown is (a), an unstructured topology, (b), a non-rigid structured topology, and (c), a rigidly structured topology. Solid arrows indicate TopologyConnections that are mutually contingent for node 'x'. Dashed arrows show a path originating from the node at the end of one TopologyConnection (north) and ending at node x at the end of a reciprocal to a mutually contingent TopologyConnection (east) on a rigidly structured topology.*

An example of an unstructured topology is the star, illustrated in Figure 4.1 (a): leaves can connect to a root node without having to consider any other TopologyConnection; root nodes can connect to a leaf node without having to consider any other TopologyConnection. (Generally, trees are also unstructured topology structures.) An example of a non-rigid topology structure, illustrated in Figure 4.1 (b), is one in which a black node must connect to both a white and grey node in order to connect to the configuration (i.e. $\{\mathrm{to\_white}, \mathrm{to\_grey}\} \subseteq OPT$ [54]), although there is no other specified relationship between the white and grey nodes. An example of a rigid topology structure is a planar mesh, e.g. of degree four, illustrated in Figure 4.1 (c), where a node cannot connect to a node to its north position (indicated by its 'northTopologyConnection', where d='north'), and then connect to a completely arbitrary node with its eastTopologyConnection (whose d='east'), since there is a defined topographical relationship between north, south, east and west. (In order for the topology structure to remain viable, it must connect to a node in its east position that is south and east of the node it connected to in its north position, and so on.) The contingent relationship can be described by $\{north, south, east, west\} \subseteq OPT$ [55], although rigid structures require additional logic in their definition, specifying the spatial relationship between the contingent TopologyConnections. That is, the topological framework in which the d in S is meaningful.

---

54 to_white, to_grey are names we give to TopologyConnections in S.

The IDAF specifies that there be a mechanism for describing the relationship between mutually contingent TopologyConnections on a rigidly structured topology. (A topographical, coordinate based approach is used by the IDAS: see Sections 5.3.2 for the IDAS Locator class that provides this logic in our prototype implementation.)

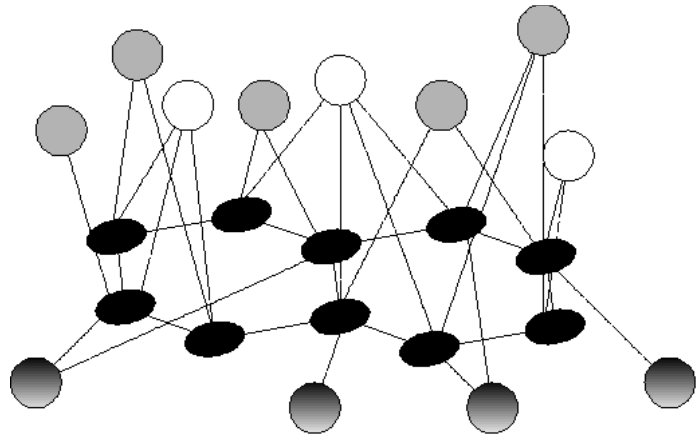

*Figure 4.2.: Rigidly structured topology (with 'non-rigid' elements). Black nodes are in rigid CGTi (middle); white and grey nodes are in non-rigid CGTi for black nodes (top); gradient nodes are non-contingent OPT TopologyConnections for black nodes (bottom).*

### 4.1.3  Building Configurations from Topologies.

We now examine how configurations are built. By building a configuration we mean how a particular configuration evolved or grew into its current state. P', from equation (2) (Section 4.1.1, p.42), is the set of all configuration connections. This set can be built up in two ways:

1. Continuously forming supersets from other configurations. This would be specified by, $P' = A_n \cup (A_{n\text{-}1} \cup (A_{n\text{-}2} \cup (A_{n\text{-}3} \cup ...)))$ where each $A_i$ is an existing configuration.

2. Forming a single set, where nodes are added one by one, specified by $P' = \{sc(a_n)\} \cup (\{sc(a_{n\text{-}1})\} \cup (\{sc(a_{n\text{-}2})\} \cup (\{sc(a_{n\text{-}3})\} \cup ...)))$ , where each

---

$a_i$ is a node, and the function $sc(a_i)$ gives the connections that are satisfied when $a_i$ joins P'.

3. Continuously forming supersets from other configurations *or* from the satisfied connections of nodes added one by one, specified by $P' = \{A_n \text{ or } sc(a_n)\} \cup (\{A_{n-1} \text{ or } sc(a_{n-1})\} \cup (\{A_{n-2} \text{ or } sc(a_{n-2})\} \cup ...))$ , where $A_i$ , $a_i$ and $sc(a_i)$ are defined as above. (Effectively allowing P' to be formed by either the process outlined in 1. or that outlined in 2.)

Scenario 1 suffers from a number of major drawbacks: a) In the case of topologies that have a fixed or a maximum number of nodes of given NodeType, which could be duplicated across any configurations A and B, integration is problematic, since one or both configurations must be substantially altered, b) there is never a definitive value of P', since any potential P' may ultimately be part of a union with another set, forming a new P' (making configuration optimisation difficult, see Section 4.3.1), c) For two structures to combine according to TopologyConnections in S, non-local knowledge may be required. (e.g. Take any two planar mesh configurations A and B of large extent. In order for edges to be placed between any two nodes bridging the two, their rigidly contingent connections must be also be satisfied, along with the contingent connection sets of every node on the unbounded face of either mesh.) We have already stated that all configuration connectivity should be subject to local knowledge only (Section 4.1.1, p.41).

Scenario 2 has the advantage of being simple, in that there is always a defined value for P' at all times. Nodes may be added one by one using local knowledge only. Building a configuration node by node is more intuitive, and likely to be far less complicated. Also, in practice, since configurations operate on the basis of local knowledge only, nodes may be added simultaneously at different locations on the configuration, since, in any case, it is impractical to monitor events at arbitrary positions on the configuration. What is important is that each node considers only one TopologyConnection at a time.

Scenario 3, by accommodating the formation of P' via either the process outlined in Scenario 1 or the process outlined in Scenario 2 suffers from the drawbacks of 1 while not benefiting from the advantages of 2.

For these reasons, the IDAF builds configurations as outlined in scenario 2.

Since TopologyConnections are defined from the perspective of a node n1 of a given NodeType, there is always a 'reciprocal' TopologyConnection defined from the perspective of a node n2 of given NodeType that refers to the same edge on the configuration. If a node of NodeType A has a TopologyConnection X that connects to a node of NodeType B, and the node of NodeType B has a corresponding TopologyConnection Y that connects it to the node of NodeType A, we say that Y is the reciprocal of X, and X is the reciprocal of Y.

Having the IDAF build a functional and logical configuration from a topology can therefore be thought of as an inductive process, starting from the first node and building from there:

1.  The first node in a configuration must be of a NodeType that has no required TopologyConnection (REQ={}), since there are no other nodes to connect to.

2.  The second node in a configuration must be of a NodeType that has a REQ or OPT in which there is a TopologyConnection that allows it to connect to the first node of given NodeType. Conversely, the NodeType of the first node must have a TopologyConnection in OPT that is the reciprocal of this TopologyConnection.

3.  The nth node in a configuration must be of a NodeType that has a REQ or OPT in which there is a TopologyConnection that allows it to connect to at least one of the n-1 nodes of a given NodeType. As in 2, a reciprocal connection must also exist from the node it connects to.

Therefore, at a minimum, the topology must specify, a) what NodeType a node must be if there are no other nodes on the configuration, b) what NodeType a node must be if there is only one pre-existing node (of given NodeType) on the configuration, and c) what NodeType a node must be if there are n-1 pre-existing nodes (of given NodeTypes) on the configuration, where n>2. (There are topologies in which b) can be omitted and c) made valid for n>1, i.e. a star. Likewise, topologies can be conceived which might have k 'b)' points, before the inductive c) where n>k.)

In summary, the IDAF builds the configuration from P'={} by adding nodes (by satisfying their REQ and OPT TopologyConnections) one by one, giving rise to a dynamic, though continuously viable P'.

The pertinent aspects of a topology's structural specification are listed below:

1.  The sets T (T1=T2=T because of reciprocal connections), O (ConnectionTypes), D (directions), and R (CGTi references),  must be defined[56].

---

2. A relation S must be defined, which describes the topology structure. The full extended S is given by S=(t1,t2,o,d,q,f,r) over the sets T1, T2, O, D, Q, F, R. (Equation 11, p.47).

3. For each NodeType in T ( $T1 \cup T2$ ) two sets REQ and OPT must be defined, to indicate the required and optional connections, respectively, that nodes of a given NodeType must satisfy[57].

4. Sets $CGT_i$ (subsets of a NodeType's REQ or OPT) must be defined for structured topologies, indicating sets of mutually contingent TopologyConnections[58].

5. The logic for selecting NodeTypes for nodes connecting to the configuration should be consistent with the inductive process outlined above.

Any topology following this specification can be implemented, and used by an implementation of the IDAF, provided that local knowledge is sufficient for connections to be made. (See Section 6.4.2 or more on scalability.) 'Local knowledge only' is an important restriction on topologies to help ensure that topologies can accommodate an arbitrarily large number of nodes, which they must do (Section 1.5.1 (4)). However, it is still possible to create topologies that do not scale, even when obeying this requirement (see Section 6.4.2).

An example of a node attempting to connect to a (rigidly structured topology) configuration is given below. Full specifications for sample topologies are given in Section 6.2.

We now examine how a node connects to a planar mesh of degree 6. (See Figure 4.3.) Here P' will have 6 contingent connections added: 3 from the perspective of the joining node, and 3 reciprocal connections (represented by the edges XA, XB and XC), owing to the admission of a single node to the configuration.

---

[57] The REQ and OPT for each NodeType can be obtained from S; or f in S can be obtained from REQ or OPT, see definitions, p.45.

[58] CGTi can be generated from S; or r in S can generated from the set of CGTis in a topology, see definitions, p.47.

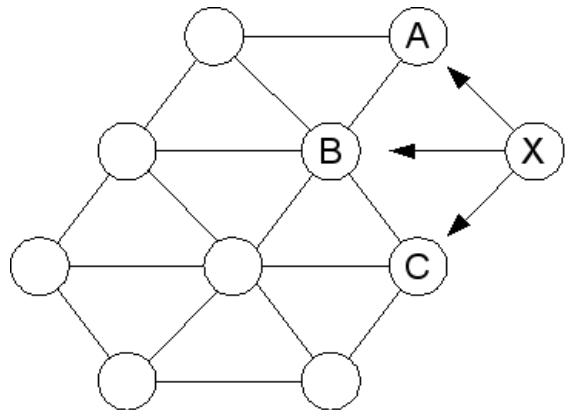

*Figure 4.3.: Regula, planar mesh of degree 6.*

In Figure 4.3 a node X attempts to attach to the configuration. It will try to satisfy 6 optional rigidly contingent TopologyConnections ( $CGT_0$ where $CGT_0 \subseteq OPT$ ), which we can think of as radiating outward at the following angles, $0°$, $60°$, $120°$, $180°$, $240°$ and $300°$. (All nodes are of the same NodeType. TopologyConnections are optional because nodes may not be available on the configuration to satisfy all of them, as is the case when $CGT_i \subseteq OPT$ .) X tries to connect to the configuration at the point illustrated. For argument's sake, it begins by trying to satisfy its TopologyConnection at $120°$ by attempting to connect to node A. Because all TopologyConnections are mutually contingent, in order for one to be satisfied so must the others, provided the nodes exist to connect to. X sees that, of the TopologyConnections contingent on $120°$ to A, nodes B and C exist to satisfy TopologyConnections at $180°$ and $240°$, therefore it must satisfy all three TopologyConnections simultaneously. Once X has satisfied these three topology connections, and A, B and C have satisfied their reciprocal topology connections (to X), X is said to be joined to the configuration. (The IDAF does not specify exactly how the relationship between rigidly structured, mutually contingent TopologyConnections is defined, however a topographical, (local) co-ordinate based approach is used by the IDAS. (See Section 5.3.2.)

## 4.2  The IDAF Layer Model

In the preceding sections, we used terms such as overlying and underlying in order to refer to the relative level of design functionality we were operating on. These levels can be seen as layers, analogous to OSI Layers [Zimmerman '80]. (In fact they can be viewed as extensions of, or additional layers upon, this model.)

### 4.2.1   The Network Layer

The network layer conveys traffic between computers participating in IDAs. It is the lowest layer in the IDAF layer model, and the layer that resides just above the OSI Model Application layer. (The OSI Model Application layer, in the current incarnation of the internet, corresponds to the IP layer.)

The network layer conveys traffic between computers by utilising the underlying IP/TCP/UDP network protocols. It puts computers on a level playing field by transparently bypassing firewalls and other network constrictions.

In Section 3.2, we gave the case for using P2P as a platform for the IDAF. The P2P overlay network comprises the bulk of the network layer in the IDAF. It gives the network layer desirable features, such as automated peer discovery and peer view, resource searching, and recently, features such as swarm file download and multimedia multicast streaming. (Though these features do not yet exist in a single overlay network.)

### 4.2.2   The Topology Layer

In Section 4.1.1 we introduced the concept of the topology. Topologies dictate how a configuration of nodes inter-connect over the course of the running of an IDA. The topology layer is responsible for exercising the logic within the topology to generate the configuration. This layer resides above the network layer and organises the unstructured network of peers provided by that layer.

Whereas all IDAs are intended to use the same network layer, each IDA will have its own configuration, and hence its own context within the topology layer.

### 4.2.3   The Application Layer

The application layer is the highest layer in the IDAF, and resides on top of the topology layer. It is at this layer that the logic for the actual specific application is executed. For example, even though SETI@home and climateprediction.net, if they were to be implemented within the IDAF, would both share the same P2P overlay network within the network layer, and use an identical topology (though obviously two different configurations with different participants would be formed for each), a star, they would differ in their application layer implementation. One would use a star topology to help search for extraterrestrials; the other would use a star to model climate change. (Therefore both applications differ only at the application layer.)

Functionality at this level accesses topology layer functionality, which acts as the infrastructure for the application layer.

Application layer functionality is implemented as follows. The IDA is comprised of the sum of all micro-interactions between nodes on the configuration. Each node has a duty to perform on the configuration, just as each NodeType plays a role within the topology. Each node executes a software component, which we call a Nodelet. Each NodeType in the topology has a Nodelet mapped to it, which controls the actions of each node of that NodeType. Thus, the application layer functionality comprises a set of Nodelets, NLD, where |NLD| = |T|, and there is a one to one mapping between NLD and T[59].

Nodelets use connections (based on TopologyConnections) to communicate with other Nodelets. (See Chapter 5 for the IDAS, a prototype implementation of the IDAF.)

Design at the application layer, therefore, consists of reducing a problem to a form where it is represented by this set of Nodelets. (A process which is analogous to designing individual bees: worker, queen and drone, in order to give rise to the hive.) (Application layer IDA implementations are given in Chapter 6.)

### 4.2.4   Summary of IDAF Layers

The full IDAF layer model, is illustrated below in Figure 4.4.

---

59  T is the set of NodeTypes in the topology (p.42).

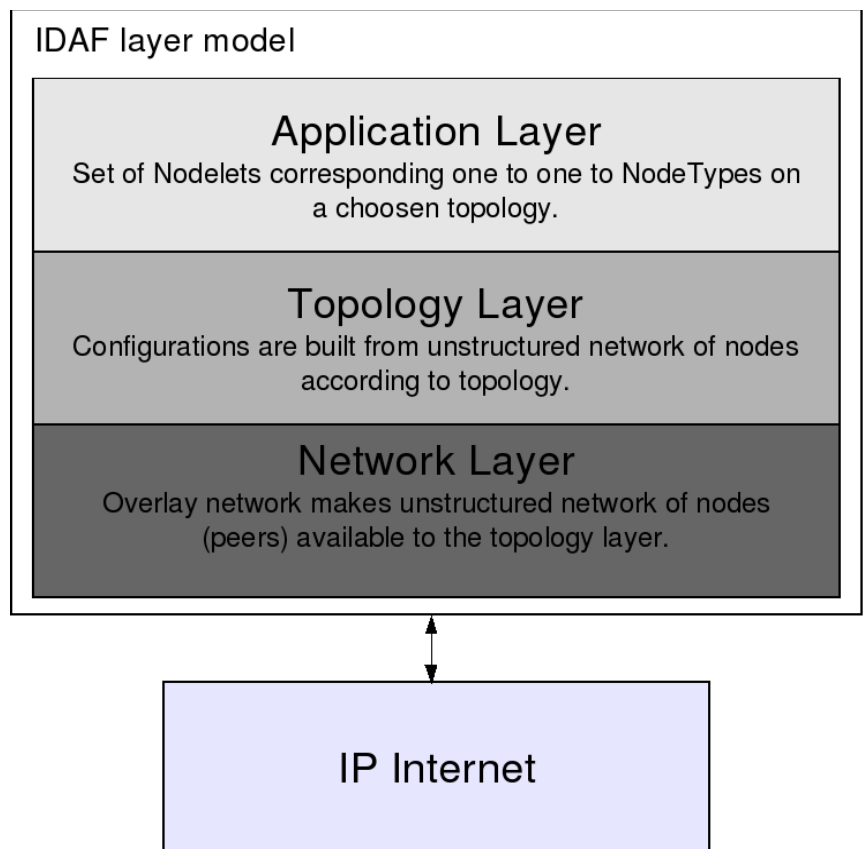

*Figure 4.4.: IDAF layer model*

It is anticipated that a system based on this IDAF layer model will provide the following gains while running a number of IDAs, when compared with a number of uncoordinated simultaneously running monolithic IDAs:

1. A common network layer, which can be used by all IDAs, and the potential for caching within the P2P overlay network.

2. Re-usability of topologies. Once a topology is built to support one IDA, it can be re-used for others.

3. Separation of network layer development, topology development and application layer (Nodelet set) development. (Allows developers to develop in a modular fashion and also to develop in areas in which they are competent, or interested.)

4. The barrier to IDA development is reduced. With a common network layer, and a number of topologies already in existence (in time), all that remains is for the application layer functionality to be built. This allows scientists and enthusiasts to develop IDAs quickly without having to divert themselves significantly from their full-time activities.

## 4.3 IDAF Constraints

We outline here constraints that any IDAF implementation ought to satisfy, further to those already stated.

### 4.3.1 Primary IDA Management Mechanisms

In Section 4.1.1 we introduced the concept of the topology. We specified that in addition to describing the topology structure, topologies must also define the logic to maintain, refine and evolve the configuration. These three processes, along with the process of joining a node to a configuration, are crucial to IDA management. The IDAF uses the logic in the topology to specify how an IDA should be managed for the duration of its execution. Since the IDAF specifies that many IDAs can be run simultaneously, it must conduct these management processes in separate runtime contexts.

In each context the following processes can be run (the first two are required):

- The Join Process: Executed once when the node joins the configuration. Starts the maintenance/optimisation process.

- The Maintenance/Optimisation Process: Runs continuously to ensure the configuration remains viable and optimised/refined. It is convenient to combine the two concepts into a single set of instructions or guidelines.

- The Evolution Process: Runs once triggered by a condition detected by the maintenance/optimisation process. (This process is considered optional.)

Clearly, individual nodes cannot maintain or evolve the configuration independently. They merely take their part in a global effort to achieve the required effect. It is intended, particularly in the case of the maintenance/optimisation process, that iterative mechanisms be used on each node to achieve this global affect. We have not looked at the evolution process in any great detail. It may be very difficult to implement in practice, particularly while maintaining the requirement for 'local knowledge only' (see Section 4.1.1 p.41). In any case, we retain the concept, while mandating that, as with the maintenance/optimisation process, the bulk of the logic comes from within the topology itself; thus the difficulties are outsourced to the topology developer, who will nonetheless have the option of implementing it if required.

### 4.3.2   From the Developer's Perspective

Each IDAF layer will be separated from the other layers by an API. Higher layers will access lower layers through these APIs. These APIs must be developed carefully so that changes to them are rare. In this way, modifications to functionality in one layer requires no modification of functionality in the other layers.

These APIs must be flexible, and easy to use for those developing at the next layer up, who may have limited understanding of the mechanisms at work at the given layer.

### 4.3.3   From the User's Perspective

The IDAF mandates a common network layer for all running IDA Nodelets. This ensures that the total bandwidth and system resource usage consumed by all Nodelets is minimised, which, if not a specific requirement, is certainly one of the goals of the IDAF.

The IDAF requires that there be a software component which allows the user to publish and discover (the existence of) IDAs, and to manage running Nodelets. This component will also allow the user to participate in, or withdraw from, the running of discovered IDAs, by means of a graphical user interface (GUI). It will optionally give the user additional information on the status of running IDAs, and a summary of system resource usage.

This software component will also allow running IDA Nodelets to present the user with a GUI component in order for the user to interact with a running Nodelet, and hence a running IDA.

In this chapter we discussed the IDAF. We introduced the topology structure description, the IDAF layers and the IDAF constraints In the next chapter, we will discuss an implementation of the IDAF, the IDAS.

## Chapter 5   An Internet Distributed Application System

The IDAS is a software prototype implementation of the IDAF. The IDAS implements the core functionality of the IDAF. Where functionality has not been implemented, this is indicated. We discuss the IDAS in this chapter.

## 5.1   IDAS Design & Implementation

The work of every software project can be broken down into 'design' and 'implementation'. Chapter 4 gives design detail as part of the IDAF description. Additional design detail, specific to the IDAS, is presented in this chapter. All implementation detail for the IDAS can be found in this chapter. We clarify below where design and implementation detail for the IDAS can be found.

The IDAF describes the core aspects that each implementing system will inherit. It describes the main components and processes, but is not otherwise prescriptive where it is felt that aspects of the design should be devolved to specific IDAF implementations Elements of the design, such as the desktop GUI software, or the communication mechanisms used between connected peers on a configuration, are not considered critical to the IDAF, and therefore are given in this chapter, in the context of the IDAS.

The IDAF gives the layer model design in Section 4.2. Implementation detail for the layer model can be found in Section 5.3.

The IDAF gives the design detail on 'processes' in the IDAF in Section 4.3.1. Implementation detail for the processes is given in Sections 5.4.2 and 5.4.3.

The software client, introduced in Section 4.3.3, is described in Section 5.2. Here, design and implementation detail is merged, since although the software client is significant from an operational standpoint, it is not significant in terms of proving the IDAF.

## 5.2  Software on the Client

The IDAS was written in Java [Java '06] (J2SE 1.4.2) and uses JXTA [JXTA '06] (J2SE 2.3 "Jambalaya") as the P2P Overlay Network for its network layer. (JXTA has a Java and a C implementation.) At the time of writing, the IDAS comprised 60 java classes, giving a total of 14,849 lines of code.

### 5.2.1  Software Overview

In Section 3.1.1, we introduced the concept of the generic software component that is capable of participating in the running of arbitrary IDAs. We give an overview of this software component here.

Since the software is written in java, it can be run on any platform that is capable of running a java virtual machine. The software component can be launched in the same way as any other java application. (See supplementary CD for instructions.) We will refer to this software component, from now on, as 'the client'. (Although the term 'client' originally referred to a client in a client-server configuration, it has also come to mean any software component on a users computer that interacts with an application that runs across more than one computer.)

Aside from the JVM installation, which contains the standard java classes, or Java Development Kit (JDK), the JXTA and JXTA related jar files, and the IDAS jar file, must be installed. These are included and incorporated into the client by means of the java 'classpath'.

The client, when run, begins by bootstrapping the JXTA P2P overlay network. After the overlay network is bootstrapped it begins to interact with the P2P network. The user is then presented with a dialog showing a list of IDAs that the client knows about: i.e. the ones which it has prior knowledge of. Other IDAs are added as they are automatically discovered. (See Figures 5.1 and 5.2.) Users can then select an IDA and choose to participate in the running of it, or later, to withdraw its participation.

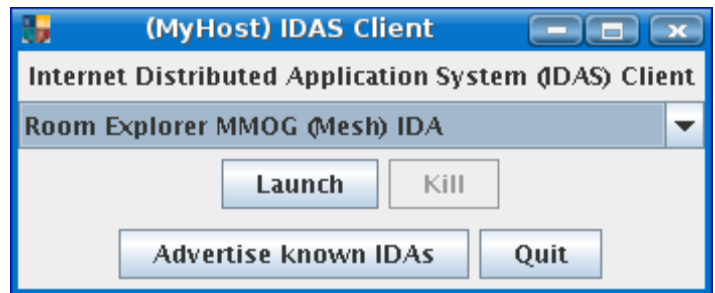

*Figure 5.1.: IDAS Client GUI*

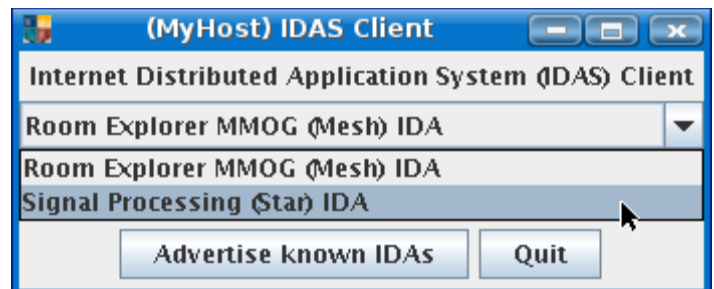

*Figure 5.2.: IDAS Client GUI with pull down list of known IDAs*

Each running IDA component, corresponding to an IDA the user (via the client) has decided to participate in, will normally have a GUI of its own that allows users to interact with the IDA. An example of an IDA GUI is shown in Figure 5.3.

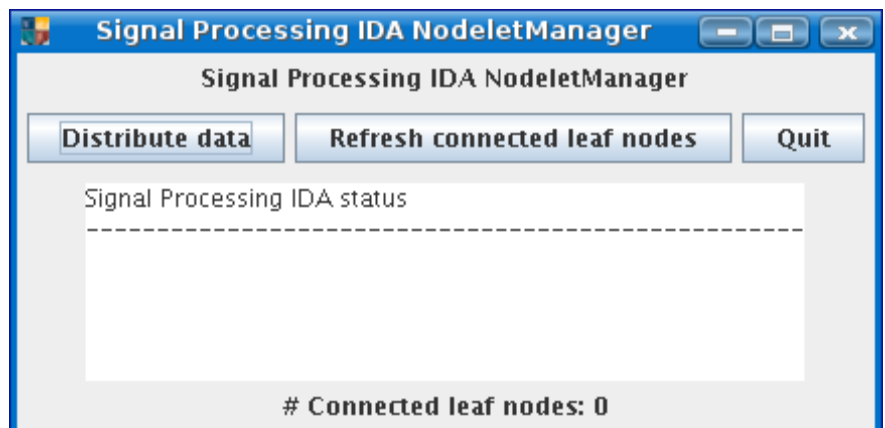

*Figure 5.3.: IDA GUI spawned separately by client*

### 5.2.2   Java Class Structure

The IDAS java code is divided into 8 categories of functionality, or 8 java packages. These packages reflect the main foci of this IDAF implementation. (The full java package name for each of the packages below can be obtained by prefixing 'ie.nuim.eeng.IDAS.' to the underlined names below.) For a functional description of the IDAS software, see Section 5.2.3. We will make use of java design pattern [Cooper '06] terminology here. Pattern names will be in italics.

## Configuration

The Configuration package contains classes that maintain and record the state of a given running configuration. It contains the ConfigurationStabiliser class, which is a Thread that is tasked with joining, refining and evolving running configurations (i.e. running the processes outlined in 5.4.2 and 5.4.3).

## Network

The Network package contains classes that govern the network layer (Section 4.2.1). In the IDAS, the chief function of the Network layer is to provide a *façade* and *adapter* API that bridges the gap between the IDAS network processes and its JXTA core. (This is accomplished by the NetworkListener class.) It also contains further *façade* classes that simplify access to the network layer for classes in other packages.

## Nodelet

The Nodelet package contains the Nodelet (see Section 4.2.3) specification classes. From the IDAF, all Nodelets should extend the Nodelet class, however, in this implementation the Nodelet class provides the Nodelet thread, and Nodelets are created by extending the NodeletEngine and NodeletNodeManager (and optionally Interpreter) classes. (The difference is due to recent revisions to the IDAF not being reflected in the IDAS at the time of writing. Both approaches are, however, functionally equivalent.)

## Platform

The Platform package contains the macro components of the client. The Platform class represents the client software: it is a thread that launches the GUI, and starts the threads that establish the overlay network and instantiate the Container class thread. The Container class manages running IDA software component instances.

## Processes

Processes used by the ConfigurationStabiliser are implemented here. They extend the Process class. The only process currently implemented is the JoinProcess, with stubs for the MaintainProcesss and EvolveProcess, specified by the IDAF, but which were not implemented at the time of writing.

## Testbench

The Testbench package contains classes which are used to test the IDAS. The PrintState class is used to produce output to the screen or to a file that can be used for debugging.

## Topology

The Topology package contains the topology specification (see Sections 4.1 and 4.2.2) classes. Each topology must, at a minimum, extend the Topology class. Rigidly structured topologies (see Section 4.1.1, p.47) need also refer to an implementation of the Locator class, which can specify arbitrary rules for topology topography, and a localised relative co-ordinate system. (See Section 5.3.2 for more on the Locator class.)

## Util

The Util package contains miscellaneous utility classes, used by the IDAS.

(On a supplementary CD accompanying this work, the entire source code for the IDAS is given along with javadoc documentation.)

### 5.2.3   IDAS Execution and Threading Model

The IDAS client is a multi-threaded[60] application. This is because many tasks must be accomplished simultaneously. Clarifying what tasks these threads perform, and how these threads interrelate, is necessary for describing how the software elements of the IDAS client function.

When the client is bootstrapped, one thread is launched. This is called the Platform thread. The Platform thread launches two threads: the Network thread and the Container thread. After launching these threads, it launches the IDAS client GUI, and serves as a listener for this GUI. The Network thread is responsible for instantiating the JXTA overlay network. The Platform and Container threads wait until the overlay network has been fully bootstrapped, and the client has joined a special JXTA PeerGroup for IDAS clients. When this process has been completed, the GUI widgets, that were formerly shaded, become active and allow the user to participate in the running of IDAs; the Container is also then ready to manage IDA software components.

---

60  We are referring to Threads [Drake '06] in the Java programming language.

When the user participates in the running of an IDA, the Container launches two new threads (for that IDA): the IDA specific Nodelet thread and the IDA specific ConfigurationStabiliser thread. The ConfigurationStabiliser is first responsible for initiating the join process, joining the IDA software component to the IDA configuration, and then for running the maintain process (and possibly the evolve process). (Since the latter two processes are not implemented, the thread loops here.) (There is a discussion of these two processes in Section 5.4.3.) The Nodelet thread waits until the ConfigurationStabiliser thread has joined the IDA software component to the IDA. Once joined, the Nodelet begins to run in earnest, performing whatever tasks are associated with the node's NodeType.

IDA Nodelet threads may launch GUIs of their own, and may in themselves be multi-threaded applications[61].

The interaction between client threads in the IDAS is given graphically in Figure 5.4. The scope of client threads in the IDAS is given graphically in Figure 5.5.

---

61  In this prototype, no facility was made for the main client GUI to incorporate IDA GUI components. However, in a production system, this might be desirable

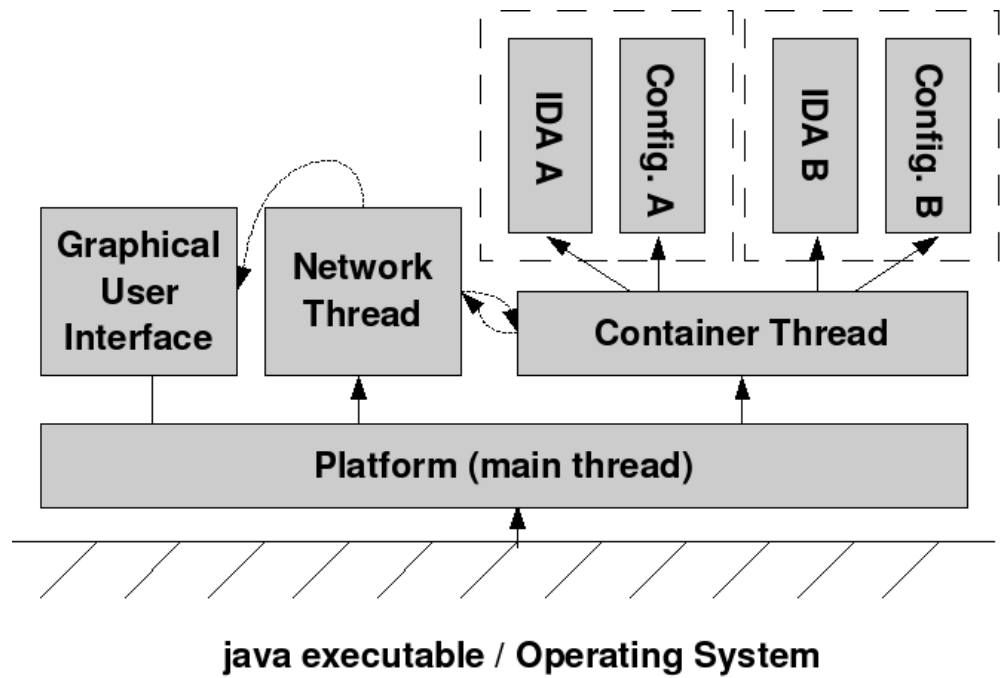

*Figure 5.4.: IDAS client thread interactions*

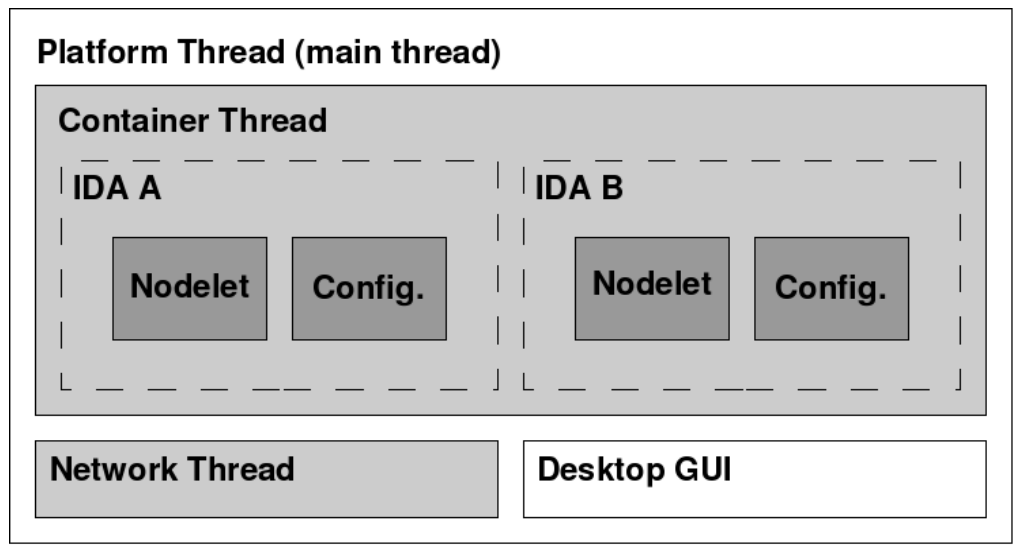

Contiguous areas of shading indicate the scope of a given thread. Threads within that scope are "owned" by that thread. The 'dashed' rectangles surround threads that are closely related.

*Figure 5.5.: IDAS client thread scope*

## 5.2.4   Inter-node Communication

Aside from the underlying overlay network node interaction, conducted by JXTA, nodes communicate in two ways in the IDAS:

1. When running a process, such as the join process.

2. When Nodelets running on different nodes interact during the course of the running of an IDA.

JXTA uses pipes (see Section 2.7) to open lines of communication between peers[62] on the P2P overlay network.

Each peer on the network maintains a CommsPipe, which is a JXTA InputPipe that listens on the network at all times for general inter-peer communication. Peers send data, or (in the IDAS) 'communiques', using a temporary JXTA OutputPipe, which can establish a connection with a CommsPipe, allowing a one-way stream of traffic between peers. During the join process, and during other processes, peers use temporary OutputPipes and the CommsPipe to send and receive messages between other peers. While peers will mostly communicate with peers that they share an IDA configuration with, during the join process peers must communicate with other peers participating in particular IDAs to which they do not yet have a connection (at least in the context of the particular IDA). See Section 5.4.2 for a discussion on the join process.

When a node is connected to an IDA configuration, it will have all of its REQ TopologyConnections and any number of its OPT TopologyConnections satisfied (see Section 4.1.2). For each satisfied TopologyConnection, and hence for each connection on the configuration, a node will maintain a JXTA InputPipe that can receive data transmissions from the node at the other end of the connection. Temporary JXTA OutputPipes are uses to send data. The attribute ConnectionType of a TopologyConnection (or element of S in Section 4.1.1) would specify the properties of these pipes, however this functionality is unimplemented, and essentially all TopologyConnections have the same ConnectionType, which corresponds to a permanently open InputPipe being connected to by temporary OutputPipes at the other end. (Specific TCP/IP characteristics of the connection are left to the JXTA pipe implementations.)

---

62  We use 'peers' here, rather than 'nodes', because we are referring to peers on the network that may or may not be participating in the running of IDAs, rather than nodes on a configuration, which is the sense that we have used 'node' in, for the most part, until this point.

Therefore, aside from maintenance communications, sent to a peer's CommsPipe (during the running of a process), peers also communicate with their 'neighbour' nodes on each IDA they are participating in the running of. Nodes on IDA configurations are characterised by continuously running Nodelets, receiving and processing asynchronous data communications from neighbour nodes on the configuration, and by sending asynchronous communications of its own to these neighbours.

### 5.2.5   Extending the IDAS with Topologies and Applications

A discussion on Topologies and application layer Nodelets can be found in Sections 5.3.2 and 5.3.3 respectively. (They are specified by the IDAF in sections 4.1 and 4.2.3 respectively.)

Topologies and application level Nodelet sets, which we will call Nodeletsets from now on, can be packaged as java jar files or class bundles. Each topology is described by a JXTA Advertisement. Each IDA, which specifies or references a topology, and includes a Nodeletset, is also described by an advertisement. Advertisements are objects that describe distributed resources on a JXTA P2P network, and which are used to advertise their existence on the network. (Advertisements translate into XML messages when shared on the network.)

These advertisements can contain arbitrary data fields (in addition to required fields, such as 'name' and 'ID'): it is anticipated in the future that a URL or unique PeerID would be specified in an advertisement describing where a topology or an IDA can be obtained. (The IDA advertisement should include the topology advertisement of the IDA's topology.) Advertisements for topologies and IDAs can then be propagated throughout the JXTA network and users can discover them. A user will want to discover IDAs in order to participate in the running of them, whereas new (separately advertised) topologies may only be of interest to IDA developers developing at the application layer.

One viable work flow for discovering and participating in IDAs is as follows: 1) the user discovers the IDA (by advertisement) from the network; 2) the client automatically downloads the associated Nodeletset jar, and, via the topology advertisement contained within the IDA advertisement, the topology jar; and 3) with both these jars installed, the user can then attempt to participate in the running of the IDA. In the current implementation, only IDAs have advertisements, and all classes must be pre-installed, however, IDA advertisements are discovered by clients from the network. The above work flow would be trivial to implement. (Requiring only additional (URL/PeerID) information to be provided by IDA and topology advertisements, and the hosting of IDA and topology jars either on the P2P network or on a central server.)

It is through the use of JXTA advertisements, and IDA advertisements in particular, that the automated discovery of IDAs from the network, discussed in Section 3.1.1, is made possible.

## 5.3  IDAS Layer Model Specifications and APIs

In this section, we will discuss the most pertinent aspects of the layer model specifications and APIs.

### 5.3.1  The Network Layer

The network layer API (used by the topology layer) is provided by the NetworkMediator class. It is through this class that other parts of the IDAS software interact with the network. The NetworkMediator class mediates between these other parts and the TrafficListener class, which directly instantiates and maintains the JXTA network (including the network thread, see Section 5.2.3).

The NetworkMediator class contains methods for starting and stopping the JXTA network; joining the special IDAS JXTA PeerGroup, in which all IDAS peers reside, which we call the ApplicationGroup; and for testing the status of the network.

Important methods are listed below for illustrative purposes:

The NetworkMediator is responsible for publishing and discovering advertisements, to and from the ApplicationGroup. The three methods below are self-explanatory, except that the argument for the second one allows a search mask to be set (search by attribute name and value), allowing for more specific advertisement searches. (See Section 5.2.3 for more on advertisements in the IDAS.)

```
void publishToApplicationGroup(Advertisement advertisement)
void discoverRemoteApplicationAdvertisements(String name, String attr)
void discoverAllRemoteAdvertisements()
```

The NetworkMediator must deal with pipe creation and the sending and receiving of messages. The following methods are responsible for this.

```
boolean startInputPipeWithinApplicationGroup(PipeAdvertisement pipeAdv)
boolean startOutputPipeWithinApplicationGroup(PipeAdvertisement pipeAdv)
Vector getPipeMessagesForApplicationGroup(PipeAdvertisement pipeAdv)
void sendMessageToApplicationGroupPipe(PipeAdvertisement pipeAdv, Message message)
```

The NetworkMediator is responsible for providing methods that instantiate Processes, such as the JoinProcess. Processes define a main thread of execution, and a set of CommandSequences, which are sessions of questions and answers between two or more peers for the purposes of negotiating information, or affecting some (iterative) change to the configuration. Fundamental process CommandSequences, used by the join process are 'JOIN_PEER' and 'DO_YOU_HAVE_A_FREE_TOPCON' (to use their descriptions in the code base). Although some CommandSequences are built into the Communicator class, they are generally defined in the process implementation (which are extensions of the Process class, see Section 5.2.2). The NetworkMediator uses the following method to initiate process CommandSequences. ('destination' is the first destination peer: the CommandSequence may send queries along a chain of peers before the initiating peer receives a response (from a peer on the chain).)

```
Communicator.Ticket initiateProtocolCommandSequence(
                        PeerAdvertisement destination,
                        Process process,
                        String command,
                        Properties overridingArguments)
```

The NetworkMediator is also responsible for returning and updating lists of peers that are on the P2P network, or those that are participating in a particular IDA. The methods below are two such examples. IDAs are referred to as 'Services' in the IDAS code base.

```
Enumeration getAllPeerAdvertisements()
void updateListAppGroupParticipatingPeers(ServiceAdvertisement service)
```

## 5.3.2  The Topology Layer

TopologyConnections that have been satisfied by a node on a configuration are accessible from the application layer by Ports (from the Port class). Ports allow Nodelets to listen for incoming transmissions on a configuration connection, or to send transmissions to a connection. Ports are automatically generated and made available to Nodelets once the node has joined the configuration. Therefore Ports comprise the API between the topology and application layers.

The topology layer itself is responsible for running Processes and, across all nodes, establishing and maintaining configurations (see Section 4.2.2). The topology layer requires only a topology specification for each configuration to accomplish this. Topologies are implemented by extending the Topology class. In order to create a meaningful topology, instances of the Topology inner classes, NodeType and TopologyConnection, are referenced by the topology. Rigidly structured topologies must also refer to a class that extends the Locator class, which provides the topography for the structure. The Locator implementation makes concepts such as TopologyConnections at 0˚, 60˚, 120˚, 180˚, 240˚ and 300˚ (in the example in Section 4.1.3, p.52) meaningful. It also allows us to reference nodes that are 120˚ from a neighbour that is 60˚ from us, for example. A localised co-ordinate system is built around such precepts.

Important Topology methods to be overridden by Topology implementations are given below for illustrative purposes:

```
String getName()
NodeType[] getNodeTypes()
TopologyConnections getNodeConnections(NodeType nodeType)
HashMap getNodeConnectionWiring(  NodeType remoteNodeType,
                                  NodeType localNodeType)[63]
```

The methods above illustrate the fact that every topology requires a name, a list of NodeTypes in the topology, a list of TopologyConnections that must be satisfied for every NodeType, and a mapping between TopologyConnections and their reciprocal connections, which is termed ConnectionWiring in the code base (See Section 4.1 for the formal development of these concepts.)

---

63  Dynamic modifiable data structures such as Vectors and HashMaps are sometimes used for convenience in the IDAS, where a custom fixed structure (new java class) would be more logical from a memory allocation point of view. For example, ultimately a ConnectionWiring class or 'array of ConnectionWires' will be implemented and used in place of a HashMap here.

```
Locator.Address[] getAddressesContingentOn(
                        NodeType source,
                        TopologyConnection sourceCon,
                        NodeType destination,
                        TopologyConnection destinationCon)
```

Rigidly structured Topologies require mutually contingent TopologyConnections and for each TopologyConnection to be addressable from any of the others. Inner classes representing the concepts Address, Coordinates and Direction are given by the Locator class (see below).

```
TopologyMetrics getTopologyMetrics()
NodeParticular[] getNodeParticulars()
Restriction[] getRestrictions()
```

TopologyMetrics, defined by the topology, are used to specify attributes of a potential node for a given free TopologyConnection that are desirable for a given node to connect to. NodeParticulars give the relevant attributes for each node and TopologyConnection that are used to evaluate the best combination (based on TopologyMetrics). Restrictions rule out node connections where minimum attributes (determined by NodeParticulars) are not met. The functionality for proofing node connections based on Restrictions and selecting node connections based on TopologyMetrics is only partially implemented.

```
HashMap getNodeTypeChangeMappings()
```

The above method gives a map of the possible NodeType transformations a node can make over the course of its participation in a configuration based on the topology. It exists to support functionality in the (unimplemented) MaintainProcess that may specify that a node changes it's NodeType. For example if a node were to be 'promoted' or 'demoted' to or from super node status on a mesh with some sort of hierarchy. Another, somewhat less flexible strategy, that does not require such logic to be built into the topology, is the idea of having joining nodes merely choose a NodeType based on the configuration scenario that is presented to it on joining, and not changing it for the duration of its participation. This strategy, although not allowing nodes to change type, may be sufficiently dynamic where the rate of node attrition is high enough to track the desired changes in the configuration.

Important Locator methods to be overridden by Topology implementations are given below for illustrative purposes:

```
Class getAddressClass()
Class getDirectionClass()
Class getCoordinatesClass()
```

These methods reflect the fact that each Locator class must refer to extensions of the inner classes, Address, Direction and Coordinates. TopologyConnections, in the case of rigidly structured topologies, have a 'direction' associated with them. Reciprocal TopologyConnections are 'opposite' in direction to these. Each neighbour node a given direction from a given source node, 'along' the source node's TopologyConnection (graph edge), is represented by coordinates, where the source node is the origin. Coordinates can be 'added' together (coordinates are additive), such that, given a particular origin node, another node arbitrarily distant on the configuration can be referenced. Coordinates can also be converted to (or returned by) 'addresses', which also specify a TopologyConnection at the destination node. Obviously coordinates and addresses are only relative to particular nodes (origins). These classes are facilitated by the Origin and Location inner classes, which are combinations of origin NodeType and TopologyConnection and destination (or location) NodeType and TopologyConnection respectively.

Other methods provided by the Locator class translate between these three classes, and facilitate the following two basic arithmetic operations between addresses:

Address add(Address A, Address B)
Address inverse(Address A)

Addresses are used to contact nodes at contingent TopologyConnections in a rigidly structured topology. (An address locates a given destination from a given origin. The inverse of that address locates the origin from the destination.)

### 5.3.3   The Application Layer

The application layer comprises a Nodeletset. There should be one Nodelet in the Nodeletset for each NodeType in the topology that it is designed to run on (see Section 4.2.3). The current implementation, based on an older working of the IDAF, uses an equivalent but deprecated mechanism. The Nodelet class provides the thread that runs what we have termed the Nodelet functionality. The actual Nodelet functionality itself is contained within three classes that are extended for each NodeType, NodeletNodeManager, NodeletEngine and, optionally, Interpreter. The NodeletNodeManager is responsible for processing and queueing incoming and outgoing inter-node communications, and provides input for the NodeletEngine. The NodeletEngine runs its own thread and performs the core functions of the Nodelet, as described in the IDAF. The Interpreter is an optional class that provides for the interpretation of instructions passing between nodes.

Important methods that must be overridden by implementations of NodeletNodeManager are:

NodeletEngine getNodeletEngine()
void manageDataFlows()
void releaseDataFlows()

The first method indicates the associated NodeletEngine. The second and third manage flows of incoming and outgoing traffic from and to other nodes. Data flows must be released while the NodeType, and hence the Nodelet, is being changed, by a potential MaintainProcess.

Important methods that must be overridden by implementations of NodeletEngine are:

void run()
void finish()
void addToIncomingQueue(Object object)
void addToOutgoingQueue(Object object)
Object takeFromIncomingQueue()
Object takeFromOutgoingQueue()

The first method executed and performs the logic of the NodeletEngine thread, while the second causes the thread to end gracefully. The next four methods manage internal queues of objects that may feed, or be fed by, the NodeletNodeManager.

## 5.4  IDA Context and Regulatory Processes

### 5.4.1  IDAS Context

As is evident from the IDAS threading model, in Section 5.2.3, each IDA configuration (in the topology layer) is maintained in a separate executable context, and likewise, each Nodelet on such a configuration (in the application layer) also operates in a separate executable context to other Nodelets on other configurations, running on the same peer.

All state information pertaining to configurations and Nodelets, such as objects and variables, persist within their own context, and therefore there can be no interference between running IDA components. This also ensures that IDA management becomes no more complex as the number of IDAs that the user simultaneously participates in increases. What minimal management is required across IDAs is provided by the Container, which is the environment in which all IDAs are instantiated and moderated (see Section 5.2.2 and 5.2.3).

For these reasons, the processes described in Sections 5.4.2 and 5.4.3 run in the context (executable and otherwise) of particular IDAs.

### 5.4.2   The Join Process

The join process is responsible for satisfying the REQ, and possibly the OPT, TopologyConnections (see Section 4.1.2) of a joining node of a particular NodeType.

The join process is conducted by the JoinProcess class, in conjunction with functionality built into the Communicator class. (See section 5.2.2 for IDAS class structure.)

The join process performs the following steps:

1. Pertinent details from the topology (software) specification (see Section 5.3.2) are queried, including the default NodeType for a potential new node (or a joining node) on the configuration. (The 'default' NodeType can either be dynamically determined from configuration state information or can be hard coded into the topology.)

2. The join process requests a list of 'participating peers' from the NetworkMediator class. These are peers participating the running of the IDA that the joining node is attempting to connect to.

3. The join process then attempts to satisfy, first the REQ TopologyConnections, and then the OPT TopologyConnections, of the default NodeType. It does this by completing the following steps for each TopologyConnection:

   (a) From the list of participating peers, it finds peers which can satisfy the TopologyConnection. (Peers with a free reciprocal connection.)

   (b) Those peers are ranked in order of 'desirability'. (Not all nodes/peers are equal, so the best are chosen Based on TopologyMetrics and NodeParticulars. See Section 5.3.2.) Retaining the top 5 nodes (adjustable), since there may be many candidates.

(c) If the TopologyConnection has no contingent connections, then the process goes straight to step (d). Otherwise, the join process checks to see if the contingent TopologyConnections can also be satisfied. If not, the candidate node is removed from the list. The remaining candidates are ranked again taking into account their sets of contingent peers.

(d) A formal join is then attempted for that TopologyConnection, and any mutually contingent TopologyConnections, with the best candidate node(s). If, for some reason, the join cannot be made, the next best candidate, or candidate connection set, is chosen from the list. (One reason why the nodes may be unavailable for a join, is if another join, concerning one or more of these nodes, is currently in progress.)

4. If all REQ connections cannot be satisfied, then already satisfied connections are released, and the join is deemed to have failed. (With the option of repeating the process.) Otherwise, the node becomes a full participant in the IDA and has secured a place on the configuration.

This algorithm is unlikely to be optimal for discovering the best combination of nodes for a joining node to connect to. For example, outside of each individual TopologyConnection, or mutually contingent set of TopologyConnections, candidate nodes are not chosen based on what is best for all TopologyConnections, since they are satisfied individually, or in mutually contingent sets, without regard for the other TopologyConnections or sets (in REQ or OPT). It is also true that more intelligence could be built into this algorithm in order to make it more persistent in attempting to join the configuration. For example, it may be beneficial to go back to an early step in the process when something goes wrong, or when a poor choice has been made (such as when no ranked nodes will accept a join request), however, the present algorithm has only the option of restarting the join process. The process could perhaps benefit from being based on a finite state machine model, rather than merely a linear sequence of steps.

However, the join process, as it is currently implemented, is capable of reliably establishing configurations in test environments (see Section 6.4) and therefore suffices for the IDAS prototype.

### 5.4.3 Other Processes

The IDAF introduces three processes in Section 4.3.1: the join process, the maintain/optimise process and the evolve process. Processes are implemented by extending the Process class (see Section 5.2.2). Only the join process was implemented by the IDAS at the time of writing (see Section 5.4.2), however we discuss the latter two here.

The three main concerns of a maintain/optimise process implementation is as follows:

1   The state of nodes of given NodeTypes may need to be redundantly held by other nodes in the event of sudden failure. (Nodelets could, in the future, implement dumpState() and restoreState() methods, allowing a Nodelet's state to be captured, transferred and restored, perhaps on another node. Two running Nodelets with identical 'state' information would be said to be identical.)

   1.1   If a node is suddenly removed from the network, its last recorded state can then be restored onto another node: either a node with 'less important' state, or a new joining node. Alternatively, this state, if it comprised a 'partially worked-out problem' might be amenable to being merged with the state of another node, or to being subdivided, where its parts could be merged with the states of a number of other nodes. (Whether or not this is possible, and how this would be achieved, would be dependent on the Nodelet, and could be ascertainable from implementations of new Nodelet methods, such as boolean isStateDivisible(), StatePart[] divideState(int num), void mergeState(StatePart part), and so on.)

2   In the event of sudden node failure or disappearance, there may be REQ TopologyConnections on a node on the configuration that are no longer satisfied. This type of failure could be tackled by using strategies such as:

2.1 'Promoting' a node on which no other node (or less nodes) rely, to satisfy REQ TopologyConnections in the new 'position' on the configuration (albeit at the expense of other less important nodes having their REQ connections severed[64]). (Once the node has been promoted, it could optionally have the previous node's last known state restored to it, as describe in 1 and 1.1 above.)

2.2 Each node that has a REQ TopologyConnection that is no longer satisfied could cease operating normally until a new node joins the configuration that could replace it. (This would not be suitable for topologies where these 'frozen' nodes may in turn be satisfying REQ connections for other nodes, which would be obliged to 'freeze' themselves; this affect, recursively applied, could freeze a large part, or all, of a configuration.)

3 It may be possible for a configuration to become more optimised by moving nodes throughout the structure, as we mentioned in Section 4.3.1. For example, a type of tree topology might be considered to be more optimised, at times, by becoming more 'broad' or more 'deep', adjustable by the level of branching at each node. Such macro effects must be brought about by the actions of this process running on each node. For example, to make a tree more deep, each node may reduce its number of children by picking a child node at random, and compelling it to re-attach to the tree, and reducing its maximum number of children. (Nodes will have to re-attach at a point on the tree further away from the root.) To make a tree more broad, each node might accept an additional child node, and child nodes might be obliged to detach themselves from their parent node, in favour of a parent node closer to the root of the tree, bringing their children with them. In each case, nodes want to attach as close as possible to the root, allowing a maximum number of child nodes to attach to it.

3.1 If, as is the case above, the problem of optimisation can be re-stated in terms of the individual wants and needs of nodes of given NodeTypes, then optimisation can occur completely in parallel and is trivial. Where this is not the case, optimisation may be considerably more complicated, or (if it cannot be achieved with local knowledge only (see Section 4.1.1, p.41)), impossible. (Whether or not optimisation can be re-stated in such terms is wholly dependent on the topology structure.)

---

64 If there are no nodes that can be promoted to replace the failed node, nodes with unsatisfied REQ connections will either be forced to leave and reconnect to the configuration, or to follow a strategy similar to the one outlined in 2.2 below.

Suggested maintenance and optimisation strategies are given for testcase IDA topologies in Section 6.3.

The evolve process is introduced in the IDAF in Section 4.3.1. It is currently unimplemented and has not been investigated in any great detail. It is mentioned only as a place holder for future discussion on configuration evolution and where it might link into the IDAS or another IDAF implementation. The current thinking is that the process would be triggered by the maintain/optimise process, and would draw heavily from functionality taken from the topology design (see Section 4.3.1).

The evolve process would conduct a change in the structure of a configuration based on any two before and after structures. An example where this might be useful is as follows. In a Gnutella v0.4 [Gnutella '06] type graph structure, where each node connects to k other nodes, the network starts to become unusable as the number of nodes on the network, N, becomes large (hence the update to the protocol). The updated protocol [Kirk '06] makes the graph structure hierarchical by introducing 'super nodes' which normal node queries go through. It might be desirable for an IDA with a gnutella-like topology to begin by using the simple v0.4 protocol for the configuration structure, but to specify that the configuration be 'evolved' to the latter structure when N becomes large. ([Sacha '06a] [Sacha '06b] offers an alternative to the strict super-node/normal node divide, using a gradient of 'utility' rated nodes that assume a level of responsibility relative to their rating.)

It is not anticipated that the evolve process will be realisable or practical for all topologies, since many transitions, or evolutions, will not be possible in the context of nodes having local knowledge only.

In this chapter we described the IDAS, a prototype implementation of the IDAF. In the next chapter we introduce sample topologies and IDAs and use them to test the IDAS and draw conclusions about the IDAF.

# Chapter 6   Testing and Sample IDAs

The IDAS was built as a prototype to verify the IDAF. In this chapter we will describe the approaches taken for showing that an implementation of the IDAF, outlined in Chapter 4, is both realisable and useful, by reference to the IDAS, outlined in Chapter 5. The IDAS also constitutes a significant 'use case' for the JXTA (Section 2.7) P2P framework, which, although a technically mature P2P platform [Kaul '06], has seen slow adoption outside of the area of instant messaging and filesharing until now.

The approaches used to test the IDAS, are given in Section 6.1. In order to test the IDAS, it was necessary to develop sample (testcase) IDAs to would run on it. In order to run these IDAs, topologies had to be developed to support them. The topologies that were developed are discussed in Section 6.2. The IDAs (Nodeletsets) are discussed in Section 6.3. We give conclusions in Section 6.4.

## 6.1  Approach to Testing

### 6.1.1  Overview

Initially, it was necessary to test the JXTA platform by writing test code utilising the JXTA classes that would submit and discover advertisements to and from the network, join special PeerGroups, discover peers through their advertisements, make connections between peers using pipes, and propagate IDA (custom) advertisements throughout the network in order to advertise IDAs. This testing was aided by the JXTA Shell [JXTA Shell '06], an application that interacts with the P2P network (joining PeerGroups, discovering advertisements, establishing pipes, etc.), that was used as a diagnostic tool. This test code developed into the TrafficListener class, which in turn became the core of the IDAS network layer. (Thus the network layer was quite well tested at an early stage.)

After the JXTA code was working satisfactorily, IDAS development began in earnest. During its early development, it was tested by installing JXTA and the test code on to two machines on a Local Area Network (LAN) in the Engineering Department on the NUI Maynooth campus. The first machine would initiate an IDA and hence an IDA configuration, while the second machine would attempt to connect to the first, hence joining the configuration.

In the later phases of development, five machines were used for testing. One machine bootstrapped JXTA and loaded the shell application; the other four machines ran the IDAS client. It was found early on in testing that using the machine running the JXTA shell as a rendezvous peer [JXTAv2.0 '06]; a special peer that generates a PeerView and shares it with other peers; allowed the other peers to discover each other quickly and deterministically. (Otherwise it took peers varying amounts of time to discover other particular peers, either via a local network broadcast, or via large rendezvous peers on the internet. However, this problem now seems to be resolved, see Section 6.1.3: Advertisement Propagation.)

Ultimately, the IDAS (and therefore the validity of the IDAF) was tested by running testcase IDAs, using developed topologies, on the platform. These tests were carried out using the five machines, as above, with the occasional addition of machines from off campus, that were behind firewalls and NAT devices, in order to verify the viability of the JXTA based IDAS across a non-ideal network. (See Section 3.1.2 for the network difficulties associated with direct host to host communication.)

Tests were carried out on Linux, Windows and Mac OS X. There was no variation in the operation of the IDAS client based on computer platform or operating system.

The five standard machines used for the bulk of our testing were identical 2.4 GHz Pentium 4 desktop computers with 512Mb RAM.

### 6.1.2  Objectives

The IDAS was tested to fulfil the the following objectives:

1. To verify the JXTA P2P overlay network, and its suitability for use with a potential IDA platform.

2. To develop and debug the IDAS, which is a contribution of this work (see Section 1.3).

3. To verify the power and flexibility of the IDAF, which is a contribution of this work (see Section 1.3), by exhibiting credible demonstration IDAs, using developed topologies, built and run on a prototype system (the IDAS).

### 6.1.3 System Performance

We measured the performance of the IDAS on the five machines in our test environment. In the start-up measurements, the IDA clients would seem to perform quite poorly. However, performance was adversely affected here (and elsewhere) due to the fact that the graphical environments for the five machines were exported, via Unix/Linux X forwarding [Linuxquestions '06], across the network so that they could be remotely controlled, thus making testing easier and more systematic.

<u>Client Start-up Measurements</u>

Here we measure the time it takes for the IDAS client to bootstrap. After bootstrapping users can participate in IDAs. In order to join the JXTA P2P network, the user must authenticate part of the way through the bootstrapping process. Therefore we give the start-up times in the table below for before and after the password is entered. To give an indication of normal (non X forwarded) performance, the start-up times are also given for a laptop computer with typical (2006) specifications.

| Measurements | | **Test** | | **Laptop** | |
|---|---|---|---|---|---|
| | | ToPass | AfterPass | ToPass | AfterPass |
| (In milliseconds) | 1 | 22126 | 3302 | 4743 | 2723 |
| | 2 | 21877 | 5025 | 4723 | 1412 |
| | 3 | 22943 | 3019 | 4666 | 1085 |
| | 4 | 22751 | 4124 | 4801 | 3302 |
| | 5 | 23094 | 3838 | 4766 | 2097 |
| | 6 | 21625 | 3218 | 4607 | 2000 |
| | 7 | 22042 | 4390 | 4868 | 1931 |
| | 8 | 23992 | 3578 | 4993 | 1994 |
| | 9 | 21498 | 3246 | 4991 | 1877 |
| | 10 | 21928 | 4366 | 4942 | 1469 |
| Average (sec) | | **22.4** | **3.8** | **4.8** | **2.0** |

*Test*: Desktop machine: Pentium 4, 2.4 GHz, 512Mb RAM.
*Laptop*: Laptop computer: Mobile Pentium 1.4 GHz, 512Mb RAM.

We see that although the client was slow to start on the machines in our test environment: just over 26 seconds; this can be attributed to the X forwarding. A 'typical' client computer took just under 7 seconds in comparison.

## IDA Start-up Measurements

Here we measure the time taken for a node to join an IDA configuration, when it is the first node (i.e. to start the configuration), and the time taken for a node to join an IDA configuration when it is the second node (i.e. joins to connect to the first node). We used the star topology, given in Section 6.2.1, to obtain our measurements. (We ran the distributed computing IDA outlined in Section 6.3.1, but this is not relevant to our results since the measurements end at the point where the IDA component begins to function.)

The measurements are given in the table below. Once again, these times will have suffered somewhat from the forwarding of X sessions (see above).

| Measurements | | Node 1 | Node 2 | |
|---|---|---|---|---|
| (In milliseconds) | 1 | 9379 | 9977 | (Node 1 = Root) |
| | 2 | 10007 | 7438 | (Node 2 = Leaf) |
| | 3 | 8182 | 12903 | |
| | 4 | 8413 | 10223 | |
| | 5 | 8591 | 11156 | |
| | 6 | 9160 | 8082 | (* = These times are |
| | 7 | 13263* | 10176 | higher for node 1 because |
| | 8 | 12605* | 9081 | it is attempting to make |
| | 9 | 11738* | 9394 | contact with a client that |
| | 10 | 13345* | 10711 | has recently shut down -- |
| | | | | unbeknownst to it.) |
| Average Times (sec) | | **10.5** | **9.9** | |

As measurements for Node 1 for measurements 7-10 indicate (see note in table), clients in the IDAS current takes some time trying to contact nodes that have recently disappeared. Aside from that, the joining times for Node 1 and Node 2 are quite similar.

## Advertisement Propagation

Efficient advertisement propagation is an important aspect of JXTA's performance, and hence that of the IDAS. (See Section 2.7 for an introduction to JXTA.) (JXTA's general performance is examined in [Halepovic '03] and its communication layers in [Antoniu '05] and [Seigneur '03].) We sought to measure the the time taken for messages to propagate from one peer to another on our test network. On an older version of JXTA that we used, advertisement propagation and discovery was quite hit and miss. However, in recent tests, on JXTA v2.3, and using an 'always on' rendezvous peer (running the JXTA shell) for stabilising message passing, message propagation appeared instantaneous

We took two machines from our test bed, deleted their advertisement cache, and started the IDAS client on both machines. IDA advertisements (in our demo setup) are either already in the peer's advertisement cache or they are explicitly advertised by one peer and discovered by the others. So for our test, we sought to time how long it took for an IDA advertisement to be advertised by one peer and discovered by an other. An IDA advertisement is seen to be discovered when it appears in the pull down menu of available IDAs on the peer's client GUI. The results for the IDA advertisement would apply to all JXTA advertisements, since all advertisements are handled equivalently.

What we discovered was that advertisements were discovered almost instantaneously, and in some cases were discovered on the 'discovering' peer even before they were discovered on the 'advertising' peer. This counter-intuitive result stems from the fact that loops in the IDAS code that listen for advertisements were not tight enough to catch the new advertisements before they had propagated to the second peer and had been caught by that peers listener code first. (The main culprit here being the 200ms loop that updates the IDA list pull-down menu.) We therefore concluded that advertisement discovery was not an issue on LANs with a stabilising rendezvous peer.

We repeated the same tests without the rendezvous peer and obtained the same results. We conclude, therefore, that JXTA advertisement propagation on LANs has become a non-issue since we began this project in 2003.

<u>Data Transmission</u>

In the IDAS, data transmission occurs over unidirectional JXTA pipes. Considerable experimental data is given in [Antoniu '05] on JXTA pipes and data transmission.

In the IDAS, OutputPipes have a time-out of 400ms. This is because we had experienced problems with pipes hanging indefinitely when they failed to reach the corresponding InputPipe, or taking too long polling a 'dead' peer. (This former issue may be resolved in the current code base.) If an OutputPipe fails to connect after 400ms, it makes two more attempts before giving up. We have noticed that when the peers are overloaded with work, or when the network is congested (when external peers are involved), that occasionally messages are not successfully sent. However, this is quite unusual, particularly for small messages. (We have found that messages with payloads of 16k or less usually fine.) More work could be done here on tuning message sizes and OutputPipe time-outs, or using a more sophisticated pipe implementation.

## 6.2  Developed Topologies

The following topologies were built to facilitate the testcase IDAs given in Section 6.3. The topology structure specification given for the star, in Section 6.2.1, and the mesh, in Section 6.2.2, are built into extensions of the Topology class, and, in the case of the mesh (a rigidly structured topology) also the Locator class (see Section 5.3.2).

For both topologies, the topology implementation is given formally, in terms of the treatment in Sections 4.1.1, 4.1.2 and 4.1.3 (see pertinent aspects of the topology's structural specification, p.51).

### 6.2.1  The Star

The structure of the star topology in given in Section 3.3, p.34.

Defined sets (1):

T = { root, leaf }, O = { 'JXTA temp. pipe' }, D = { }, R = { }

S (in tabular form) (2):

| $S$ | t1 | t2 | o | d | q | f | r |
|------|------|------|------|------|----------|-------|------|
| R_to_L | root | leaf | pipe | null | $\infty$ | false | null |
| L_to_R | leaf | root | pipe | null | 1 | true | null |

REQ and OPT are given each NodeType below (in tabular form) (3):

| root | | leaf | |
|------|-----|------|--------|
| OPT | REQ | OPT | REQ |
| R_to_L | | | L_to_R |

Contingent TopologyConnection sets (4):

The star defines no sets of contingent connections.

<u>Logic for selecting NodeTypes for joining nodes (5):</u>

The first node to join the configuration is assigned the 'root' NodeType. All nodes thereafter are assigned the 'leaf' NodeType.

<u>Miscellaneous</u>

Since the maintain/optimise and evolve processes (Section 4.3.1) are not implemented, there is no functionality built into the star to take advantage of them. If the root node fails, the IDA fails. If a leaf node fails the configuration continues as normal, unless the root node relies on a response from that particular leaf node.

### 6.2.2   The Mesh

The structure of a planar mesh topology in given in Section 3.3, p.34.

The mesh will be a regular, planar mesh of degree 4.

<u>Defined sets (1):</u>

T = { node }, O = { 'JXTA temp. pipe' }, D = { north, south, east, west }, R = { cont_ref }.

<u>S (in tabular form) (2):</u>

| *S* | t1 | t2 | o | d | q | f | r |
|---|---|---|---|---|---|---|---|
| north | node | node | pipe | north | 1 | false | cont_ref |
| south | node | node | pipe | south | 1 | false | cont_ref |
| east | node | node | pipe | east | 1 | false | cont_ref |
| west | node | node | pipe | west | 1 | false | cont_ref |

REQ and OPT are given each NodeType below (in tabular form) (3):

| *node* | |
|---|---|
| OPT | REQ |
| north | |
| south | |
| east | |
| west | |

Contingent TopologyConnection sets (4):

The mesh defines one set of contingent connections identified by cont_ref as { north, south, east, west }.

Logic for selecting NodeTypes for joining nodes (5):

Every node is of 'node' NodeType.

Miscellaneous

Since the optimise and evolve processes (Section 4.3.1) are not implemented, there is no functionality built into the mesh to take advantage of them. If a node fails, the structure is damaged, and the IDA ceases to function properly.

## 6.3  Testcase IDAs & Demonstrations

We outline here the testcase IDAs that were developed to verify the IDAS.

In the IDAS, three classes are used in place of the Nodelet (see Section 5.3.3), introduced in Section 4.2.3. However we use Nodelet here to refer to the combined functionality of all three, i.e. that functionality that runs on the application layer, on each node, that is determined by the NodeType. See Section 4.2.3 for a discussion on Nodelets and Nodeletsets, NLD (p.55).

### 6.3.1 Star-based Distributed Computing IDA

A star based distributed computing testcase IDA was chosen for the following reasons: a) star topology structures are quite prevalent in existing IDAs, and so it is important to demonstrate one here; and b) to show that distributed computing IDAs (see Section 2.2) such as SETI@home and climateprediction.net can be built modularly, with low risk, and with relative ease, using a system based on the IDAF. (In addition, other advantages of the IDAF/IDAS, such as automated peer discovery and IDA advertisement, can also be leveraged.)

Description

The star based distributed computing IDA is called the 'Signal Processing' IDA. It performs a signal processing task. The first node (the root) manages the application. As nodes join, they automatically configure themselves into a configuration based on a star topology. Once the IDA is started, the system begins to process data (the workload is divided amongst the leaf nodes). Once all of the data has been processed, a result is produced. This scenario is suitable for all tasks that can be accomplished by one computer, but which can be subdivided arbitrarily into smaller tasks and run on a number of computers in parallel, such that the overall task is completed more quickly. Such tasks are termed 'trivially parallelisable'. This type of IDA is designed to mimic the operation of IDAs such as SETI@home and climateprediction.net.

In this case, the data is signal data. A discrete Fourier transform (DFT) is performed on this signal, which allows the spectrum of the signal to be analysed (as in SETI@home). The data is divided into blocks for distributed processing, and the result is then assembled from these processed blocks. ([National Instruments '06] gives an introduction to FFT based (block) signal analysis.) The testcase IDA merely performs a DFT of the signal blocks and provides no further analysis.

The IDA has two Nodelets in its Nodeletset, corresponding to the two NodeTypes in the star topology, root and leaf.

```
NLD = { rootNodelet, leafNodelet }
```

The rootNodelet

The rootNodelet presents a GUI to the user (see Figure 6.1). The GUI lists the number of leaf nodes that are attached to the root node. Clicking the 'Distribute data' button, divides the data into 'blocks', and distributes the blocks equally among the attached leaf nodes. Once the leaf nodes have processed the data and sent back the results, the rootNodelet assembles the data into a meaningful form. (In this case, the results can be examined to show that the DFTs have been performed.)

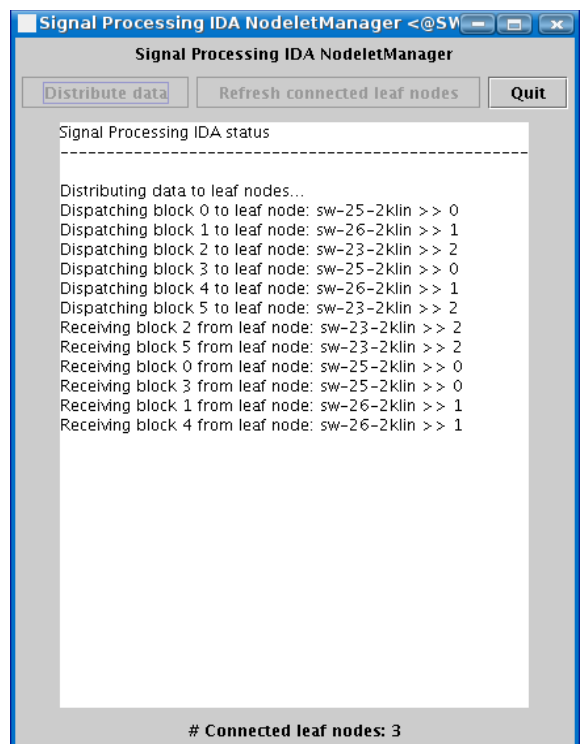

*Figure 6.1.: Star based signal processing IDA rootNodelet GUI*

The leafNodelet

The leafNodelet processes data sent to it from the root node. It simply performs a DFT on the data and sends it back. In the style of SETI@home, a GUI could be provided (not currently implemented) which would give the user information on the processing that is taking place (this information is available from examining text files). Additionally, updates to the entire process could be sent to leaf nodes from the root node on an occasional basis.

## Demonstration

The IDAS was bootstrapped on 4 machines (a fifth machine ran a JXTA shell, as described in Section 6.1.1).

Machines were joined to the 'Signal Processing IDA' one at a time. The first machine became the root node, while the subsequent 3 joining machines became leaf nodes. Clicking 'Refresh connected leaf nodes' on the rootNodelet GUI indicated that there were three leaf nodes attached (see Figure 6.1).

Clicking 'Distribute data', subdivided the data into blocks and sent them to the leaf nodes one at a time in a round robin fashion (as in CPU scheduling or pairing in games such as chess (tournaments) [Wikipedia '06e]). Each leaf node processed the data blocks and returned the results. Figure 6.1 shows a record of which peer was sent which data block. It also records the receipt of returned result blocks. Note that, as might be expected when sending data for processing to different machines, across the network, the result blocks are returned out of sequence. The blocks, sent in the order 0,1,2,3,4,5; are returned in the order 2,5,0,3,1,4.

The results can be verified easily by performing a separate DFT of the signal blocks and comparing them to the returned results. The IDA was given a simple signal to process: one composed of just 3 frequency components. We take the sum of 3 sine waves of differing frequency but equal phase. These three frequency components are represented by three impulses in the frequency domain (or rather 6 impulses, with the second three a reflection of the first through the point (F/2,0), where F is the number of frequency samples in the spectrum) [Smith '03]. (Where the signal is in discrete time, and where it is divided into blocks, there can be distortion [National Instruments '06].) Our sample signal is given below by x(t), while its frequency domain equivalent is given by X(f).

$$x(t) = \sin(\omega_0 t) + \sin(\omega_1 t) + \sin(\omega_2 t) \tag{12}$$

$$X(f) = 1/2\mathrm{j}(\delta(f - \omega_0) - \delta(f + \omega_0) + \delta(f - \omega_1) - \delta(f + \omega_1) +$$
$$\delta(f - \omega_2) - \delta(f + \omega_2)) \tag{13}$$

Graphically, the IDA can be seen to have performed correctly, by examining with DFT results, and comparing them to results obtained from a program such as Matlab or Octave. In the case above, where the input signal is a merely the sum of 3 frequency components, a visual examination is sufficient.

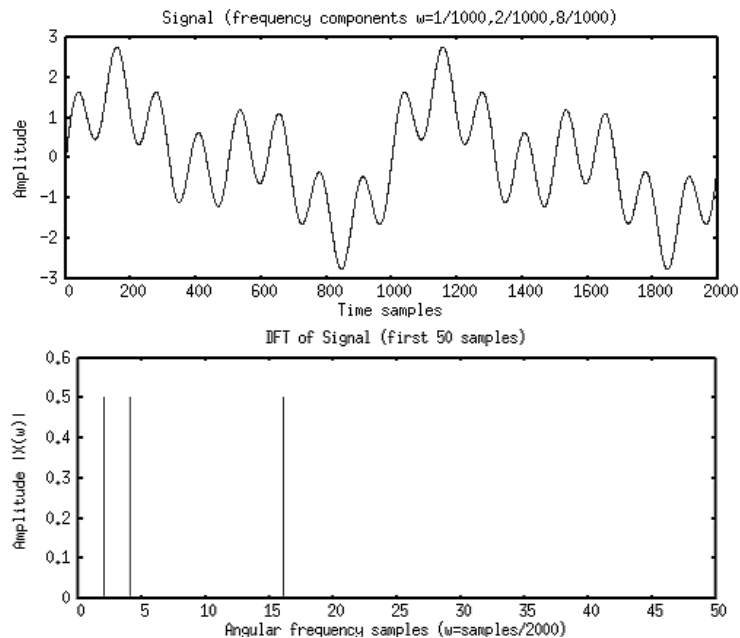

*Figure 6.2.: The input (time) signal, and (a portion of) the magnitude of the output (freq.) produced by the Signal Processing leaf node DFT.*

Given in Figure 6.2 is a block (each block is identical for the purposes of the test) of the original signal that is sent to the IDA (above). The (low) frequency components of this signal in the frequency domain, given by the DFT (below), can be observed within the first 16 samples.

Given in Figure are time measurements taken for how long it took the signal processing IDA to accomplish the same task natively, with one (leaf) machine, with two machines and with three machines. 10 blocks of 2,000 samples were distributed among the leaf nodes in each case in order to have a DFT performed on it. Each DFT was performed 8 times in order to intensify the computational task such that it was worthwhile to distribute it. (Simply increasing the size of the blocks would have been more natural, however large block seizes give rise to large JXTA messages which caused problems for message transmission. See Section 6.3.1 'Data Transmission' for issues with data transmission.)

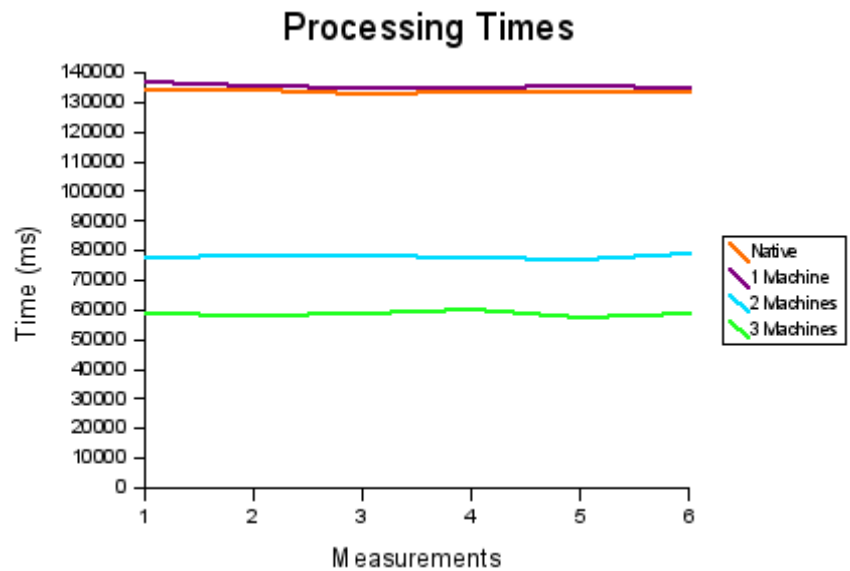

*Figure 6.3.: Processing time by number of computers for signal processing task*

This demonstration shows that signal processing applications which have hitherto been programmed separately, or (recently) for a platform such as BOINC [Anderson '04], can be built and run on a system implementing the IDAF with relative ease.

### 6.3.2 Mesh-based MMOG IDA

A mesh based MMOG testcase IDA was chosen for the following reasons: a) a mesh may provide a good basis for MMOGs in the future; and b) a mesh is a radically different structure to a star, and so provides a good demonstration of the versatility of the IDAF/IDAS.

<u>Description</u>

The mesh based MMOG testcase IDA is called 'Room Explorer'. It imitates the inter-client behaviour of an MMOG. Each node is a 'room' in an imaginary game environment. 'Characters' in the game environment navigate from room to room by leaving (a room) in the direction of north, south, east or west (although, naturally, the game environment is bounded by the outer edge of the mesh). Characters have sample actions available to them: *move*, *stay*, *pick-up* cookie and *drop* cookie.

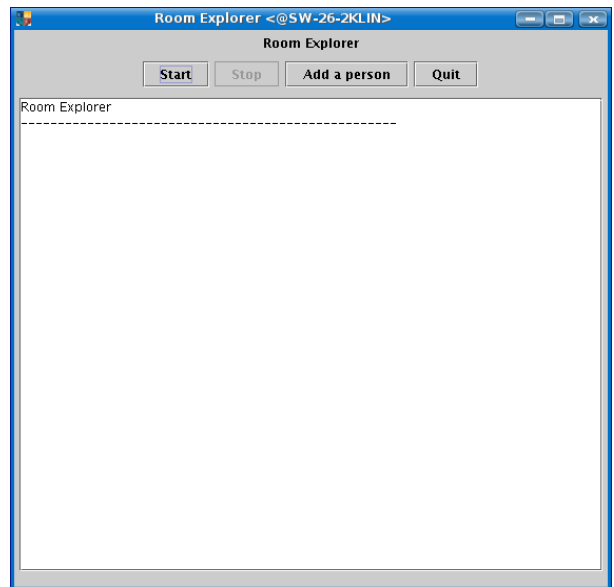

*Figure 6.4.: Room Explorer client GUI*

In each room there is a 'cookie jar', which can hold 'cookies'. Cookies are tokens that the characters carry with them while moving around the game world. The *move* action causes the character to move to another room in the game world, the *stay* action means that the character does nothing, the *pick-up* cookie action causes the character to pick up a cookie from the cookie jar (in a room) if there are any cookies in it, and the *drop* cookie action allows the character to place one of his cookies (if he has any) in the cookie jar.

To facilitate testing, the users do not control the characters in the game. Instead, the characters execute a random one of the four actions described, at a fixed time interval.

The IDA has one Nodelet in its Nodeletset, corresponding to the single NodeType in the mesh topology.

`NLD = { normNodelet }`

The normNodelet

The normNodelet presents a GUI to the user (see Figure 6.4). Aside from starting and stopping the client's interaction with the game (stop freezes the local game environment), the user can add a new character, or 'Person', to the game. That character can interact with the game as normal.

The normNodelet is responsible for the local game environment: that is the game environment within its 'room'. It is also responsible for transmitting characters moving through the game world to other nodes, or rooms. (Persons, along with the objects they carry (cookies), are serialized (Java Serialization API [Greanier '00]) and transmitted to other nodes using a north, south, east or west TopologyConnection.

In addition to managing the local game environment, the normNodelet also generates actions for the characters currently residing in its domain (at regular intervals).

### Demonstration

The IDAS was bootstrapped on 4 machines (a fifth machine ran a JXTA shell, as outlined in Section 6.1.1).

Given only 4 machines, we programmed a bias into the join selection process, such that the machines, as they were added, would connect themselves into the configuration as illustrated in Figure 6.5 a. (Figure 6.5 b. and c. give other possible configurations of nodes arising from the mesh.) All machines assumed the same NodeType.

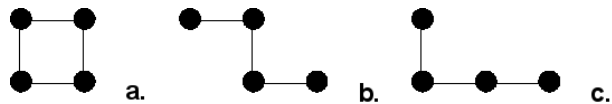

*Figure 6.5.: Possible configurations of 4 nodes arising from mesh*

For testing purposes, it is convenient for us to deal with one particular configuration. The configuration illustrated in a. also shows that the logic for establishing contingent connections is working normally, since the last joining node must connect to two nodes simultaneously.

The initial game conditions were as follows:

- All cookie jars (in all four rooms) were empty.

- One Person begins the game in each room.

- Each Person enters the game with 5 cookies.

The 'start' button on each of the clients was clicked, thus activating each part of the game environment. Each Person begins to complete actions at random, alternating between moving to other rooms, staying still, picking up cookies and dropping cookies. The 'add a person' button creates another person and locates him in the room corresponding to the node the GUI client is running on.

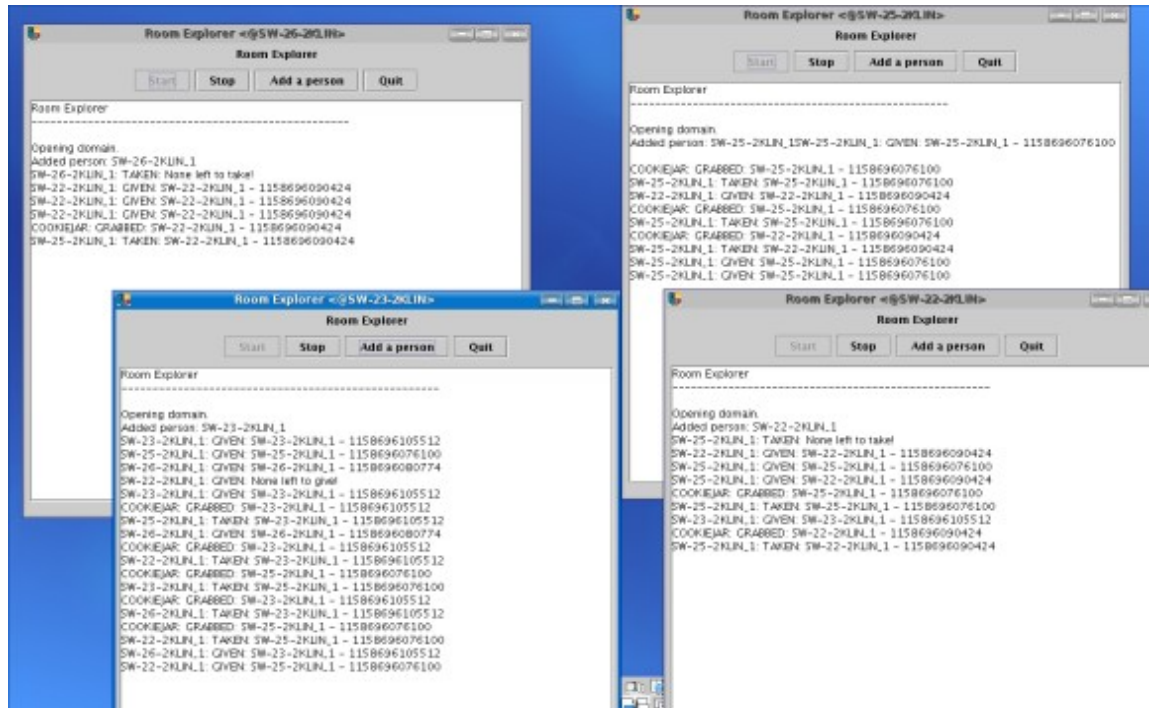

*Figure 6.6.: Screenshot of four clients remotely controlling Room Explorer software components running on 4 machines*

Figure 6.5 shows the four Room Explorer client GUIs running remotely on a single desktop, made possible by X forwarding. In each client window there is a log of the activity within the 'room' running on that node (which is also stored as a file). It is therefore possible to track the activities of each Person in the game, where they go and how the cookies distribute themselves throughout the game world, as they are carried by the characters from cookie jar to cookie jar. Thus a consistent game environment can be demonstrated.

This demonstration shows that mesh based MMOG applications are possible, and may be a strong prospect for the future. It also gives a further indication as to the versatility of the IDAF/IDAS, in its ability to support a range of IDAs with disparate topology structures.

## 6.4 Conclusions

### 6.4.1 General Conclusions on the IDAF/IDAS

The demonstrations in Section 6.3 demonstrate the following:

1.  The IDAS is performing its functions correctly.

2.  The IDAF is an implementable framework. Although the maintain process (Sections 4.3.1 and 5.4.3), necessary to preserve the configuration in uncertain network conditions, was not implemented, similar techniques apply to maintaining a configuration as to those used to maintain P2P overlay networks. ([Porter '06] discusses generic techniques that may be applicable to both.)

3.  The IDAS is capable of running IDAs with vastly different network configurations.

4.  It is possible to break down the design of an IDA into network, topology and application layers, thus making overall design easier and more extensible. Overlay network design, the definition of S for topology structures (see Section 4.1.2, p.47), and modular Nodelets; perform logical and powerful roles within the three layers, respectively.

5.  Since the demonstrations described in Section 6.3 can be run simultaneously, the following is true:

    a)  A single generic software component can run and manage a number of IDAs running simultaneously.

    b)  A single P2P overlay network can be used to accommodate a number of IDAs.

## 6.4.2   IDAS Scalability

Scalability in the IDAS is based on two factors: a) whether the network layer P2P overlay network can scale arbitrarily, and b) whether the topology can scale arbitrarily. Both the overlay network and the topology are pluggable components in the IDAF/IDAS.

a) should be one of the primary technical considerations of any P2P overlay network, or certainly of any that is suitable for use with the IDAS. This is the case with JXTA [Heiss '05], which we have used for the IDAS. [JXTAv2.0 '06] introduced enhancements to greatly improve scalability [Li '03], which was lacking in the 'flat' (no super peers) structure of JXTAv1.0. [Jan '06] recently performed a successful large scale (580 nodes at 6 sites) experimental evaluation of JXTA (March 2006).

b) Topology scalability is an issue for the specific topology. In Section 4.1.1, p.41, we said that topologies, and hence configurations, must operate on the basis of local knowledge only. By this we mean that each node should be concerned only with its only locality on the configuration. If we look at distances between nodes in terms of hops, where each neighbour is one hop away, and the neighbour of a neighbour is two hops away, and so on; then what should not increase as the configuration grows, is the 'radius' in hops away from it on a configuration, that a node must be concerned with. This is what we mean when we talk about operating with local knowledge only. However, even operating on this basis, does not mean that all topologies will scale. Within this 'radius' of nodes, the number of nodes may still become very large and untenable.

The mesh (k=4) is an example of a topology that can scale arbitrarily. It need only be concerned with its immediate neighbours ('radius'=1); that is, only 4 nodes, no matter how large the configuration grows. The star is an example of a topology that cannot scale arbitrarily. Although the root node in the star need only be concerned with its immediate neighbours ('radius'=1), the number of these immediate neighbours grows with the size of the configuration (in fact the entire configuration is its locality).

Since the IDAF/IDAS is merely an enabling platform for IDAs, and prescribes neither a custom P2P overlay network, or any topologies as part of its design, it does not make sense to 'measure' the scalability of the IDAF/IDAS.

### 6.4.3  Code Efficiency

In Section 4.2.4 we talk about the re-usability of code on the topology layer, as well as the advantages of using a common network layer for all IDAs. Here we look at the 'code efficiency' that arises from this, with reference to the demonstrations outlined in this chapter. What we mean by code efficiency is the savings for the developer in building his or her IDA on a system based on the IDAF rather than coding the IDA from scratch.

The IDAF/IDAS gives the glue that links together common components of all IDAs: the network layer, the topology structure code (the topology), and the code that actually performs the task for which the IDA was designed (the Nodeletset). We argue that this 'glue' will be required for all IDAs, in order to link the higher level functionality to the lower level functionality. Therefore, a portion of the overall work is already completed on the developer's behalf.

More importantly, however, it is anticipated that as time goes on, and as a system based on the IDAF is adopted increasingly, very few IDAs will be developed from scratch. This is because layer re-use is expected to be the norm rather than the exception. With the IDAF in place, and the ability to re-use a previously developed network or topology layer, most IDAs will require significantly less code to develop, because large portions of the overall code will have already been written by others.

As it stands, the IDAS uses a flexible P2P overlay network, in JXTA. This should suffice for most applications, and can be replaced by another if a more suitable one can be found. Having a suitable P2P overlay network that can be 'plugged' into an IDAF implementation allows for massive code savings, since good overlay networks are both difficult and time consuming to build. Furthermore, as topologies appear, the number of IDAs that can be built using existing topologies grow larger and larger.

Ultimately, if a developer can avoid implementing both the P2P overlay network and the topology, then he or she can focus solely on programming the actual task of the IDA (the Nodeletset), which could be a game or a computation. The main aim of a any supporting framework is to allow the developer to focus on the code that is relevant to their application. With sufficient uptake, the IDAF can make this possible.

In addition to the possibilities of modular code re-use outlined above, there is also scope for adapting code in one topology or Nodeletset for use in a new topology or Nodeletset. This is particularly promising in the case of Nodeletset development, where small modifications to code can produce large changes in IDA function. In Appendix A, we give an example of this, using the code for the signal processing IDA , given in Section 6.3.1 (shown are the leaf Nodelet and star topology, the root Nodelet merely obtains the input data and saves the results, and is not shown). Most of the code for the leaf Nodelet is re-used, and only a small code change is required to create an application that performs a completely different processing task.

In summary, the provision of a flexible and logical IDA platform based on the IDAF, in combination with the pluggability of network and topology layer components, and the potential re-usability of code across topologies and Nodeletsets; means that it will almost always be more efficient to implement an IDA on an IDAF implementation rather than from scratch, provided there is good uptake on an IDAF based system. (Unless there emerges a category of IDA which, for whatever reason, is not easily accommodated within the terms of the IDAF.)

To help illustrate the power and flexibility of the IDAS, we give the number of lines of java code required to implement the demonstrations given in Section 6.3. The small amount of code required for these demonstrations gives an indication as to how much work/code the IDAS saves the programmer.

1. The Star topology was implemented in only 270 lines of code.

2. The Mesh topology was implemented in only 315 lines of code. The accompanying Locator (that is required to provide the topographical definition in rigidly structured topologies, see Section 5.3.2), which can be re-used by other, similar topologies, was implemented in only 356 lines.

3. The Signal Processing Nodeletset was implemented in only 1,314 lines of code.

4. The Room Explorer Nodeletset was implemented in only 1,008 lines of code.

In this chapter we outlined a series of tests and demonstrations to verify that the IDAS functions correctly and that the IDAF is an implementable framework. In the next chapter we summarise the contributions of this work, and draw overall conclusions on and the future development of IDAs.

# Chapter 7    Summary & Conclusions

This chapter summarises the contributions of this work, identifies avenues for future work, and offers general conclusions on current and future IDA development.

In Section 1.3, we laid out the main contributions of this work. In Section 7.1 we give a summary of how these contributions were realised. During the course of the completion of this work, we identified areas of work which naturally follow on from it. We discuss possible future work in Section 7.2. During the course of our research we have analysed IDA technologies. We have formed certain conclusions on the state of this area, and on how IDA development should progress from here. In Section 7.3 we give these conclusions.

## 7.1  Summary of Work

Headings in this section refers to contributions of this work as stated in Section 1.3.

### 7.1.1  Analysis of Existing IDAs

In Chapter 2, we analysed important existing IDAs with a view to finding commonalities between them, and to harmonising what are often considered different technologies, or even fields. This process enabled us to develop a generic platform that is capable of running arbitrary IDAs. In Chapter 3, we took this analysis further in considering an IDA framework.

In Section 7.3.2 there is a note on the consolidation and synthesis of existing fields and technologies.

### 7.1.2   The IDAF

The IDAF was developed in Chapter 4, arising from analysis performed in Chapters 2 and 3. The IDAF gives us a new way of describing IDAs. Roughly speaking an IDA can be broken down into three intuitive blocks:

- An overlay network.

- A graph description of the topology structure (plus logic to maintain it).

- A set of Nodelets, one for each NodeType in the topology.

The IDAF describes, and outlines a set of constraints for, a generic platform that is capable of incorporating and running a number of IDAs simultaneously.

### 7.1.3   The IDAF Layer Model

The division of the IDAF model into three separate logical layers is a major advance on previous practice, where IDAs were developed as single monolithic software packages. Only with the introduction of BOINC [BOINC '06], midway through this project, was the first move toward generic platforms made, albeit for a range of distributed computing IDAs. The IDAF layer model simplifies IDA development and isolates logically unrelated functionality, so that each layer can be worked on independently. Developing for three layers (instead of one) also reduces IDA development project risk and increases code re-usability. Thus, the overall barrier to IDA development is greatly reduced.

### 7.1.4   The IDAS

The IDAS was outlined in Chapter 5, and is a prototype implementation of the IDAF. It was developed as a proof of concept for the IDAF. The pertinent aspects of the IDAF are implemented in this prototype, which is capable of running and managing arbitrary IDAs.

Sample IDAs, described in Chapter 6, were developed in order to test the IDAS, and prove the IDAF.

### 7.1.5 Analysis of the Field and Increased Edge Peer Participation

IDA technology has been extensively analysed over the course of this project. This analysis has informed the development of the IDAF, but has also culminated in a vision for the future of IDA development. This analysis is given in Section 7.3 and concludes this work.

## 7.2 Future Work

### 7.2.1 Enhancements to the IDAF/IDAS

In order to advance the IDAF as a viable framework for IDA platforms, it is necessary to present an industry strength system to the public. This system could be derived from the IDAS. Currently, the IDAS (being a prototype) lacks a number of the features and refinements required by a production system. These are listed as follows:

- The client GUI is not of professional standard, and only has very basic features. The GUI should give the user more useful information about the system status and more options for managing IDAs. IDA Nodelet GUI integration should be possible.

- The maintain process (see Sections 4.3.1 and 5.4.3) should be completed. The scheme outlined for the IDAS, on p.77, or an alternative scheme must be implemented.

- The system as a whole requires extensive testing, and documentation should be provided.

### 7.2.2 Mature Topologies

Topologies perform a significant role in the functioning of IDAs. They mandate not only the configuration structure, but also the logic for joining, maintaining and evolving the configuration (see Section 4.3.1). For this reason, topologies have a huge impact on the stability of a configuration, and hence on the stability of an IDA running on it. We have demonstrated that basic topologies can be provided with very little investment in time and resources, however, production grade IDAs will require mature, production grade topologies. For this reason, a given topology may go through a large number of iterations before it is considered 'mature' and is seen to give rise to robust configurations. Therefore topology design and implementation is seen as an area that will require a considerable amount of ongoing future work.

One of the key advantages of the IDAF is that Nodeletsets can be built for an IDA that uses a topology that is undergoing rapid development. Provided the structure of the topology remains unchanged ('S' in Section 4.1.2, p.47, along with the logic for evolving the structure, if implemented) the topology can be continuously upgraded without requiring any change to the Nodeletset.

### 7.2.3   Nodeletset Development

Nodeletsets perform the business of the IDA, with the infrastructure for running it provided by the network and topology layers. It is envisaged that as a fully featured and stable P2P overlay network becomes a slowly changing element of an IDAF implementation, and as a number of robust topologies become available; that much future work will be conducted at the application layer, on Nodeletsets.

Nodeletset development is intended to be extremely flexible for the scientist, application developer and enthusiast. It is hoped that a great many individuals with minimal programming skills will contribute at this level.

## 7.3   Conclusions

### 7.3.1   The Development of IDA Technology

Over the last number of years, the popularity of IDAs has been increasing strongly. The trend has been for a 'killer application' to appear, for it to enjoy extraordinary growth and popularity, and ultimately for it to be imitated and improved.

In the mid to late 90's, a number of landmark IDAs, or killer (internet) applications, appeared. Prominent examples include, instant messaging, which began to enter the mainstream in 1996 with ICQ [ICQ '06]; Ultima Online, arguably the first popular massively multiplayer online game, which appeared in 1997 [Kent '03];  Napster, the first file-sharing application [King '02], which appeared in 1999; and SETI@home, the first internet distributed computing project to enter the mainstream, which also appeared in 1999 [SETI@Home '06b]. These IDAs all spawned numerous successors: instant messaging is ubiquitous today, with many clients running many protocols; file-sharing applications have continued to prove extremely popular, accounting for an estimated 71% of all internet traffic [Ernesto '06]; and SETI@home (the software) was succeeded by BOINC [BOINC '06] and joined by numerous other BOINC based applications. Recently, internet telephony has become another 'killer application', with the introduction of Skype [Skype '06].

What is notable about the IDAs listed above is that, currently, each is often considered a separate technology. Each 'technology' tends to be the brainchild of an individual or a small group of people who have the vision to drive research in a specific direction, or to secure venture capital in order to realise their ideas. Two of the most popular IDAs in their time, Kazaa [Kazaa '06] and Skype, were developed by the same two men, Janus Friis and Niklas Zennstrom [Charny '03]. David Anderson, who developed SETI@home, also developed BOINC, 7 years later. We argue the fact that similar IDAs are currently considered separate technologies, that IDAs have emerged sporadically over the past decade, and the fact that a relatively small number of people have been able to make inordinately high contributions to IDA development, indicate that the IDA 'space' is still at an immature stage of its development.

One of the things that we have tried to do in this work is to define the IDA space and to contribute infrastructure that can help it move beyond this early stage of development. In this work, we have considered all of the technologies mentioned above to be facets of a single overall technology. However, the current environment is not conducive to the rapid and seamless proliferation of IDAs. The killer applications that we have discussed have failed to trigger mass IDA development. A small number of people have been successful in developing IDAs that have proved hugely popular, but IDA development, in a broader sense, has not achieved the critical mass that sees new applications appear on a continuous basis.

Considering IDA development as a unified technological field is valid from a technical perspective. We show this in Chapters 2, 3 and 4, where we give the commonalities between existing IDAs and ultimately develop a framework and platform on which they can be built and run. Further to this, we show that there is huge economy in treating IDAs as software bundles that can be run on generic software components, that can use the same P2P overlay network, and that can be managed, advertised and discovered by a single desktop client.

A useful analogy to the state of the IDA space at present is to liken it to the internet prior to the introduction of the WWW. Pre-WWW, a number of disparate internet services, such as email, newsgroups, public FTP site search tools and bulletin board systems were popular 'killer applications' for the internet. Although popular, these applications were considered separate, and there was no context in which to conceive of them as being part of an overall software offering. Arguably as a result of this, the internet changed very little for many years. With the arrival of the 'web', these applications eventually re-invented themselves within the new paradigm; web email became an alternative to 'regular' email; the text-based Archie and gopher gave way to Yahoo and Google; newsgroups and BBSs were largely replaced by message boards; and so on. However, in addition to these applications being thought of in a new wider context, the straight forward technology, and transparent goals, of the web, created a good breeding ground for other applications that could not have been previously foreseen. Examples include the online auction site, Ebay [EBay '06], a site where one can view movie trailers [Apple '06], and the interactive atlas, google earth [Google Earth '06]. It is our contention, that IDA technology still languishes at this earlier stage of development where the technology is uncoordinated and the the overall landscape is ill defined. This provides a de facto barrier to IDA development.

We have attempted as part of this work to outline the technical components and behaviour of IDAs, and the space that they inhabit. We have defined IDAs precisely in Section 1.5.1. We have detailed how each IDA can be broken down into network, topology and application layers. We have shown that the value of P2P is not in the absence of structure, but in the capability that it gives us to strip back the arbitrary structure of the TCP/IP internet, and replace it with a structure that is relevant and useful to the IDA. We have shown that network, topology and application layer developments can happen in parallel, that the advancement or redevelopment of one area ought not to impinge on the viability of the others. Significantly, overlay networks, topologies and Nodeletsets can be continuously reused. Where previously the IDA space was occupied by a number of monolithic, killer applications such as Napster or Bittorrent, which were developed individually from the ground up, and added only minimal definition to the space; it will now be possible to talk in terms of 'star based file sharers' or 'mesh based MMOGs'.

When IDAs can be described in terms of three well defined layers, it is envisaged that such powerful and transparent building blocks will enable IDA technology to go beyond a set of killer applications, and will give rise to a diverse eco-system of IDAs, similar to what has occurred with web applications on the WWW. When the space is well defined and the technology is made accessible, applications are far more easily conceived and developed. The scene is then ripe for IDAs to proliferate.

### 7.3.2 The Move towards an Edge Peer Empowered Internet

The majority of IDAs available on the internet today are provided by third party commercial interests or public agencies. Each message board must be hosted on a central server. Traditional Bittorrent requires a tracker server. MMOGs require game servers. Only in the case of P2P IDAs is a third party not strictly required. However, even in this case, users must often obtain proprietary software in order to gain access to a P2P network. Closed source clients must often be used in order to protect third party investment. These may sold for a price (particularly with 'premium' editions of the software), or they may be vehicles for advertising or spyware [BizReport '05]. The user may also have to interact with the third party website on a semi-regular basis in order to download never versions of the client.

We envisage an internet where users are free to build, run and distribute IDAs themselves, without any obligation to a third party for a service or for software (except their internet service provider). Most ordinary users[65] currently have poor all round internet connectivity, having no fixed IP address and having to go through a NAT interface to reach the internet (see Section 3.1.2), amongst other possible restrictions. They can consume internet content by downloading it, but it is difficult for them to participate fully in the internet's service offering. We say that ordinary users are therefore operating on the 'edge' of the internet, while servers, with dedicated internet connections, static IP address, and so on, are toward the 'centre' of the internet. Overlay networks, P2P and otherwise, are putting users back into the position of being able to provide content as well as being able to consume it, (where many of the disadvantages mentioned above are overcome). We call this process 'edge peer empowerment'.

---

65 For the purpose of this discussion, the ordinary user is one who connects to the internet via a service offered by an internet service provider (ISP), such as Digital Subscriber Line (DSL).

We argue that while IDAs are tethered to service providers, the IDA space will not achieve its full potential. The practical necessity for the third party in having to secure revenue for their P2P product, which would otherwise be free, forces the third party to insert itself between the user and the IDA network. We see this as being a retarding influence on the IDA space.

In this work we have introduced a framework for IDA development that we anticipate will make development and distribution logical and straight forward (see discussion in Section 7.3.1). In order for this framework to be viable, its layer APIs (see Section 4.2), at the very least, must be open source (definition: [Perens '97]), and the IDAF implementation 'network' must not be controlled by a third party. The IDAF can become a powerful platform given a high level of adoption. However, because the incentive for using an IDAF based system will be low in the beginning (when topologies are immature and Nodeletsets are few), proprietary control of system elements would create a significant barrier to popular public engagement with it. For this reason, we believe that the code base of an IDAF implementation should be open source. We believe that the IDA space, given that it is still at an early stage of its development, has not benefited from third party 'tolls', be they monetary, adware/spyware or restriction to the underlying source code and technology.

We envisage that a framework for IDA development (the IDAF), made possible by a cogent description of the IDA space, combined with the absence of third party control or interference, will give rise to a situation where ordinary internet users will provide a major potion of the IDA software that is developed. This will entail generating (or modifying) IDA Nodeletsets, (since overlay networks and topologies are expected to be in situ). The huge number of people that have designed their own web site using technologies such as HTML and JavaScript, give us significant grounds for optimism that the same phenomenon will occur with IDAs. It is also likely that rapid IDA (Nodeletset) development software could further simplify this process, in the same way as web authoring tools (e.g. Dreamweaver [Dreamweaver '06]) have simplified web development.

IDAs can be shared with others via advertising on the underlying P2P network (see Section 3.1.1). This form of distributed advertising of IDAs is an entirely new way of advertising IDAs. Advertisements may be published or discovered from peer groups (either the JXTA implementation or otherwise) and may be based on common interests or locality. There is therefore, significant potential for online communities [Inderscience '06] to form from groups of users with common interests, defined by their common participation in categories of IDAs. This would be similar to those that are formed around internet discussion boards dedicated to particular interests (e.g. Forum for Irish Anglers [Irish Angling Forum '06]).

In conclusion, we believe that the internet will become a lot less centralised in terms of service provision in the future. It will retain its centralised resources, but will also benefit from the 'empowering' of nodes at the edge of the internet. With many more nodes becoming involved in the internet's service offering, less internet capable computers will sit wastefully idle while their users are not busy interacting with them. Ordinary users will be able to design, develop, co-host and co-run their own IDAs, on an IDAF based network. They will advertise and discover IDAs in peer groups representing groups of users having common interests. This process will likely give rise to diverse online communities, which discuss, produce and consume categories of IDAs.

# Appendix A   IDA Code Re-usability

Leaving out the network layer implementation, which is based on JXTA, we give here the topology and application layer portions that make up the sample distributed computing IDA given in Section 6.3.1.

The code for the Star topology, given in Section 6.2.1, and the Signal Processing IDA, given in Section 6.3.1, can be found below. We show that, by altering only a very small portion of the code, the IDA can perform a completely different distributed computing task. Naturally, the topology remains unchanged, so there are no changes to Star.java. Similarly, there are no changes to the FFTNodeManager.java, which queues incoming and outgoing, data and results, to and from the root. The changes are confined to FFTEngine.java, which performs a simple DFT on blocks of time domain samples (each sample is a 'double' array of size 2, where index '0' gives the real component and index '1' gives the complex component). This class returns blocks of frequency domain samples.

## Star.java

```
/*
 * Star.java
 *
 * Created on January 12, 2005, 1:16 PM
 *
 * v0.9 on June 19, 2005, 11:09 AM
 */

package ie.nuim.eeng.IDAS.Topology.impl.Star;

import ie.nuim.eeng.IDAS.Topology.*;
import ie.nuim.eeng.IDAS.Topology.Topology.*;

import java.util.Properties;
import java.util.HashMap;

/**
 *
 * @author  merlin
 */
public class Star extends Topology {

    /*
     * Name of Topology.
     */

    private String name = "Star";

    /*
     * Node types for Star topology
```

```
     */

    protected static final String[] nodesTypesStrings =
                                        { StarConstants.ROOT_NODE,
                                          StarConstants.LEAF_NODE };

    public final NodeType rootNode = new NodeType(StarConstants.ROOT_NODE);
    public final NodeType leafNode = new NodeType(StarConstants.LEAF_NODE);

    public final TopologyConnection unlimitedToLeaf;
    public final TopologyConnection singleRequiredToRoot;

    protected final NodeType[] nodeTypes = { rootNode, leafNode };

    /*
     * Topology connections for NodeTypes
     */

    private final TopologyConnections rootConnections;
    private final TopologyConnections leafConnections;
    /*
     * Connection Wiring
     */

    private final HashMap rootToLeafWiring;
    private final HashMap leafToRootWiring;

    /*
     * Node change mappings
     */

    private final HashMap nodeChangeMappings;

    public Star() {

        /****************************************************************
         * CONFIGURATION OF ROOT NODE
         ****************************************************************
         */

        /*
         * Root nodes can have unlimited numbers of one generic connection
         */

        rootConnections = this.new TopologyConnections();

        unlimitedToLeaf =
                new GenericTopologyConnection( StarConstants.UNLIMITED_TO_LEAF_TYPE,
                                               new NodeType(StarConstants.LEAF_NODE),
                                               null,
                                               null,
                                               null,
                                               true);

        rootConnections.addConnection(unlimitedToLeaf, false);

        /****************************************************************
         * CONFIGURATION OF LEAF NODE
         ****************************************************************
         */

        /*
         * Any number of nodes may make one connection to the root
         */

        leafConnections = this.new TopologyConnections();

        singleRequiredToRoot =
                new GenericTopologyConnection(
StarConstants.SINGLE_REQUIRED_TO_ROOT_TYPE,
                                               new NodeType(StarConstants.ROOT_NODE),
                                               null,
                                               null,
                                               null,
                                               false);

        leafConnections.addConnection(singleRequiredToRoot, true);
```

```
        /*****************************************************************
         * CONNECTION WIRING
         *****************************************************************
         */

        rootToLeafWiring = new HashMap();
        leafToRootWiring = new HashMap();

        /* Local ROOT_TYPE to remote LEAF_TYPE */

        rootToLeafWiring.put(unlimitedToLeaf, singleRequiredToRoot);

        /* Local LEAF_TYPE to remote ROOT_TYPE */

        leafToRootWiring.put(singleRequiredToRoot, unlimitedToLeaf);

        /*****************************************************************
         * NODE CHANGE MAPPINGS
         *****************************************************************
         */

        nodeChangeMappings = new HashMap();

        nodeChangeMappings.put(leafNode, rootNode);
        nodeChangeMappings.put(rootNode, leafNode);
    }

    public String getName() {
        return name;
    }

    public NodeType[] getNodeTypes() {
        return nodeTypes;
    }

    public TopologyConnections getNodeConnections(NodeType nodeType) {
        TopologyConnections cons = null;

        if(nodeType.equals(rootNode))    // i.e. ROOT_TYPE
            cons = rootConnections;

        if(nodeType.equals(leafNode))    // i.e. LEAF_TYPE
            cons = leafConnections;

        return cons;
    }

    public HashMap getNodeConnectionWiring(NodeType remoteNodeType, NodeType
localNodeType) {
        HashMap connectionWiring = null;

        if(remoteNodeType.equals(rootNode) && localNodeType.equals(leafNode))
            connectionWiring = leafToRootWiring;

        if(remoteNodeType.equals(leafNode) && localNodeType.equals(rootNode))
            connectionWiring = rootToLeafWiring;

        /*
         * Other combinations of NodeType have null connectionWiring
         */

        return connectionWiring;
    }

    public Object[] getObjectsToBeCached() {
        return null;
    }

    public Restriction[] getRestrictions() {
        return null;
    }

    public NodeParticular[] getNodeParticulars() {
        return null;
    }

    public NodeType getJoinNodeType() {
        return leafNode;
    }

    public Object[] getConnectionsContingentOn(NodeType source) {
```

```
        return null;
    }

    public Locator.Address[] getAddressesContingentOn(NodeType source,
TopologyConnection sourceCon, NodeType destination, TopologyConnection destinationCon) {
        return null;
    }

    public Object[] getContingentGroups() {
        return null;
    }

    public Locator getLocator() {
        return null;
    }

    public HashMap getNodeTypeChangeMappings() {
        return nodeChangeMappings;
    }

    public NodeType getInitialConnectNodeType() {
        return rootNode;
    }

    public NodeType getFailedConnectNodeType() {
        return rootNode;
    }

    public NodeType getJoinedNodeType() {
        return leafNode;
    }

}
```

## FFTNodeManager.java

```
/*
 * FFTNodeManager.java
 *
 * Created on June 21, 2005, 11:36 AM
 */

package ie.nuim.eeng.IDAS.Nodelet.impl.SignalProcessingStar;

import ie.nuim.eeng.IDAS.Nodelet.*;
import ie.nuim.eeng.IDAS.Topology.Topology;
import ie.nuim.eeng.IDAS.Topology.Topology.*;
import ie.nuim.eeng.IDAS.Topology.impl.Star.*;

import java.util.HashMap;
import java.io.Serializable;

/**
 *
 * @author  markmclaughlin
 */
public class FFTNodeManager extends NodeletNodeManager {

    private final HashMap interpreters = new HashMap();

    private NodeletEngine nodeletEngine = null;

    public FFTNodeManager(  Nodelet nodelet,
                            Topology.NodeType nodeType) {

        super(nodelet, nodeType);

        setInterpreter(((Star)nodelet.getTopology()).singleRequiredToRoot, new
FFTInterpreter(this));
        setNodeletEngine(new FFTEngine(this));
    }

    public Interpreter getInterpreter(Topology.TopologyConnection topCon) {
        return (Interpreter)interpreters.get(topCon);
    }

    protected void setInterpreter( Topology.TopologyConnection topCon,
                                   Interpreter interpreter) {
```

```
        interpreters.put(topCon, interpreter);
    }

    public NodeletEngine getNodeletEngine() {
        return nodeletEngine;
    }

    protected void setNodeletEngine(NodeletEngine nodeletEngine) {
        this.nodeletEngine = nodeletEngine;
    }

    /*
     * Define some Runnables for our threads.
     */

    private abstract class QuitRunnable implements Runnable {
        public abstract void run();

        protected boolean quit = false;

        public void quit() {
            quit = true;
        }
    }

    /* Retrieve data from root node */

    private QuitRunnable retrieveDataFromRoot = new QuitRunnable() {
        public void run() {
            ie.nuim.eeng.IDAS.Testbench.PrintState.print(this, "FFT NodeManager retrieve
run()");

            Port portToRoot = getPortToRoot();

            Thread engineThread = new Thread(nodeletEngine);
            ie.nuim.eeng.IDAS.Testbench.PrintState.print(this, "Starting engine
thread.");
            engineThread.start();

            while(!quit) {

                /* Delay */
                try {
                    Thread.sleep(200);
                }
                catch(Exception e) {
                    e.printStackTrace();
                }

                Object[] objects = portToRoot.retrieveObjects();

                if(objects==null) continue;

                for(int i=0; i<objects.length; i++) {
                    if(objects[i]==null || !(objects[i] instanceof Serializable)) {
                        try {
                            if(objects[i]==null) {
                                throw new NodeletException("Null objects not allowed: "
                                                    + i + " of " + objects.length);
                            } else {
                                throw new NodeletException("Non Serializable objects not
allowed: "
                                                    + i + " of " + objects.length);
                            }
                        }
                        catch(NodeletException ne) {
                            System.out.println(ne.getMessage());
                            ne.printStackTrace();
                            continue;
                        }
                    }

                    //interpreter.addPayload((Serializable)objects[i]);
                    nodeletEngine.addToIncomingQueue(objects[i]);
                }

                if(quit) break;

            }

            if(quit) nodeletEngine.finish();

            ie.nuim.eeng.IDAS.Testbench.PrintState.print(this, "FFT NodeManager retrieve
```

```
run(): done");
            if(quit) return;
        }
    };

    /* Return results to root node */

    private QuitRunnable returnResultsToRoot = new QuitRunnable() {
        public void run() {
            ie.nuim.eeng.IDAS.Testbench.PrintState.print(this, "FFT NodeManager return
run()");
            Port portToRoot = getPortToRoot();

            while(!quit) {

                /* Delay */
                try {
                    Thread.sleep(200);
                }
                catch(Exception e) {
                    e.printStackTrace();
                }

                while(nodeletEngine.canTakeFromOutgoingQueue()) {
                    Object object = nodeletEngine.takeFromOutgoingQueue();
                    System.out.print("RET>>");
                    portToRoot.sendObject((Serializable)object);
                }

            }

            if(quit) nodeletEngine.finish();

            ie.nuim.eeng.IDAS.Testbench.PrintState.print(this, "FFT NodeManager return
run(): done");
            if(quit) return;
        }
    };

    private Port getPortToRoot() {
        String nodeType = StarConstants.ROOT_NODE;
        String name = StarConstants.SINGLE_REQUIRED_TO_ROOT_NAME;
        String type = StarConstants.SINGLE_REQUIRED_TO_ROOT_TYPE;

        Port portToRoot = ports.getPort(nodeType, name, type);

        return portToRoot;
    }

    private Ports ports = null;

    public void manageDataFlows() {
        ports = getPorts();

        Thread retrieveData = new Thread(retrieveDataFromRoot);
        Thread returnResults = new Thread(returnResultsToRoot);

        retrieveData.start();
        returnResults.start();

    }

    public void releaseDataFlows() {
        retrieveDataFromRoot.quit();
        returnResultsToRoot.quit();
    }

}
```

## FFTEngine.java

```
/*
 * FFTEngine.java
 *
 * Created on June 19, 2005, 11:47 AM
 */

package ie.nuim.eeng.IDAS.Nodelet.impl.SignalProcessingStar;
```

```
import ie.nuim.eeng.IDAS.Nodelet.*;

import java.io.Serializable;

/**
 *
 * @author  markmclaughlin
 */
public class FFTEngine extends NodeletEngine {

    /** Creates a new instance of FFTEngine */
    public FFTEngine(NodeletNodeManager nodeletNodeManager) {
        super(nodeletNodeManager);
    }

    private final Queue incomingQueue = this.new Queue();
    private final Queue outgoingQueue = this.new Queue();

    public void addToIncomingQueue(Object object) {
        incomingQueue.add(object);
    }

    protected void addToOutgoingQueue(Object object) {
        outgoingQueue.add(object);
    }

    private boolean timeToStop = false;

    public void finish() {
        timeToStop = true;
    }

    public void run() {
        ie.nuim.eeng.IDAS.Testbench.PrintState.print(this, "FFT NodeEngine starts.");
        while(!timeToStop) {

            while(incomingQueue.canTake()) {
                //Object[] samples = (Object[])takeFromIncomingQueue();
                Object sample = takeFromIncomingQueue();
                ie.nuim.eeng.IDAS.Testbench.PrintState.print(this, "sample class: " +
sample.getClass());
                Serializable[] samples = (Serializable[])sample;

                /* Have to unpackage doubles */

                ie.nuim.eeng.IDAS.Testbench.PrintState.print(this, "samples[0]==null: "
+ (samples[0]==null ? true : false));
                ie.nuim.eeng.IDAS.Testbench.PrintState.print(this, "samples[1]==null: "
+ (samples[1]==null ? true : false));
                ie.nuim.eeng.IDAS.Testbench.PrintState.print(this, "samples[0] class: "
+ samples[0].getClass());
                ie.nuim.eeng.IDAS.Testbench.PrintState.print(this, "samples[1] class: "
+ samples[1].getClass());

                Double[] realDouble = (Double[])samples[0];
                Double[] imaginaryDouble = (Double[])samples[1];
                /* sequenceCode passes straight through */

                int size = realDouble.length; // Arrays should be same size

                double[] real = new double[size];
                double[] imaginary = new double[size];

                for(int i=0; i<size; i++) {
                    real[i] = realDouble[i].doubleValue();
                    imaginary[i] = imaginaryDouble[i].doubleValue();
                }

                ie.nuim.eeng.IDAS.Testbench.PrintState.print(this, "DFT Start: " +
System.currentTimeMillis());

                doDFT(true, size, real, imaginary);

                ie.nuim.eeng.IDAS.Testbench.PrintState.print(this, "DFT End: " +
System.currentTimeMillis());

                /* Have to repackage doubles */

                for(int i=0; i<size; i++) {
                    realDouble[i] = new Double(real[i]);
                    imaginaryDouble[i] = new Double(imaginary[i]);
```

```
                }

                samples[0] = realDouble;
                samples[1] = imaginaryDouble;

                /* Finally we add to outgoing */

                addToOutgoingQueue(samples);

                if(timeToStop) break;
            }

            if(timeToStop) break;

            try {
                /* If there is nothing to do, wait. */
                Thread.sleep(1000);
            }
            catch(Exception e) {
            }
        }
        ie.nuim.eeng.IDAS.Testbench.PrintState.print(this, "FFT NodeEngine ends.");
    }

    protected Object takeFromIncomingQueue() {
        return incomingQueue.take();
    }

    public Object takeFromOutgoingQueue() {
        return outgoingQueue.take();
    }

    public boolean canTakeFromOutgoingQueue() {
        return outgoingQueue.canTake();
    }

    /*
     * Refills original arrays with transformed values
     */

    protected void doDFT(boolean forward, int size, double[] real, double[] imaginary) {
        double[] newReal = new double[size];
        double[] newImaginary = new double[size];
        int direction = forward ? 1 : -1;

        for (int i=0; i<size; i++) {
            newReal[i] = 0;
            newImaginary[i] = 0;
            double arg = - direction * -2.0 * 3.141592654 * (double)i / (double)size;
            for (int k=0 ;k<size ;k++) {
                double cosarg = Math.cos(k * arg);
                double sinarg = Math.sin(k * arg);
                newReal[i] += (real[k] * cosarg - imaginary[k] * sinarg);
                newImaginary[i] += (real[k] * sinarg + imaginary[k] * cosarg);
            }
        }

        /* Copy the data back */
        if (direction == 1) {
            for (int i=0; i<size; i++) {
                real[i] = newReal[i] / (double)size;
                imaginary[i] = newImaginary[i] / (double)size;
            }
        } else {
            for (int i=0; i<size; i++) {
                real[i] = newReal[i];
                imaginary[i] = newImaginary[i];
            }
        }
    }

}
```

As is evident from the highlighted (red) portion of the code above, only the portion directly pertaining to the DFT function need be changed in order to create an entirely different signal processing IDA. Below we give a method that could be substituted in for the highlighted portion above that would give an IDA that performed a mean and standard deviation of an input data block.

# Appendix A

One field where such results are useful is in the study of weather, specifically highest and lowest air temperatures in a particular locality. In this simple example, the results can be stored in two output doubles: one for the mean, and one for the standard deviation.

```
doMeanAndDeviation(size, signal);

...

    protected void doMeanAndDeviation(int size, double[] signal) {

        /* Mean calculation */
        double sum = 0;
        for (int i=0; i<size; i++) {
            sum += signal[i];
        }
        double mean = sum / size;

        /* Standard Deviation calculation */
        sum = 0;
        for (int i=0; i<size; i++) {
            sum += Math.pow(signal[i] - mean, 2);
        }
        double deviation = Math.sqrt(sum / size);

        /* Copy the data back */
        signal[0] = mean;
        signal[1] = deviation;

    }
```

The formulas used for the mean and standard deviation are:

$$Mean = m = \frac{1}{N} \sum_{i=1}^{N} x_i$$

$$Standard\ deviation = \sigma = \sqrt{\frac{1}{N} \sum_{i=1}^{N} (x_i - m)^2}$$